\documentclass[twocolumn]{aastex62}
\pdfoutput=1 %for arXiv submission

\usepackage{amsmath,amstext}
\usepackage[T1]{fontenc}
\usepackage[figure,figure*]{hypcap}

\usepackage{comment}
\usepackage{graphicx}
\usepackage{graphics}
\usepackage{color}
\usepackage[caption=false]{subfig}
\definecolor{dartmouthgreen}{rgb}{0.05, 0.5, 0.06}
\usepackage{natbib}
\usepackage{hyperref}

\def\arcsec{\hbox{$^{\prime\prime}$}}
\def\deg{\hbox{$^\circ$}}

\def\min{\textsuperscript{m}}
\def\sec{\textsuperscript{s}\hspace{-0.7mm}}

\def\nh{$N_{\rm H}$}
\def\nhm{\textit{N}_{\rm{H}}/\rm{cm}^{-2}}
\def\lx{$L_{\rm{X}}$}
\def\lxobs{$L_{\rm{X,\,Obs.}}$}

\def\lfour{$L_{22\,\rm{\mu m}}$}
\def\ltwo{$L_{4.6\,\rm{\mu m}}$}
\def\ltwel{$L_{12\,\rm{\mu m}}$}

\def\init{\hspace{0.75 mm}}

\def\nh{$N_\mathrm{H}$}

\def\chandra{\textit{Chandra}}
\def\xmm{\textit{XMM-Newton}}
\def\WISE{\textit{WISE}}
\def\nustar{\textit{NuSTAR}}
\def\swift{\textit{Swift}/BAT}
\graphicspath{ {Images/} }

\shorttitle{BASS-XXIII: Mid-Infrared Diagnostic for Absorption in AGN}
\shortauthors{R.\init W.\init Pfeifle et al.}

\begin{document}
\title{BAT AGN Spectroscopic Survey-XXIII. A New Mid-Infrared Diagnostic for Absorption in Active Galactic Nuclei}

\author[0000-0001-8640-8522]{Ryan W. Pfeifle}
\affiliation{Department of Physics \& Astronomy, George Mason University, 4400 University Drive, MSN 3F3, Fairfax, VA 22030, USA}

\author[0000-0001-5231-2645]{Claudio Ricci}
\affiliation{N\'ucleo de Astronom\'ia de la Facultad de Ingenier\'ia, Universidad Diego Portales, Av. Ej\'ercito Libertador 441, Santiago, Chile}
\affiliation{Kavli Institute for Astronomy and Astrophysics, Peking University, Beijing 100871, China}
\affiliation{Department of Physics \& Astronomy, George Mason University, 4400 University Drive, MSN 3F3, Fairfax, VA 22030, USA}

\author[0000-0001-9379-4716]{Peter G. Boorman}
\affiliation{Astronomical Institute, Academy of Sciences, Bo\u{c}n\'{i} II 1401, CZ-14131 Prague, Czech Republic}
\affiliation{Department of Physics \& Astronomy, Faculty of Physical Sciences and Engineering, University of Southampton, Southampton SO17 1BJ, UK}

\author[0000-0001-5146-8330]{Marko Stalevski}
\affiliation{Astronomical Observatory, Volgina 7, 11060 Belgrade, Serbia}
\affiliation{Sterrenkundig Observatorium, Universiteit Gent, Krijgslaan 281-S9, Gent, 9000, Belgium}

\author[0000-0003-0220-2063]{Daniel Asmus}
\affiliation{Department of Physics \& Astronomy, University of Southampton, SO17 1BJ, Southampton, Hampshire, United Kingdom}

\author[0000-0002-3683-7297]{Benny Trakhtenbrot}
\affiliation{School of Physics and Astronomy, Tel Aviv University, Tel Aviv 69978, Israel}

\author[0000-0002-7998-9581]{Michael J. Koss}
\affiliation{Eureka Scientific, 2452 Delmer Street Suite 100, Oakland, CA 94602-3017, USA}

\author[0000-0003-2686-9241]{Daniel Stern}
\affiliation{Jet Propulsion Laboratory, California Institute of Technology, 4800 Oak Grove Drive, Mail Stop 169-221, Pasadena, CA 91109, USA}

\author[0000-0001-5742-5980]{Federica Ricci}
\affiliation{Instituto de Astrof\'isica, Facultad de F\'isica, Pontificia Universidad Catolica de Chile, Casilla 306, Santiago 22, Chile}

\author[0000-0003-2277-2354]{Shobita Satyapal}
\affiliation{Department of Physics \& Astronomy, George Mason University, 4400 University Drive, MSN 3F3, Fairfax, VA 22030, USA}

\author[0000-0002-4377-903X]{Kohei Ichikawa}
\affil{Frontier Research Institute for Interdisciplinary Sciences, Tohoku University, Sendai 980-8578, Japan}
\affil{Astronomical Institute, Tohoku University, Aramaki, Aoba-ku, Sendai, Miyagi 980-8578, Japan}

\author[0000-0002-0001-3587]{David J. Rosario}
\affiliation{Centre for Extragalactic Astronomy, Department of Physics, Durham University, South Road, DH1 3LE Durham, UK}

\author[0000-0002-9144-2255]{Turgay Caglar}
\affiliation{Leiden Observatory, PO Box 9513, 2300 RA Leiden, The Netherlands}

\author[0000-0001-7568-6412]{Ezequiel Treister}
\affiliation{ Instituto de Astrofísica, Facultad de Física, Pontificia Universidad Católica de Chile, Casilla 306, Santiago 22, Chile}

\author[0000-0003-2284-8603]{Meredith Powell}
\affiliation{Department of Physics, Yale University, 217 Prospect St, New Haven, CT 06511, USA}

\author[0000-0002-5037-951X]{Kyuseok Oh}
\affiliation{Korea Astronomy and Space Science Institute, Daedeokdae-ro 776, Yuseong-gu, Daejeon 34055, Republic of Korea}

\author[0000-0002-0745-9792]{C. Megan Urry}
\affiliation{Department of Physics and Yale Center for Astronomy and Astrophysics, Yale University, 217 Prospect St, New Haven, CT 06511, USA}

\author{Fiona Harrison}
\affiliation{Cahill Center for Astronomy and Astrophysics, California Institute of Technology, Pasadena, CA 91125, USA}

%\correspondingauthor{Ryan W.\init Pfeifle}
%\email{rpfeifle@masonlive.gmu.edu}
\begin{abstract}
In this study, we use the \swift{} AGN sample, which has received extensive multiwavelength follow-up analysis as a result of the BAT AGN Spectroscopic Survey (BASS), to develop a diagnostic for nuclear obscuration by examining the relationship between the line-of-sight column densities (\nh{}), the 2--10 keV-to-$12\,\rm{\mu m}$ luminosity ratio, and \WISE{} mid-infrared colors. We demonstrate that heavily obscured AGNs tend to exhibit both preferentially ``redder'' mid-infrared colors and lower values of \lxobs{}/\ltwel{} than less obscured AGNs, and we derive expressions relating \nh{} to the \lxobs{}/\ltwel{} and \lfour{}/\ltwo{} luminosity ratios as well as develop diagnostic criteria using these ratios. Our diagnostic regions yield samples that are $\gtrsim80$\% complete and $\gtrsim60$\% pure for AGNs with log($\nhm{})\geq24$, as well as $\gtrsim85$\% pure for AGNs with $\rm{log}(\nhm{})\gtrsim23.5$. We find that these diagnostics cannot be used to differentiate between optically star forming galaxies and active galaxies. Further, mid-IR contributions from host galaxies that dominate the observed $12~\rm{\mu m}$ emission can lead to larger apparent X-ray deficits and redder mid-IR colors than the AGNs would intrinsically exhibit, though this effect helps to better separate less obscured and more obscured AGNs. Finally, we test our diagnostics on two catalogs of AGNs and infrared galaxies, including the \xmm{} XXL-N field, and we identify several known Compton-thick AGNs as well as a handful of candidate heavily obscured AGNs based upon our proposed obscuration diagnostics.\\
\end{abstract}

\section{Introduction}
\label{sec:intro}
Known to reside at the centers of most galaxies (e.g. \citealp{ferrarese2000,kormendy2013}), supermassive black holes (SMBHs) grow and evolve through periods of activity characterized by the accretion of large quantities of gas. Classically, these active galactic nuclei (AGNs) are  categorized based upon the characteristics of their optical spectroscopic emission lines, where the apparent differences between AGNs may be reconciled through a unification scheme involving a dusty obscuring torus (e.g. \citealp{antonucci1993,urry1995,netzer2015,ramosalmeida2017}), for which different inclination angles of the torus correspond to the observation of different AGN classes. 

One ubiquitous observational signature of accretion onto SMBHs is X-ray emission, produced very close to the accretion disk \citep{fabian2009nature} due to inverse Compton scattering of  optical and ultraviolet (UV) photons from the accretion disk by hot electrons in the corona \citep{haardt1991,haardt1993}. In the X-ray band, the line-of-sight gas column density, \nh{}, is largely transparent to the 2--10 keV X-ray flux even up to column densities of a few times $10^{23}$~cm$^{-2}$, however, significant attenuation and reprocessing of the 2--10 keV X-ray emission does occur for Compton-thick (CT) AGNs, which possess gas column densities of $\gtrsim10^{24}$~cm$^{-2}$ (e.g. \citealp{lansbury2014,ricci2015,bauer2015,puccetti2016,lamassa2019,toba2020}). CT AGNs even at low redshift have thus proven to be very difficult to find and characterize \citep{alexander2012} using lower energy X-ray observatories such as \chandra{} and \xmm{}, since the X-ray flux below 10~keV suffers significant photoelectric absorption and Compton scattering. This prevents the detection of some sources, and even those detected have fewer observed photons, which reduces the accuracy of spectral modeling. Important spectral signatures used to characterize the nature of AGNs can be missed without higher energy X-ray observations \citep{lansbury2015,kossapj2016,ricci2017apjs}.

%CT AGNs are particularly important among the general AGN population; CT AGNs likely represent a significant fraction of the total AGN population up to both $z\simeq0.1$ and $z\simeq1.0$ \citep[$50\pm9$\% and $56\pm7$\%, respectively,][]{ananna2019}, as such large fractions of CT AGNs are required to reproduce the observed Cosmic X-ray Background \citep[CXB,][and references therein]{ananna2019,buchner2015}. 

CT AGNs are particularly important among the general AGN population, as large fractions of CT AGNs are required to reproduce the observed Cosmic X-ray Background \citep[CXB,][]{gilli2007,buchner2015,ananna2019}. CT AGNs likely represent a significant fraction of the total intrinsic AGN population in the local Universe ($\sim$20\%, \citealp{burlon2011}; $\sim$27\%, \citealp{ricci2015}), with the most recent SMBH synthesis model developed by \citet{ananna2019} suggesting that CT AGNs represent $50\pm9$\% and $56\pm7$\% of the total intrinsic AGN population up to both $z\sim0.1$ and $z\sim1.0$, respectively, though many previous synthesis models predicted lower intrinsic fractions of CT AGNs than this, including but not limited to \citet{gilli2007,treister2009,akylas2012,ueda2014}. Furthermore, questions remain regarding the exact nature of the obscuring structure and how it relates to, for example, the host environment: some CT AGNs could represent the evolutionary phase predicted to occur as a result of galaxy mergers \citep[e.g.][]{hopkins2008a,kocevski2015,ricci2017mnras,blecha2018}. Identifying further cases of CT AGNs is crucial for providing a full census of accreting SMBHs, placing constraints on the CXB, as well as placing constraints on evolutionary and unification models.

UV radiation from the accretion disk is also reprocessed by the obscuring dusty torus, wherein the radiation is scattered and absorbed by the dust grains and re-emitted thermally with a peak usually at mid-infrared (mid-IR) wavelengths. While the classically accepted origin of the mid-IR emission is the dusty torus itself, recent high angular resolution infrared observations of AGNs suggest that the mid-IR emission is in fact dominated by a dusty polar outflow rather than the torus itself (e.g. \citealp{hoenig2012,hoenig2013,tristram2014,asmus2016,hoenig2017,stalevski2018,hoenig2019}). Other recent studies \citep{baron2019a,baron2019b} have actually attributed mid-IR emission to dusty outflows located on the order of tens to hundreds of parsecs from the centers of the galaxies.
%though some recent studies \cite{baron2019a,baron2019b} have exami... see also \cite{baron2019a,baron2019b} which examined dusty outflows ). 

%In the X-ray band, AGNs are described as ``obscured'' if the line-of-sight gas column density, \nh{}, exceeds $10^{22}$~cm$^{-2}$, though this obscuring material is largely transparent to the 2--10 keV X-ray flux even up to column densities of a few times $10^{23}$~cm$^{-2}$

A correlation between the intrinsic (unabsorbed) hard X-ray 2--10 keV luminosity ($L_{\rm{X}}$) and mid-IR luminosity ($L_{\rm{MIR}}$) of AGNs was reported as early as \cite{elvis1978}. Universally, studies find no difference ($<0.3$ dex) in the ratios of $L_{\rm{X}}$ to $L_{\rm{MIR}}$ between Type 1 and Type 2 AGNs \citep{lutz2004,levenson2009,gandhi2009,ichikawa2012,mateos2015,asmus2015} and this ratio is also insensitive to the neutral gas column density along the line-of-sight \citep{gandhi2009,mateos2015,asmus2015}, even in the case of CT AGNs after correcting the X-rays for absorption \citep{gandhi2009}. Many previous studies have also pointed out that this relation can serve as a useful tool to select obscured, particularly CT, AGNs, because CT AGNs tend to exhibit severe deficits in their absorbed X-ray emission when compared to their mid-IR emission, and thus CT AGNs fall significantly off of the $L_{\rm{X}}$-to-$L_{\rm{MIR}}$ relation \citep{alexander2008,goulding2011,georgantopoulos2011,rovilos2014,asmus2015}. Moreover, \citet{asmus2015} also demonstrated that the ratio of $L_{\rm{X}}$/$L_{\rm{MIR}}$ can be used to \textit{predict} column densities for significantly obscured objects, deriving an equation that relates $L_{\rm{X}}$/$L_{\rm{MIR}}$ to log($\nhm{}$) for log($\nhm{})>22.8$.

It is important to gather large samples of powerful AGNs across a range of column densities to test the utility of the X-ray to mid-IR relation as a tracer of nuclear obscuration. Hard X-ray selection provides one of the least biased methods of identifying powerful AGNs, as hard X-ray emission is largely unaffected by the line-of-sight obscuration for column densities $<10^{24}$~cm$^{-2}$ (see Fig.~1 from \citealp{ricci2015}). The \swift{} ultra-hard X-ray (14--195~keV) all-sky survey has dramatically increased the number of known hard X-ray extragalactic sources \citep{baumgartner2013,oh2018}, and has therefore been the focus of a large multiwavelength follow-up campaign (the BAT AGN Spectroscopic Survey, or BASS\footnote{\url{https://www.bass-survey.com/}}) designed to characterize the most powerful AGNs in the local Universe \citep{koss2017,ricci2017apjs,lamperti2017}. A second release of optical spectroscopy (BASS DR2) will also soon be publicly available (Koss et al., in prep; Oh et al., in prep). \citet{ricci2017apjs} presented a detailed X-ray analysis of 838 ultra hard X-ray detected \swift{} AGNs, providing constraints on column densities (\nh{}) as well as the absorbed (observed) and unabsorbed 2--10~keV luminosities, while \citet{ichikawa2017} provided mid-IR to far-infrared photometric data and corresponding IR luminosities for the 604 mid-IR-detected \swift{} AGNs. These two catalogs provide precisely the information needed to test the relationship between the absorbed hard X-ray and mid-IR emission with respect to the line-of-sight obscuration in AGNs.

In this paper, we present a new diagnostic for absorption in AGNs which combines the power of the known \lx{}/\ltwel{} correlation with \WISE{} mid-IR colors. Using multiwavelength catalogs available for the \swift{} hard X-ray-selected sample of AGNs, we show this diagnostic reliably identifies the most obscured AGNs, at least for nearby X-ray bright AGN. Our proposed diagnostics could prove valuable in the search for obscured AGNs in the ongoing eROSITA survey \citep{predehl2010,merloni2012}. In Section~\ref{sec:sampleselect} we describe our sample. In Section~\ref{sec:analysis} we describe the analysis of the sample and propose our new absorption diagnostics as well as develop expressions that constrain column densities. In Section~\ref{sec:discussion} we explore the emission ratios of optically-selected star forming galaxies, we compare our diagnostics for obscuration to other recent studies, and we apply our diagnostics to the \xmm{} XXL North field \citep{pierre2016,pierre2017}. In Section~\ref{sec:conclusions} we detail our conclusions. Throughout this manuscript we assume the following cosmological values: $H_0=70$~km~s$^{-1}$~Mpc$^{-1}$, $\Omega_{\rm{M}}=0.3$, $\Omega_\Lambda=0.7$. All luminosities quoted in this work are given in units of erg~s$^{-1}$. 

\section{Sample Construction}
\label{sec:sampleselect}
We selected our sample from the 70-month \swift{} X-ray properties catalog \citep{ricci2017apjs} which details the broadband 0.3--150~keV X-ray spectral properties of the 838 AGNs detected in the ultra hard X-ray 14--195~keV band by \swift{} and reported in the 70-month source catalog \citep{baumgartner2013}. We matched this catalog to the 70-month \swift{} infrared catalog of \citet{ichikawa2017}, which provides the complete near-infrared to far-infrared photometry for 604 \swift{} non-beamed AGNs at high Galactic latitudes (|b|>10\deg{}). We refer the reader to \citet{ricci2017apjs} and \citet{ichikawa2017} for further details on the construction of these catalogs. These catalogs yielded a parent sample of 604 non-beamed AGNs; any systems flagged as beamed in the \citet{ricci2017apjs} catalog were removed during the matching process.

In order to conduct our analysis, we required X-ray and mid-IR detections in all four \WISE{} bands for the AGNs in our sample, which excluded another 78 AGNs from the final sample\footnote{While detected in the \WISE{} $W1$[$3.4\,\rm{\mu m}$] and $W2$[$4.6\,\rm{\mu m}$] bands, 67 AGNs did not satisfy the \WISE{} data quality cuts established in Section~2.2.1 of \citet{ichikawa2017}. A further 11 AGNs did not possess detections at either $12\,\rm{\mu m}$ or $22\,\rm{\mu m}$. All AGNs within the parent sample of 604 AGNs possessed 2-10~keV X-ray detections.}. We adopted the hard X-ray 2--10~keV luminosities (observed, uncorrected for intrinsic absorption) from \citet{ricci2017apjs} and the infrared luminosities in all four \WISE{} bands ($3.6\,\rm{\mu m}$, $4.6\,\rm{\mu m}$, $12\,\rm{\mu m}$, and $22\,\rm{\mu m}$) from \citet{ichikawa2017}, which are not corrected for any host galaxy contributions. We further limited the sample to $z<0.1$ AGNs to avoid redshift effects; this redshift cut removed another 70 AGNs from the sample. The \citet{ricci2017apjs} catalog includes independent estimates of \nh{} from a torus model in the event that the column density found with the phenomenological model was $\geq10^{24}$~cm$^{-2}$. In these cases, we instead use the column density inferred from the torus model, because torus models more accurately account for the 2--10 keV emission of CT AGNs. 

Following the matching of the catalogs and the application of the above requirements, the final sample was composed of 456 nearby, non-beamed AGNs with a median redshift of $z\simeq0.032$ and mid-IR luminosities in the range $1.6\times10^{39}\leq L_{12\,\mu\rm{m}}/\rm{erg~s}^{-1}\leq8.5\times10^{44}$.

\begin{figure}
    \centering
    \subfloat{\includegraphics[width=1.\linewidth]{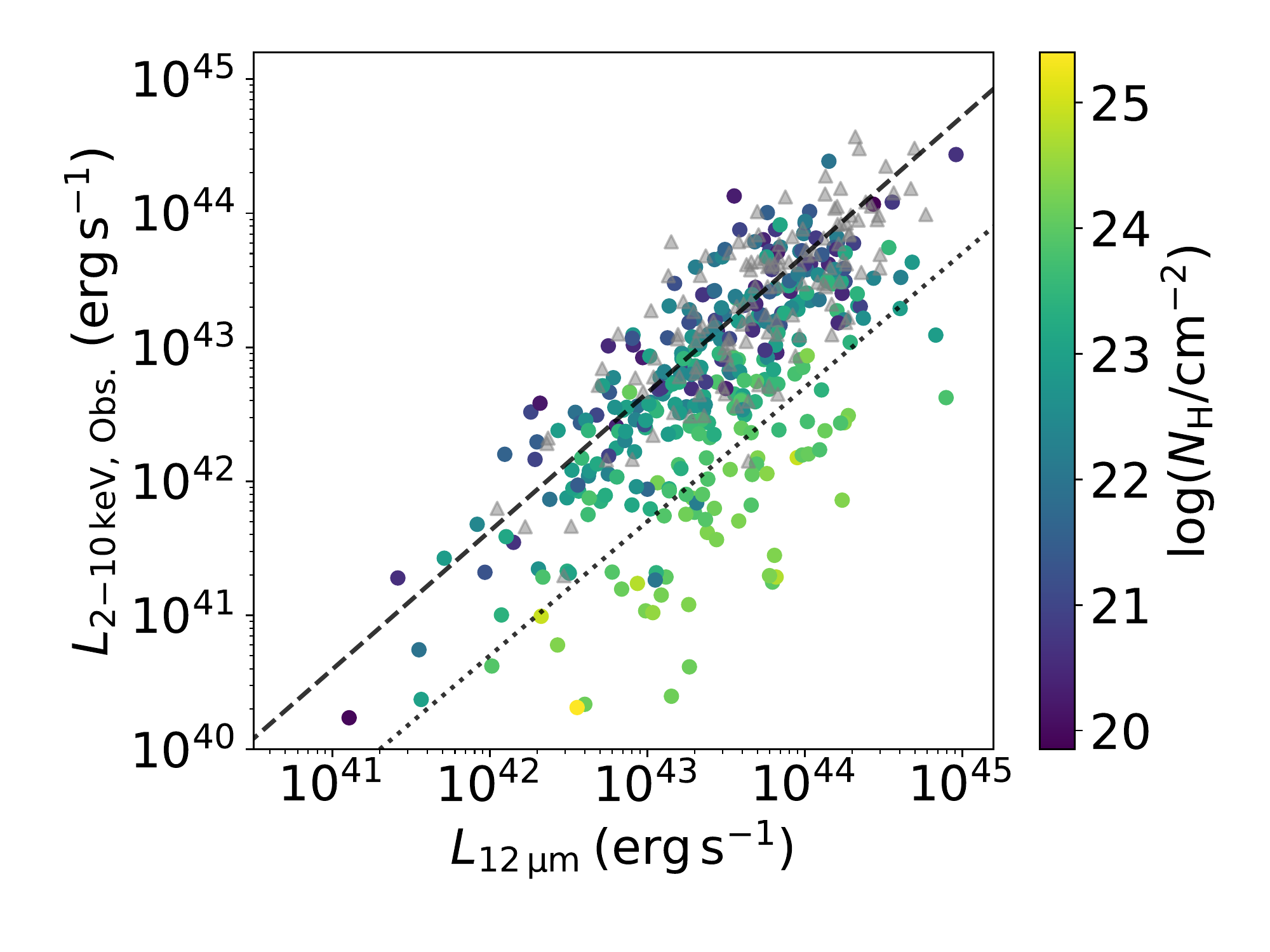}}\\ 
    \caption{The observed 2--10 keV X-ray vs. $12\,\rm{\mu m}$ luminosities. We  color code the data points using the derived column density from the \citet{ricci2017apjs} catalog, where the color map is denoted on the auxiliary axis. AGNs with \nh{} upper limits of $\rm{log}(\nhm{}) \leq 20.0$ are denoted with gray triangles. When comparing the 2--10 keV X-ray luminosity to the $12\,\rm{\mu m}$ luminosity for our sample of 456 \swift{} AGNs, we see a general decrease in the X-ray-to-mid-IR ratio with increasing column density, as expected. The relation between the AGN intrinsic 2--10~keV luminosity (corrected for absorption) and the nuclear 12\,$\mu$m luminosity derived by \citet{asmus2015} is represented by a dashed black line, whereas we plot the logarithmic ratio of log(\lxobs{}/\ltwel{}) $=-1.3$ with a black dotted line. Most heavily obscured (CT) AGNs exhibit luminosity ratios log(\lxobs{}/\ltwel{}) $<-1.3$.}
    \label{fig:LXLMIR_v_NH}
\end{figure}

\begin{figure}
    \centering
    \subfloat{\includegraphics[width=1.\linewidth]{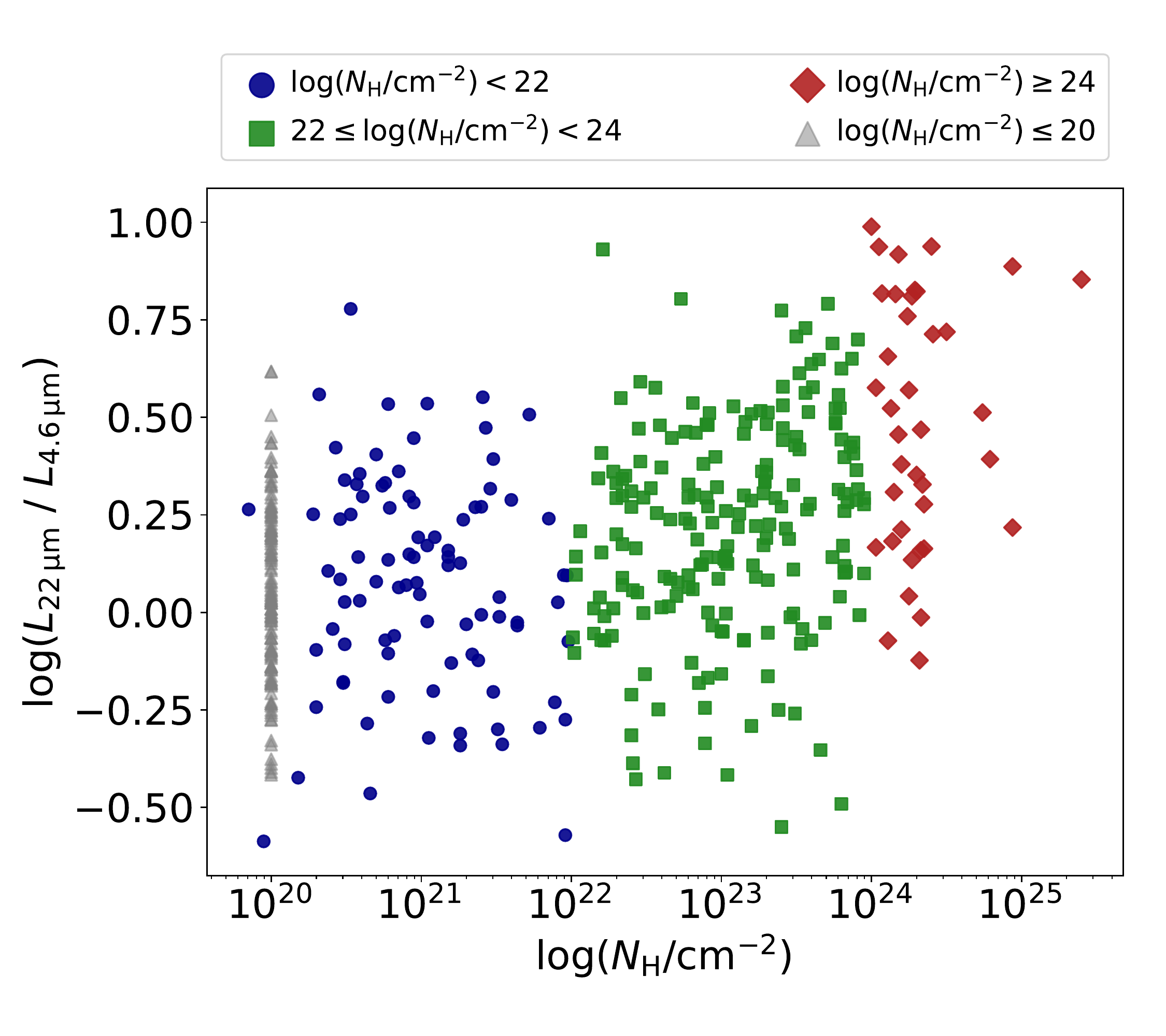}}\\
    \caption{The ratio of the $22\,\rm{\mu m}$ and $4.6\,\rm{\mu m}$ \WISE{} luminosities as a function of column density (reported in \citealp{ricci2017apjs}), and we differentiate between unobscured (blue circles), Compton-thin (green squares), and Compton-thick (red diamonds) AGNs, as denoted in the legend. AGNs with \nh{} upper limits of $\rm{log}(\nhm{}) \leq 20.0$ are denoted with gray triangles. While there is a large amount of scatter, there is a general upturn in the luminosity ratio at the highest column densities, beginning at $\sim5$--$8\times10^{23}$~cm$^{-2}$.
    }
    \label{fig:W4W2_v_NH_scatter}
\end{figure}

\begin{figure*}

\begin{minipage}[t]{0.46\linewidth}
    \centering
    \hfill \subfloat{\includegraphics[width=0.98\textwidth]{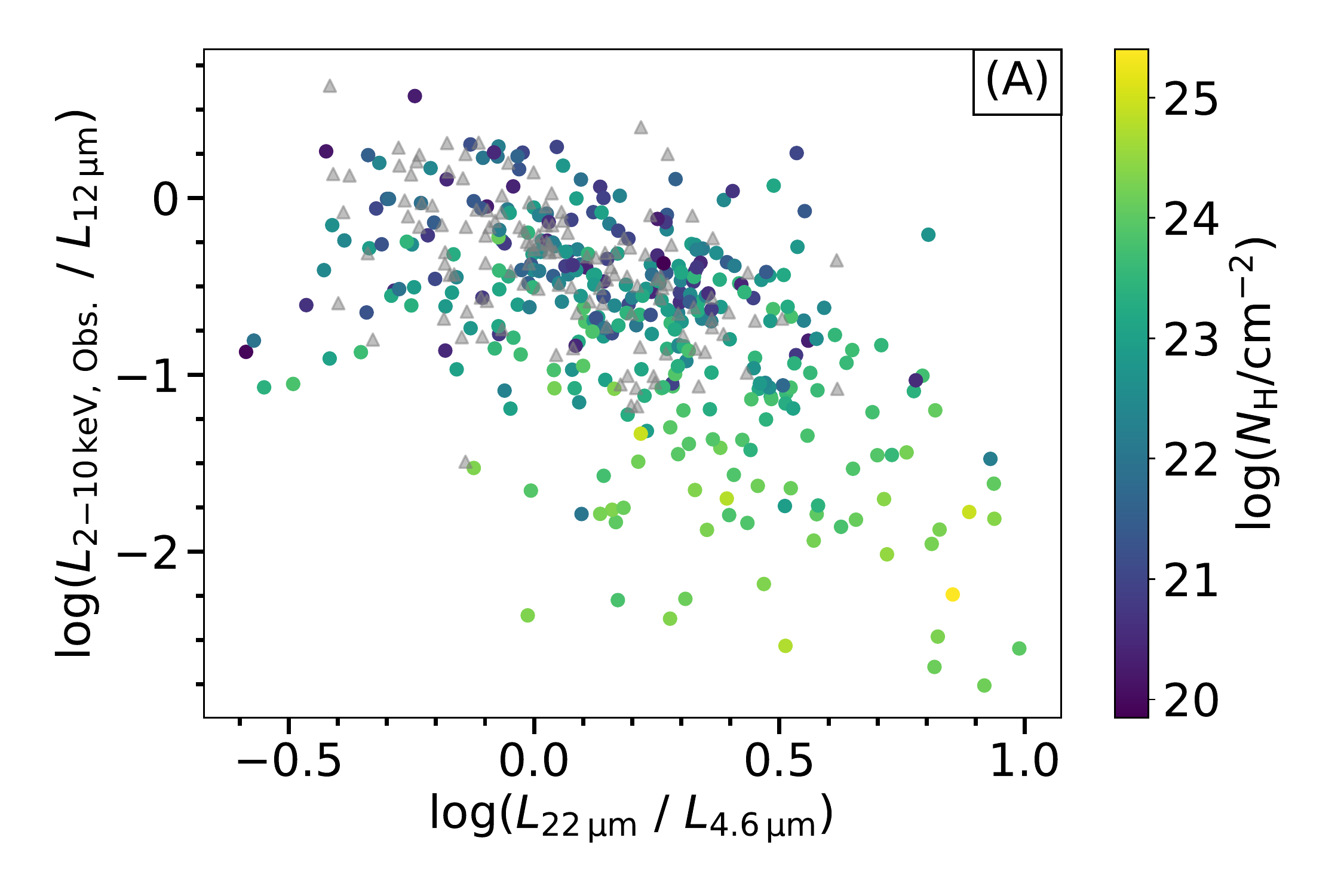}}
    \vspace{-7mm}
    \subfloat{\includegraphics[width=0.85\textwidth]{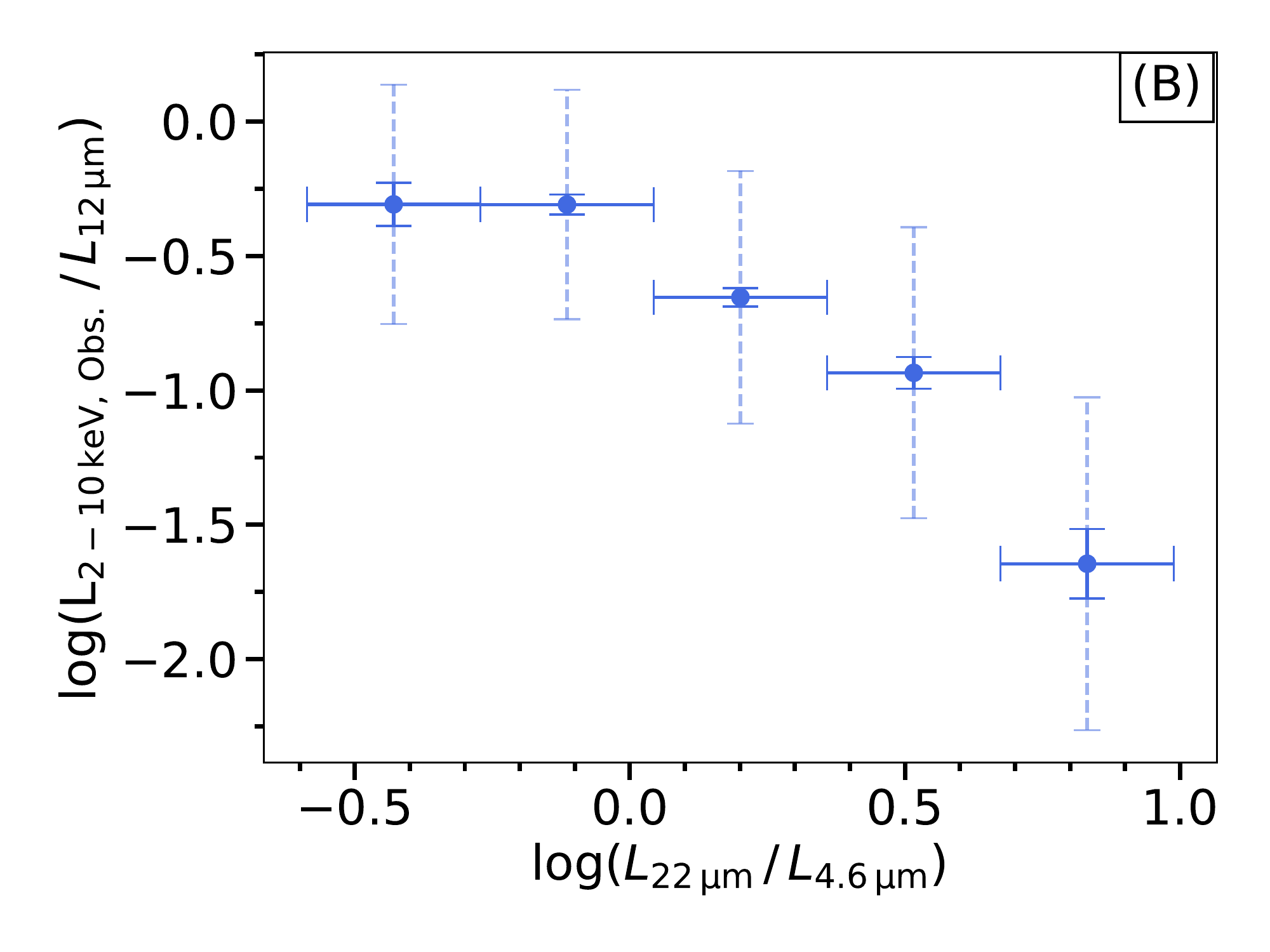}} \hfill 
    \vspace{-7mm}
    \subfloat{\includegraphics[width=0.85\textwidth]{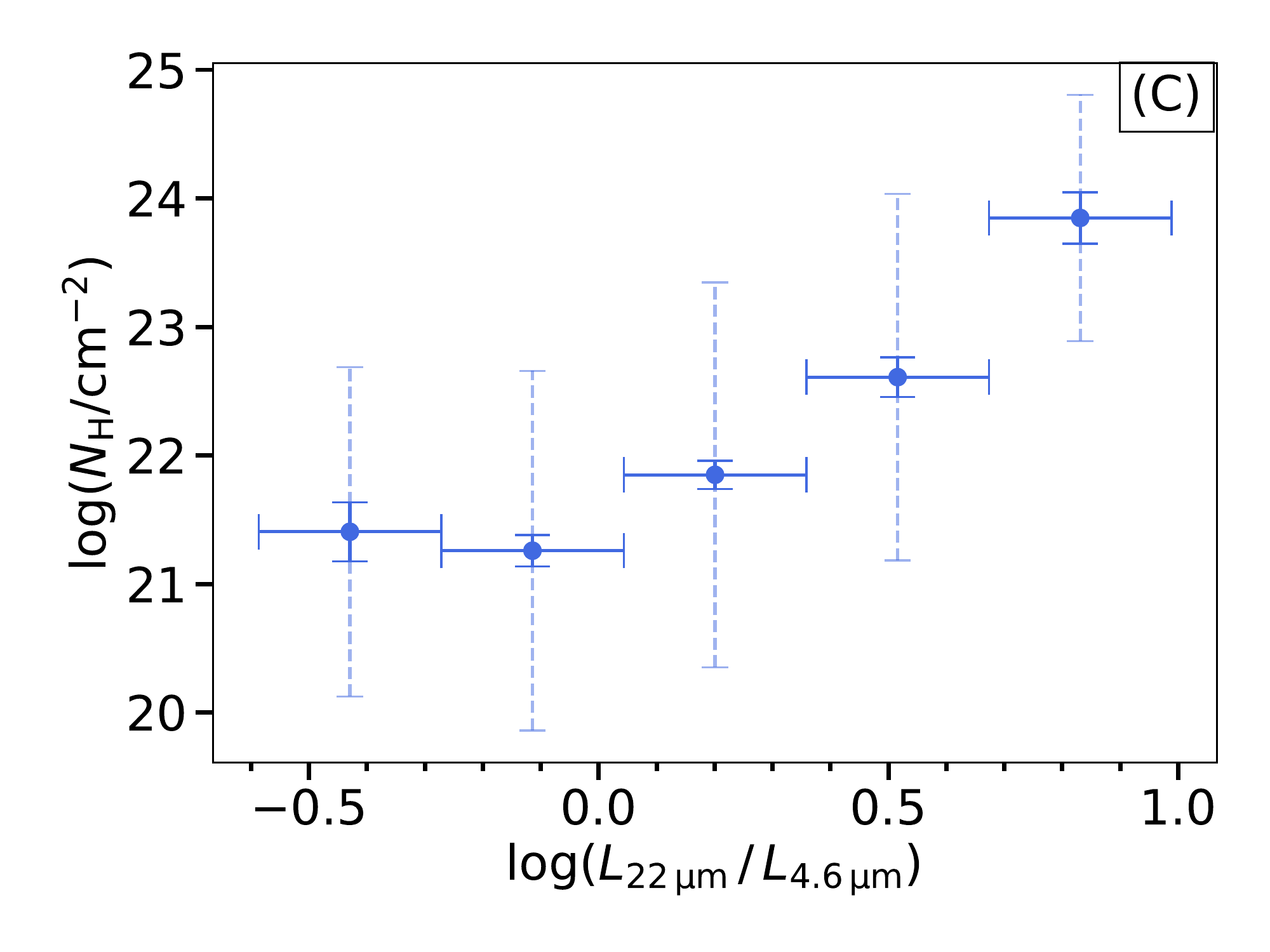}} \hfill
    \vspace{-4mm}
    \caption{The logarithmic \lxobs{}/\ltwel{} and \lfour{}/\ltwo{} luminosity ratios (Panel A) with each point color coded to indicate the column density (as indicated by the auxiliary axis). Sources with \nh{} upper limits of $\rm{log}(\nhm{}) < 20.0$ are denoted with gray triangles. Binning by the \lfour{}/\ltwo{} luminosity ratio (Panel B), we  see a general trend of decreasing X-ray-to-mid-IR ratio with increasing values of the \lfour{}/\ltwo{} ratio. Binning by the \lfour{}/\ltwo{} luminosity ratio and comparing it to the column density (Panel C), we see a general trend of increasing \WISE{} luminosity ratios with increasing column density. Solid error bars in Panels B and C represent the standard error of the mean while dashed error bars represent the standard deviation computed for the respective bin.}
    \label{fig:W4W2fullsamplediagram}
\end{minipage}
%\hspace{0.1cm}
\hfill
\begin{minipage}[t]{0.46\linewidth}
    \centering
    \hfill \subfloat{\includegraphics[width=0.98\textwidth]{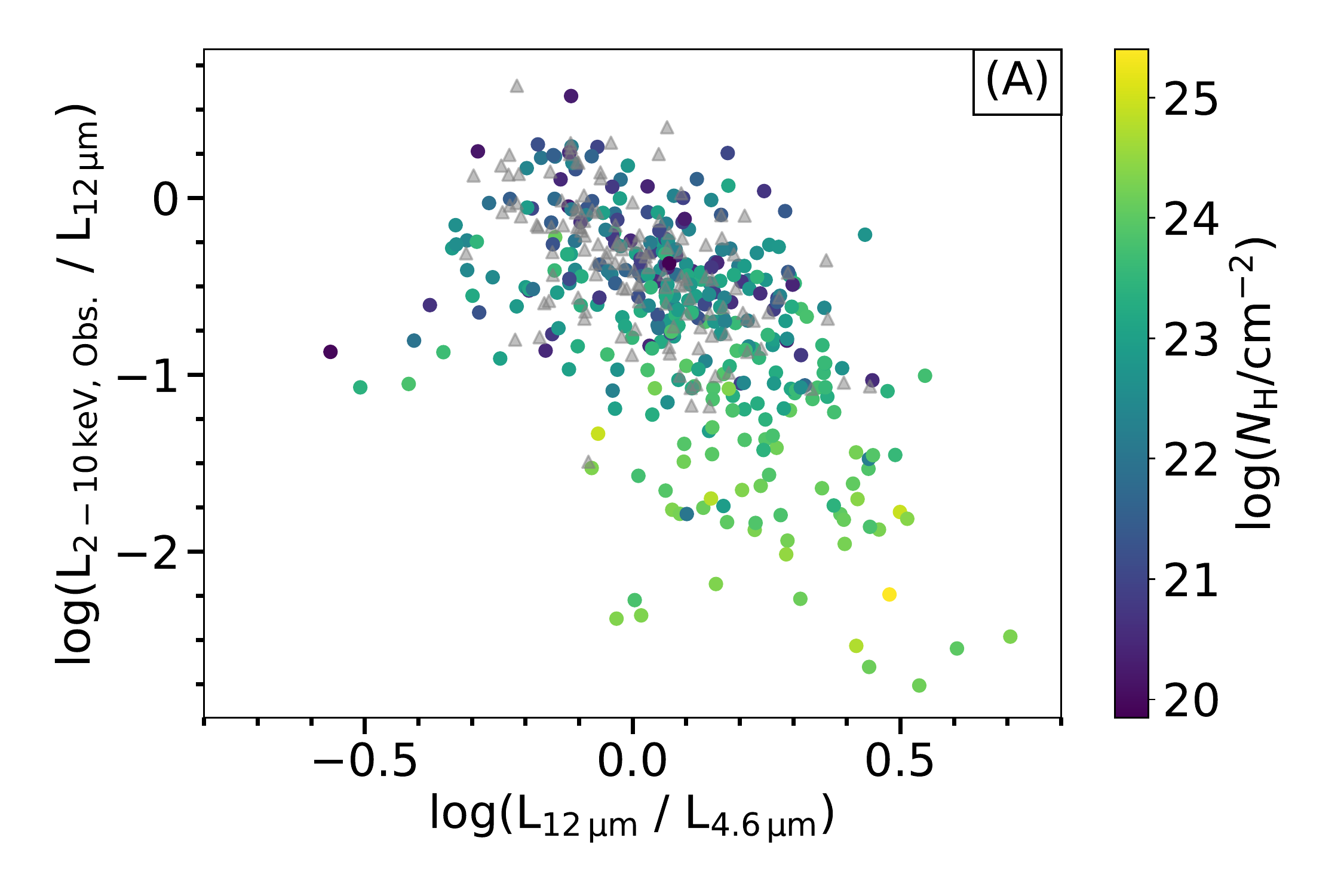}}\\ \vspace{-7mm}
    \subfloat{\includegraphics[width=0.85\textwidth]{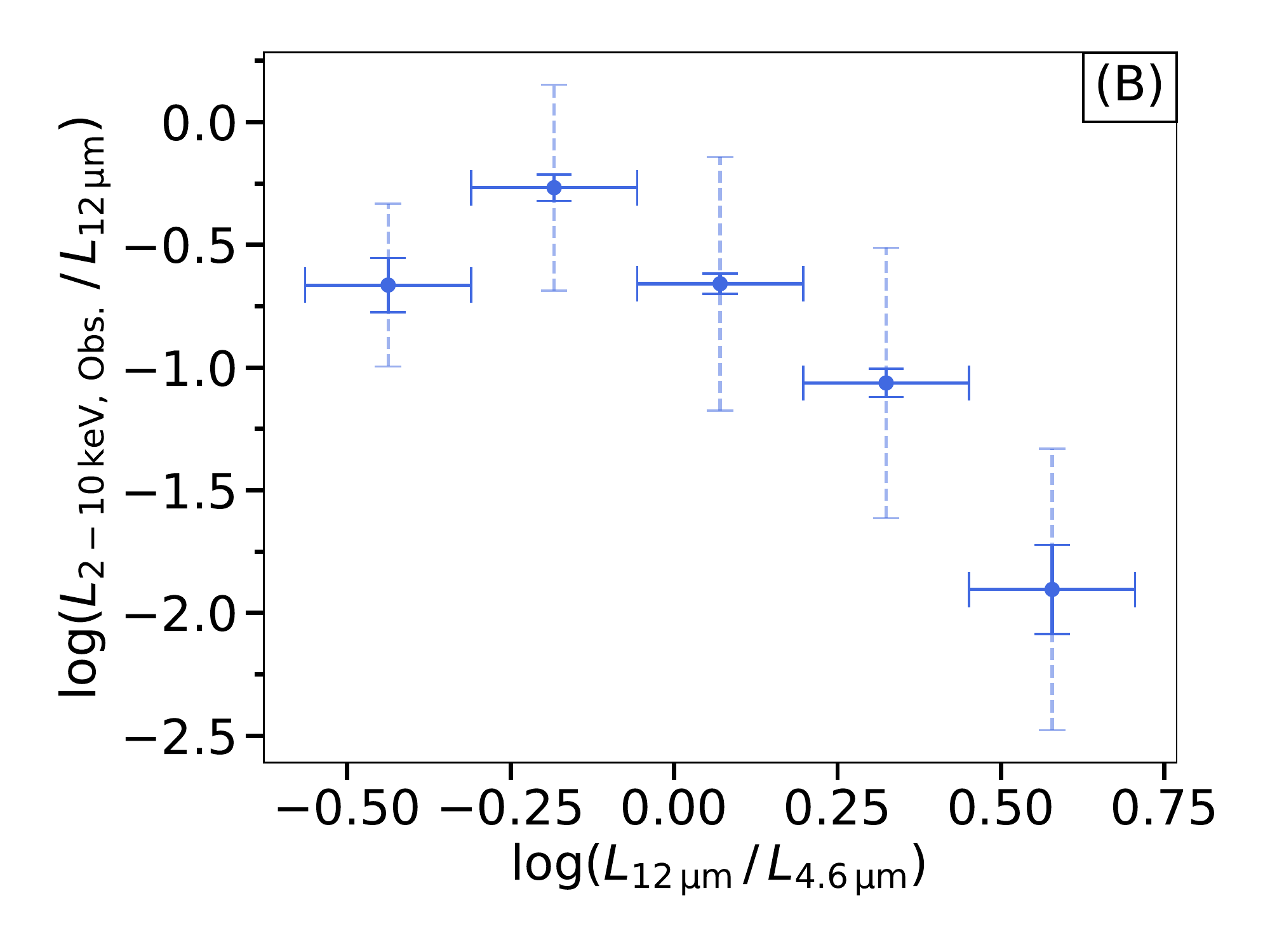}} \hfill
    \vspace{-7mm}
    \subfloat{\includegraphics[width=0.85\textwidth]{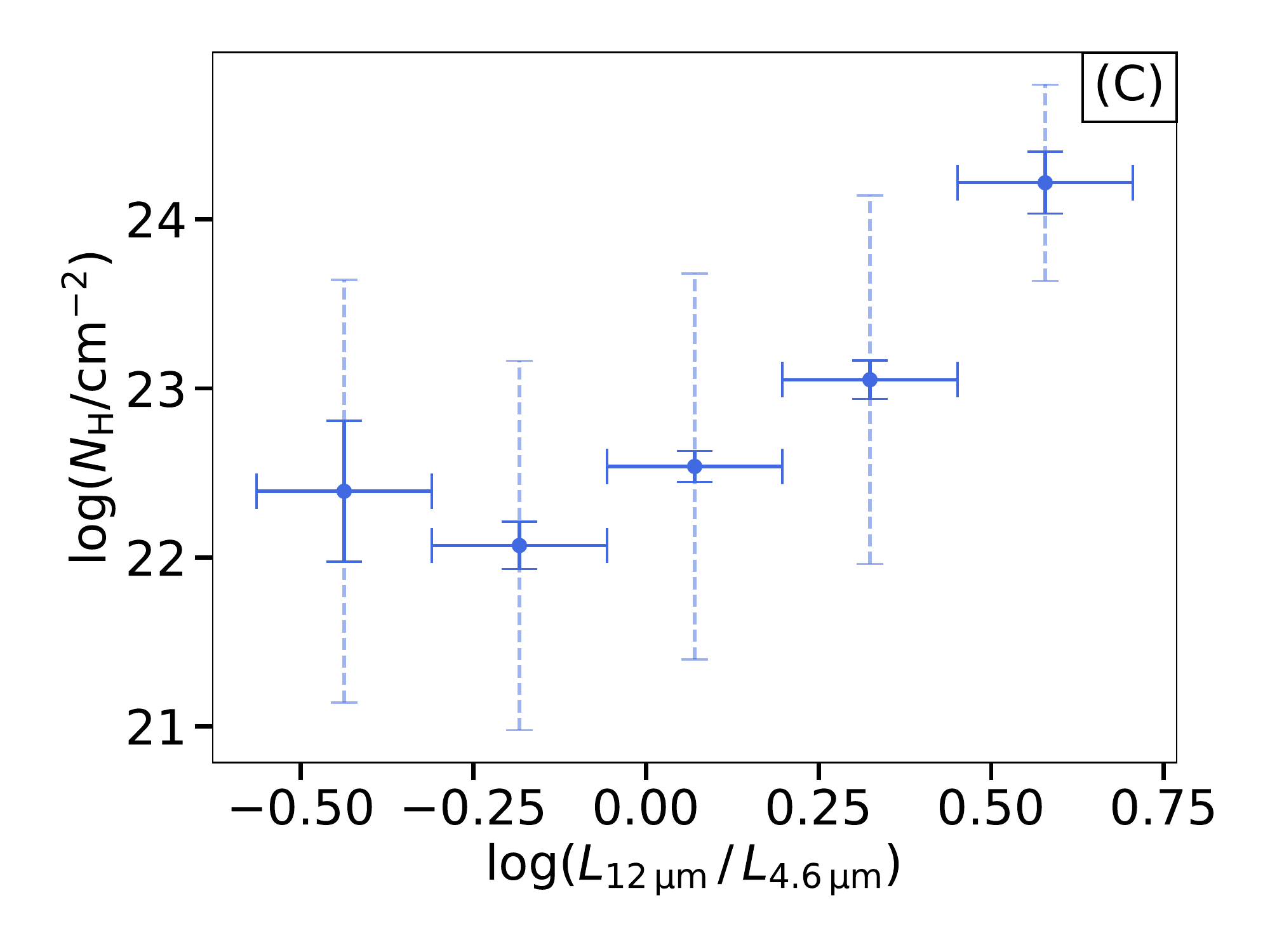}} \hfill
    \vspace{-4mm}
    \caption{In an identical fashion to that shown in Figure~\ref{fig:W4W2fullsamplediagram}, we show the logarithmic \lxobs{}/\ltwel{} and \ltwel{}/\ltwo{} luminosity ratios (Panel A) with the column density given on the auxiliary axis. We find very similar results to those shown in Figure~\ref{fig:W4W2fullsamplediagram} when binning and comparing the \lxobs{}/\ltwel{} and \ltwel{}/\ltwo{} luminosity ratio (Panel B) and when binning and comparing the \ltwel{}/\ltwo{} ratio and \nh{} (Panel C).}
    \label{fig:W3W2fullsamplediagram}
\end{minipage}

\end{figure*}

\section{Data Analysis} 
\label{sec:analysis}
Comparing the observed X-ray 2--10~keV luminosities (\lxobs{}) to the $12\,\rm{\mu m}$ luminosities (\ltwel{}), we see in Figure~\ref{fig:LXLMIR_v_NH} that unobscured \swift{} AGNs tend to follow the relation between the intrinsic (unabsorbed) 2--10 X-ray and nuclear $12\,\rm{\mu m}$ luminosities (dashed black line) established by \citet{asmus2015}, whereas obscured AGNs appear X-ray suppressed when compared to \ltwel{}. This decrease in \lxobs{} compared to \ltwel{} with increasing column density is expected, of course, since the X-ray emission will suffer greater attenuation than the mid-IR emission, and we color code the data in Figure~\ref{fig:LXLMIR_v_NH} according to the column densities adopted in Section~\ref{sec:sampleselect}. As a population, CT AGNs are generally the furthest offset from the \citet{asmus2015} relation, exhibiting luminosity ratios of log(\lxobs{}/\ltwel{}$)<-1.3$ (this was previously discussed in, e.g. \citealp{alexander2008}); we plot this ratio between log(\lxobs{}) and log(\ltwel{}) as a dotted black line in Figure~\ref{fig:LXLMIR_v_NH}. As has been done in previous works \citep[e.g.][]{alexander2008,goulding2011,asmus2015}, we can use this ratio between \lxobs{} and \ltwel{} as a diagnostic tool to differentiate between less obscured and heavily obscured or CT AGNs.

Ratios of two mid-IR luminosities, sufficiently separated in wavelength, could exhibit the same trend as is observed for the \lxobs{}/\ltwel{} ratio, in that the shorter wavelength mid-IR emission could appear suppressed compared to the longer wavelength emission due to the obscuring material surrounding the AGN. We test a new diagnostic tool based on the ratio of the \WISE{} $22\,\rm{\mu m}$ to $4.6\,\rm{\mu m}$ luminosity (\lfour{}/\ltwo{}) in Figure~\ref{fig:W4W2_v_NH_scatter}, and we find that the logarithmic ratio \textit{increases} with column density, with the most significant increase in \WISE{} ratio corresponding to the highest column densities (although with significant scatter). For example, we find a mean luminosity ratio and 1-$\sigma$ uncertainty of log(\lfour{}/\ltwo{}$)=0.50\pm0.32$ for AGNs with log($\nhm{})\geq24$, whereas we find a mean ratio and 1-$\sigma$ uncertainty of log(\lfour{}/\ltwo{}$)=0.06\pm0.25$ for unobscured AGNs with column densities log($\nhm{})<22$. This suggests that \WISE{} colors can also be used as a diagnostic tool for identifying heavily absorbed AGNs. 

%In a complementary fashion to this work, Asmus et al. (submitted) also reports on this observed trend between the column density and the \WISE{} mid-IR colors for the \swift{} AGNs, although their work focuses mainly on how the mid-IR spectral energy distributions of the \swift{} AGNs are affected by fundamental AGN properties, such as the AGN luminosity, $\lambda_{\rm{Edd}}$, and \nh{}.

% Spectral curvature might be tricky with low redshift sources. Probably just need to do  some discussion points 

We combine the \lxobs{}/\ltwel{} and \lfour{}/\ltwo{} ratio diagnostics in Figure~\ref{fig:W4W2fullsamplediagram}, plotting the two diagnostic ratios against one another in Panel A and color coding the data points by column density on the auxiliary axis. As the column density increases, the AGNs tend to exhibit lower values of \lxobs{}/\ltwel{} and higher values of \lfour{}/\ltwo{}, and thus we find that the most heavily obscured AGNs predominantly occupy the lower right portion of the parameter space (i.e. the largest X-ray deficits and highest \lfour{}/\ltwo{} ratios). In Panel B of Figure~\ref{fig:W4W2fullsamplediagram} we show this same result after binning the data by \lfour{}/\ltwo{}. In Panel C we bin by \lfour{}/\ltwo{} and examine the scatter in the \emph{column density} by bin; we notice a positive correlation between the column density and the mid-IR luminosity ratio. All solid error bars in Panels B and C represent the standard error of the mean while dashed error bars represent the standard deviation computed for the respective bin. Considering these three panels together, we may define a parameter space using the \lxobs{}/\ltwel{} and the \lfour{}/\ltwo{} luminosity ratio, which could be used to identify the most heavily absorbed AGNs. We repeated this analysis for an alternative \WISE{} luminosity ratio of \ltwel{}/\ltwo{} and found very similar results (see Figure~\ref{fig:W3W2fullsamplediagram}). We explored several AGN selection methods to potentially mitigate or at least account for the scatter observed, which we discuss in the Appendix. Ultimately, we did not apply any selection criteria to our sample of \swift{} AGNs during the analysis described below. Of important note, however, is the interesting result that the correlation between the mid-IR color and \nh{} holds true for both \WISE{} (selected via $W1-W2>0.8$; \citealp{stern2012}) \emph{as well as} non-\WISE{} AGNs.

\subsection{A Relation for \nh{} as a Function of \lxobs{}/\ltwel{}}
\label{sec:nh_v_lxl12}
Since the \lxobs{}/\ltwel{} ratio is known to correlate with obscuring column (e.g. \citealp{ichikawa2012,asmus2015,yan2019}), we derived an expression to describe this relationship using the \swift{} sample studied here. Note that, in contrast to \citet{asmus2015} (see Section~4.2 for further details), we do not exclude sources with log($\nhm{})<22.8$ from the fitting process. 

For our fitting process, we incorporate the luminosity ratios and the associated uncertainties. We pull the uncertainties in the mid-IR fluxes from \citet{ichikawa2017} and \citet{ichikawa2019}; while uncertainties in the observed X-ray luminosities were not available in the \citet{ricci2017apjs} catalog, we conservatively adopt uncertainties of log(\lxobs{}$)=\pm\,0.1$ for AGNs with log($\nhm{})<24$ and log(\lxobs{}$)=\pm\,0.3$ for AGNs with log($\nhm{})\geq24$. In order to take into account the asymmetric uncertainties associated with log($\nhm{}$) and log(\lxobs{}/\ltwel{}), we employed the following Monte Carlo fitting routine:
\begin{enumerate}
\item \textit{Bootstrap}: To fully account for the effect of outliers, we generated a new data set as a sample of the original, allowing repeats.
\item \textit{Monte Carlo}: For each point in the bootstrapped data set, we generate a new point given the uncertainties in log($\nhm{}$) and log(\lxobs{}/\ltwel{}). We incorporate asymmetric error bars by randomly drawing from a Gaussian distribution separately for the negative and positive error bars. For upper limits in log(\nh{})\footnote{Note there are no limits for log(\lxobs{}/\ltwel{}).}, we generate a new point using a uniform distribution from the limit to an arbitrarily small log($\nhm{}$) value below it. After experimenting a number of different values, each giving similar results, we settled for a lower bound of log($\nhm{}$)\,=\,19, which is only marginally lower than the smallest log($\nhm{}$) value through the Milky Way according to the maps by \citep{kalberla2005}.
\item \textit{Orthogonal Distance Regression}: We fit each Monte Carlo-bootstrapped data set with a function of the form $y=a\cdot10^{b\cdot(x-20)}+c$ using orthogonal distance regression (from the Python \textsc{scipy} package \textsc{odr} (\citealp{boggs1990}, pp. 186; \citealp{virtanen2020}). This ensures we account for minimisation in both variables during the fitting procedure.
\item \textit{Parameter Estimation}: Steps 1\,--\,3 were performed many times, giving a distribution for the parameters a, b and c. For each parameter we report the $50^{\rm{th}}$ percentile and uncertainties derived from the the $84^{\rm{th}}$ and $16^{\rm{th}}$ percentiles.
\end{enumerate}

\begin{figure}[t!]
    \centering
    \subfloat{\includegraphics[width=1.0\linewidth]{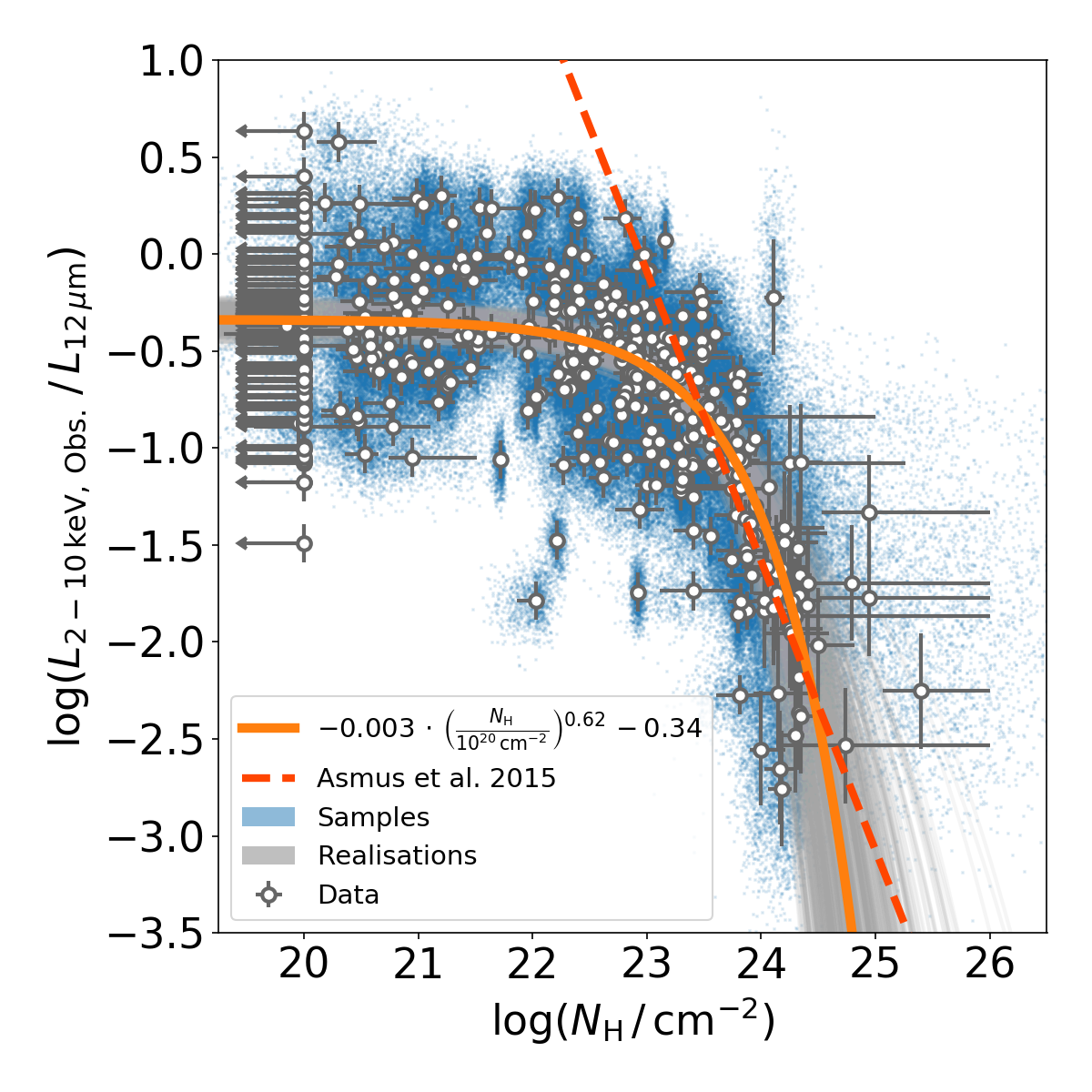}}\\ \vspace{-4mm} %png format for now, for compiling reasons
    \caption{Relation for the column density vs. log(\lxobs{}/\ltwel{}) for the \swift{} sample (white circles). New data samples simulated during the MC fitting routine are displayed as blue dots. The best-fit trendline (orange line, given by Equation~\ref{eq:lxl12_nh_eq}), represents the median of 1000 realizations (gray lines) obtained through orthogonal distance regression during the MC fitting process (see Section~\ref{sec:nh_v_lxl12}). The best-fit relation from \citet{asmus2015} is shown as a red dashed line.}
    \label{fig:LXLMIR_v_NH_fit}
\end{figure}

The relationship between \lxobs{}/\ltwel{} and \nh{} can be expressed as:
\begin{equation}
\label{eq:lxl12_nh_eq}
    \begin{aligned}
        \rm{log}(L_{X,\,\rm{Obs.}} / L_{12\,\mu m}) = (-0.34^{+0.06}_{-0.06})\; + \\ (-0.003^{+0.002}_{-0.005})\, \times\, (N_{\rm{H}}\,/\,10^{20}\,\rm{cm}^{-2})^{(0.62^{+0.13}_{-0.12})}
    \end{aligned}
\end{equation}
We plot the \swift{} sample and Equation~\ref{eq:lxl12_nh_eq} (orange line) in Figure~\ref{fig:LXLMIR_v_NH_fit}. To visualize the  uncertainty in this relation, we show the 1000 realizations of the best fit as grey lines.

We attempted to fit directly for log(\nh{}) as a function of log(\lxobs{}/\ltwel{}) but the fitting routine could not successfully converge on a reasonable fit to the data. Instead, we chose to invert Equation~\ref{eq:lxl12_nh_eq} to recover the obscuring column density as a function of the X-ray to mid-IR luminosity ratio:
\begin{equation}
\label{eq:lxl12_nh_eq_fNH}
    \begin{aligned}
    \rm{log}(N_{\rm{H}}\,/\,\rm{cm}^{-2}) = 20\; +\; (1.61^{+0.33}_{-0.31})\; \times \\\rm{log}\left(\left|\frac{\rm{log}\left(\frac{L_{X,\,\rm{Obs.}}}{L_{12\,\mu m}}\right) +(0.34^{+0.06}_{-0.06})}{(-0.003^{+0.002}_{-0.005})}\right|\right)
    \end{aligned}
\end{equation}

To better quantify the uncertainty in log($\nhm{}$) associated with this relation, we tabulated values of log(\nh{}) and the uncertainty for specific values of log(\lxobs{}/\ltwel{}), as shown in Table~\ref{table:lx_v_nh_values}. Using the parameter distributions calculated during the MC fitting routine, we derived a distribution of log($\nhm{}$) values using an expression of the form shown in Equation~\ref{eq:lxl12_nh_eq_fNH} (inverted from Equation~\ref{eq:lxl12_nh_eq}) and use the $50^{\rm{th}}$, $84^{\rm{th}}$, and $16^{\rm{th}}$ percentiles of the distribution to list the median column density and associated uncertainty for each chosen value of log(\lxobs{}/\ltwel{}). While strictly empirical, Equations~\ref{eq:lxl12_nh_eq} and~\ref{eq:lxl12_nh_eq_fNH} reproduce the observed trend of decreasing values of log(\lxobs{}/\ltwel{}) with increasing column density. We do note that this relation is largely insensitive to ratios of log(\lxobs{}/\ltwel{}$)>-0.3$, with the relation appearing nearly flat for column densities of log($
\nhm{})<22.5$. Equations~\ref{eq:lxl12_nh_eq} and~\ref{eq:lxl12_nh_eq_fNH} are most effective for obscured AGNs with column densities of log($\nhm{})\geq22.5$, and readers should  keep this in mind when using these expressions to derive estimates for column densities.

\begin{table}[t!]
\caption{log($\nhm{}$) Derived from Equations~\ref{eq:lxl12_nh_eq_fNH} and \ref{eq:asmusrelation}}
\begin{center}
\begin{tabular}{ccc}
\hline
\hline
\noalign{\smallskip}
\noalign{\smallskip}
log(\lxobs{}/\ltwel{}) & log($\nhm{}$) & log($\nhm{}$)  \\
 & (this work) & \citep{asmus2015} \\
\noalign{\smallskip}
\noalign{\smallskip}
\hline
\noalign{\smallskip}
$-0.3 $ & $21.8^{+0.7}_{-1.0}$ & $23.1\pm0.1$ \\
$-0.5 $ & $22.7^{+0.2}_{-0.3}$ & $23.3\pm0.1$ \\
$-0.75$ & $23.4^{+0.1}_{-0.1}$ & $23.4\pm0.2$ \\
$-1.0 $ & $23.7^{+0.1}_{-0.1}$ & $23.6\pm0.2$ \\
$-1.3 $ & $24.0^{+0.2}_{-0.1}$ & $23.8\pm0.2$ \\
$-1.5 $ & $24.1^{+0.2}_{-0.1}$ & $23.9\pm0.3$ \\
$-2.0 $ & $24.4^{+0.2}_{-0.2}$ & $24.3\pm0.3$ \\
$-2.5 $ & $24.5^{+0.3}_{-0.2}$ & $24.6\pm0.4$ \\
$-3.0 $ & $24.7^{+0.3}_{-0.2}$ & $25.0\pm0.4$ \\
\noalign{\smallskip}
\hline
\end{tabular}
\end{center}
\tablecomments{Column densities as a function of log(\lxobs{}/\ltwel{}). Column 1: logarithmic ratio of the 2--10~keV to 12~$\mu$m luminosities; Equation~\ref{eq:lxl12_nh_eq_fNH} is insensitive to ratios of log(\lxobs{}/\ltwel{}$)>-0.3$, which are largely exhibited by AGNs with log($\nhm{})<23$. Column 2: column density and associated error, derived by Equation~\ref{eq:lxl12_nh_eq_fNH}; the parameter distributions found during the MC fitting process described in Section~\ref{sec:nh_v_lxl12} were read into an inverted expression of the form in Equation~\ref{eq:lxl12_nh_eq_fNH}, from which we retrieved the median column density and upper and lower bounds using the $50^{\rm{th}}$, $84^{\rm{th}}$, and $16^{\rm{th}}$ percentiles of the resulting log($\nhm$) distribution. Column 3: column density derived using Equation 6 from \citet{asmus2015}.}
\label{table:lx_v_nh_values}
\end{table}

We compare these results to those found in \citet{asmus2015} in Section~\ref{sec:asmus_comp}, and we plot the best fit relation for log(\nh{}) versus log(\lxobs{}\ltwel{}) found by \citet{asmus2015} in Figure~\ref{fig:LXLMIR_v_NH_fit}.

\subsection{Relation between \nh{} and the Mid-Infrared Colors}
\label{sec:nh_v_l22l46}
\begin{figure}[t!]
    \centering
    \subfloat{\includegraphics[width=1.0\linewidth]{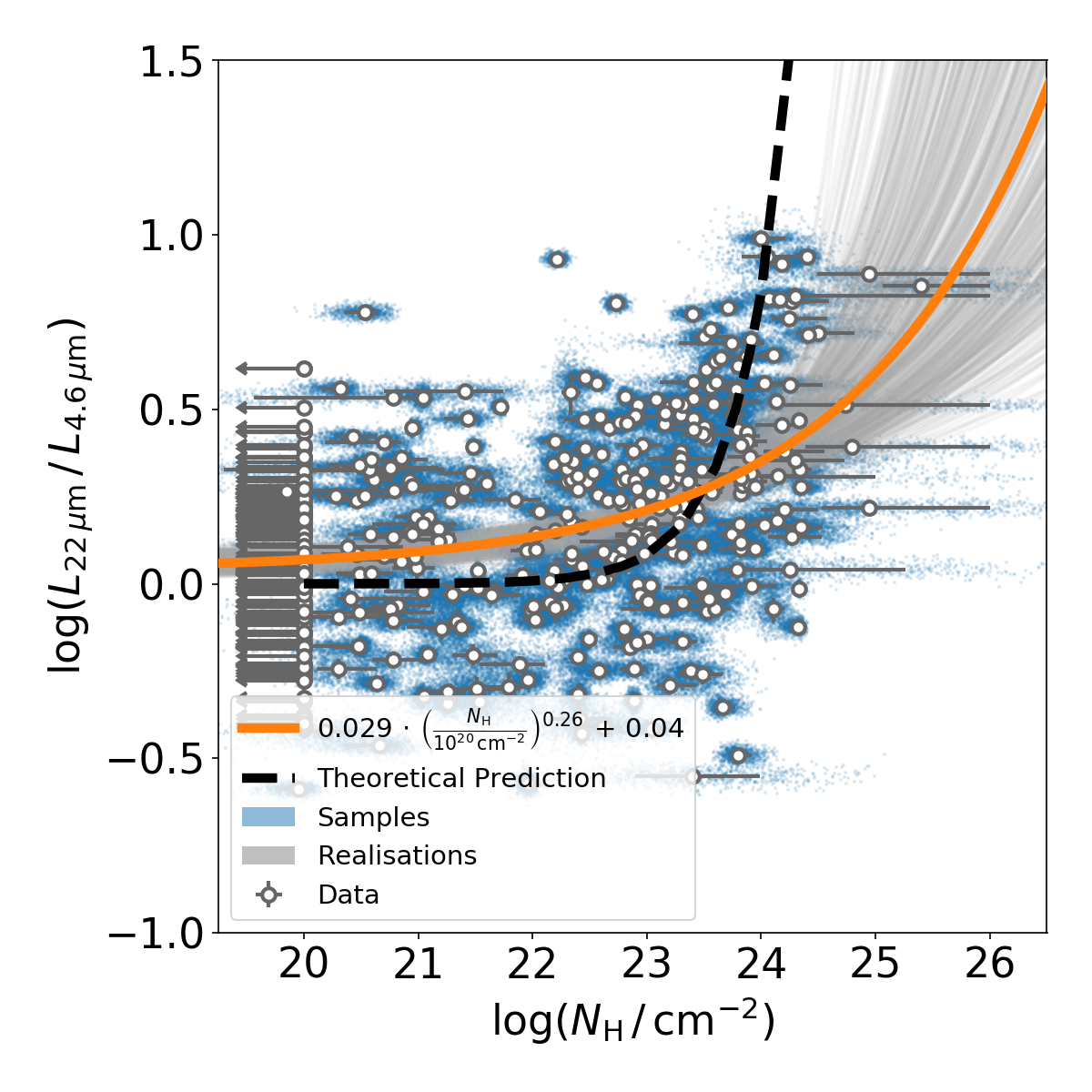}}\\ \vspace{-4mm} %png file format for now
    %NH_v_LW4LW2_medianFit.pdf
    \caption{Relation for column density vs. log($L_{22\,\rm{\mu m}}/L_{4.6\,\rm{\mu m}}$) for the \swift{} sample (white circles). The fitting process for this relation is identical to that in Figure~\ref{fig:LXLMIR_v_NH_fit}. The best-fit trendline (orange line, given by Equation~\ref{eq:lxl12_nh_eq}) represents the median of 1000 realizations (gray lines) obtained during the MC fitting process. The black dashed line represents a theoretical prediction for a simple model consisting of monochromatic radiation passing through a homogeneous screen of dust (see Section~\ref{sec:physorigin}.)}
    %We define $\alpha=\rm{log}(\nhm{})$.
    %As with Equation~\ref{eq:lxl12_nh_eq},
    %found during the fitting process
    \label{fig:L22L46_v_NH_fit}
\end{figure}

Given the correlation between \lfour{}/\ltwo{} and \nh{}, as shown in Figure~\ref{fig:W4W2fullsamplediagram}, we  followed the same fitting procedure as discussed above to develop an expression relating the mid-IR luminosity ratio and \nh{}. The equation relating these properties can be expressed as:
\begin{equation}
\label{eq:l22l46_nh_eq}
    \begin{aligned}
        \rm{log}(L_{22\,\rm{\mu m}} / L_{4.6\,\rm{\mu m}}) = (0.04^{+0.02}_{-0.02})\; +\; \\(0.03^{+0.02}_{-0.02}) \times\, (N_{\rm{H}}\,/\,10^{20}\,\rm{cm}^{-2})^{(0.26^{+0.13}_{-0.07})}
    \end{aligned}
\end{equation}
In Figure~\ref{fig:L22L46_v_NH_fit} we plot the \swift{} sample and the best-fitting trendline (orange line) along with 1000 realizations of the best fit found during the fitting procedure (shown as gray lines).

As mentioned in Section~\ref{sec:nh_v_lxl12}, we attempted to fit directly for \nh{} as a function of \lxobs{}/\ltwel{}, but the fitting routine could not successfully converge on a reasonable line of best fit. We therefore simply invert Equation~\ref{eq:l22l46_nh_eq} like before to recover the column density as a function of the mid-IR luminosity ratio:
\begin{equation}
\label{eq:l22l46_nh_eq_fNH}
    \begin{aligned}
        \rm{log}(N_{\rm{H}}\,/\,\rm{cm}^{-2}) = 20\;+\; (3.86^{+1.94}_{-1.00})\; \times \\ \rm{log}\left(\left|\frac{\rm{log}\left(\frac{L_{22\,\rm{\mu m}}}{L_{4.6\,\rm{\mu m}}}\right) -(0.04^{+0.02}_{-0.02})}{(0.03^{+0.02}_{-0.02})}\right|\right)
    \end{aligned}
\end{equation}
In an identical fashion to the calculations in Section~\ref{sec:nh_v_lxl12}, we tabulated values of log($\nhm{}$) and the associated uncertainties for specific values of log(\lfour{}/\ltwo{}) in Table~\ref{table:lw4lw2_v_nh_values}. Equation~\ref{eq:l22l46_nh_eq_fNH} is not sensitive to values of log(\lfour{}/\ltwo{}$)<0$ and, as in the case of Equation~\ref{eq:lxl12_nh_eq_fNH}, readers should use this relation cautiously and bear in mind that it is really only effective for obscured AGNs with log($\nhm{})\geq22.5$.

\begin{table}[t!]
\caption{log($\nhm{}$) Derived from Equation~\ref{eq:l22l46_nh_eq_fNH}}
\begin{center}
\begin{tabular}{cc}
\hline
\hline
\noalign{\smallskip}
\noalign{\smallskip}
log(\lfour{}/\ltwo{}) & log($\nhm{}$) \\
\noalign{\smallskip}
\noalign{\smallskip}
\hline
\noalign{\smallskip}
0.0 & $20.6^{+1.4}_{-2.2}$ \\
0.1 & $21.3^{+0.6}_{-0.8}$ \\
0.2 & $22.9^{+0.2}_{-0.3}$ \\
0.25 & $23.4^{+0.2}_{-0.2}$ \\
0.3 & $23.7^{+0.2}_{-0.2}$ \\
0.4 & $24.2^{+0.4}_{-0.3}$ \\
0.5 & $24.7^{+0.5}_{-0.4}$ \\
0.75 & $25.4^{+0.7}_{-0.6}$ \\
1.0 & $25.9^{+0.9}_{-0.8}$ \\
\noalign{\smallskip}
\hline
\end{tabular}
\end{center}
\tablecomments{Column densities as a function of log(\lfour{}/\ltwo{}). Column 1: logarithmic ratio of the 22~$\mu$m to 4.6~$\mu$m luminosities; Equation~\ref{eq:lxl12_nh_eq_fNH} is insensitive to ratios of log(\lfour{}/\ltwo{}$)>0$. Column 2: column density and associated error, derived by Equation~\ref{eq:l22l46_nh_eq_fNH}; the median value and associated uncertainties were derived in an identical fashion to that described in Table~\ref{table:lx_v_nh_values}}.
\label{table:lw4lw2_v_nh_values}
\end{table}

\subsection{Diagnostic Regions for Heavily Absorbed AGNs}
\label{sec:diagnosticregions}

\begin{figure*}[t!]
    \centering
    \subfloat{\includegraphics[width=0.5\linewidth]{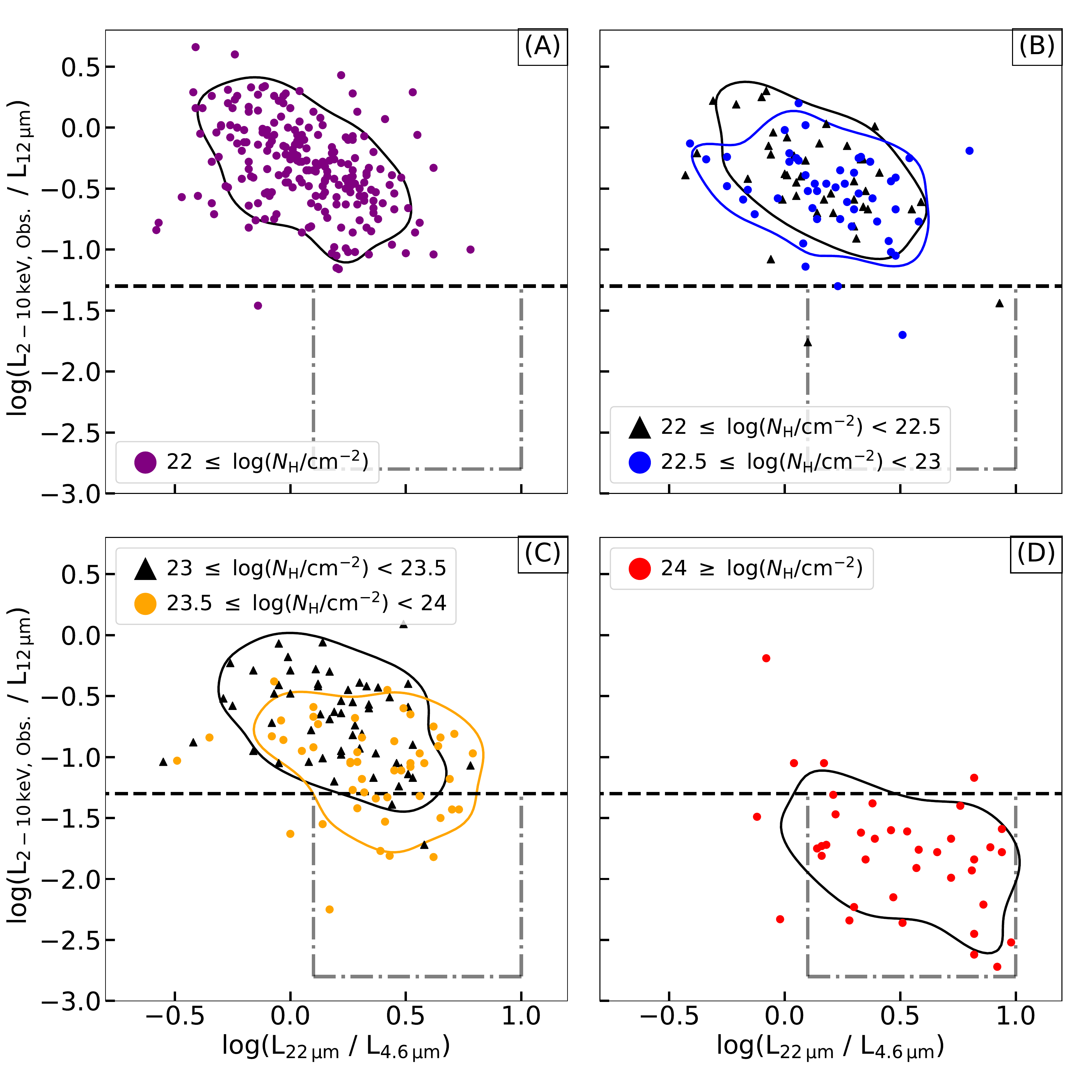}}
    \subfloat{\includegraphics[width=0.5\linewidth]{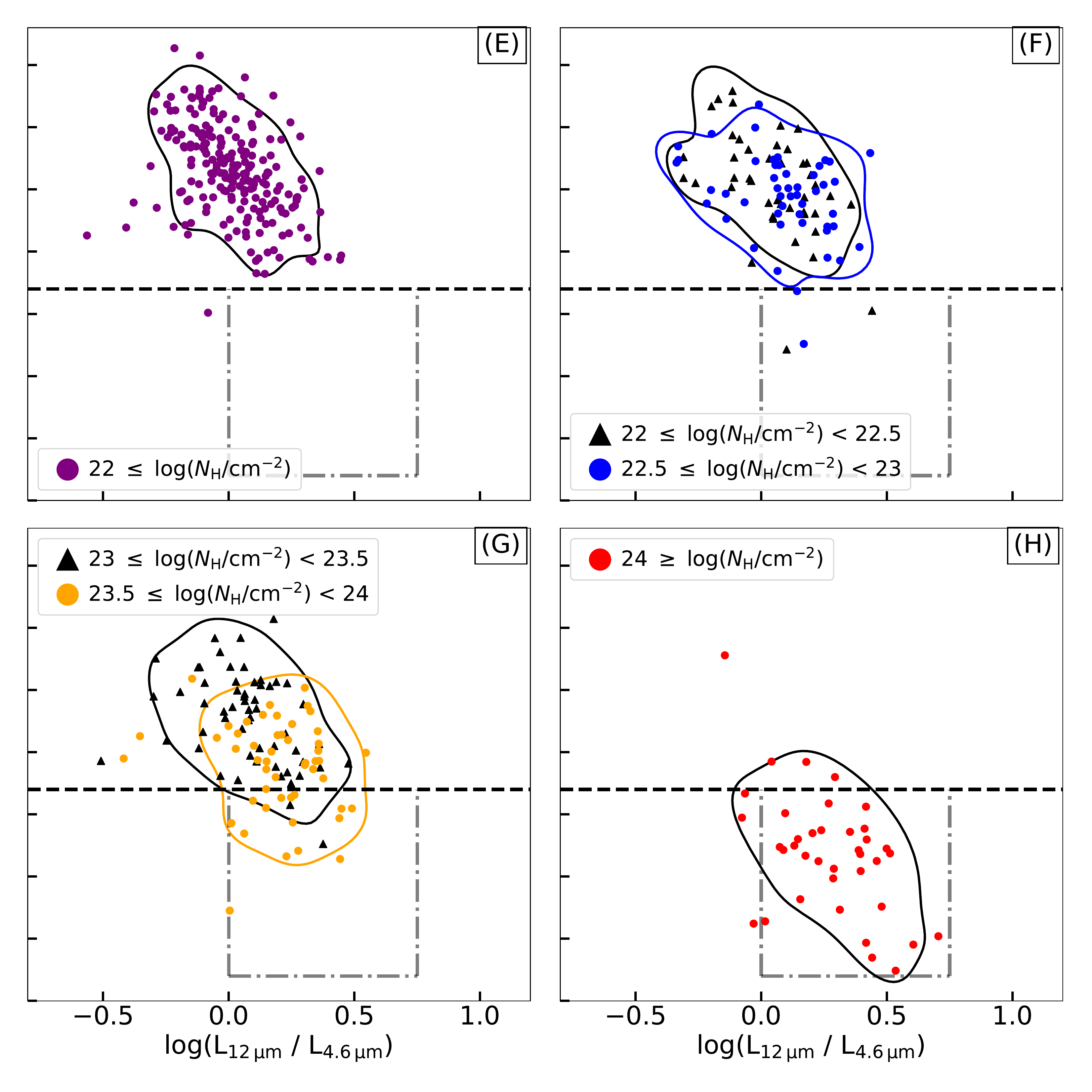}}
    \caption{\lfour{}/\ltwo{} (Panels A--D) and \ltwel{}/\ltwo{} (Panels E--H) diagnostics for AGNs in the \swift{} sample binned by \nh{}, where we have unobscured (Panel A, E; $\log(\nhm{})<22$), Compton-thin `lightly obscured' (Panel B, F; $22\leq \log(\nhm{})<23$), Compton-thin `moderately obscured' (Panel C, G; $23\leq \log(\nhm{})<24$), and `heavily obscured' to Compton-thick (Panel D, H; $\log(\nhm{})\geq24$). The Compton-thin bins are broken into two sub-bins each: $22\leq \log(\nhm{})<22.5$ (black points), $22.5\leq \log(\nhm{})<23$ (blue points), $23\leq \log(\nhm{})<23.5$ (black points), and $23.5\leq \log(\nhm{})<24$ (orange points). Contours were computed to encompass $\sim68\%$ of the population for each respective bin or sub-bin. Note \textit{in general} that the most obscured sources tend to populate a different region of the parameter space than do the unobscured sources and Compton-thin `lightly obscured' sources, while there is some overlap with Compton-thin `moderately obscured' sources.}
    \label{fig:W42_W32_subbins_box_diagram}
\end{figure*}

The correlations between \lxobs{}/\ltwel{}, $\,\,\,$ \lfour{}/\ltwo{}, and \nh{} discussed above suggest that these relationships may be combined to help differentiate between \swift{} AGNs according to the levels of obscuration. We probed this potential diagnostic for heavily absorbed AGNs by plotting \lfour{}/\ltwo{} vs. \lxobs{}/\ltwel{} and binning the full sample by absorbing column as shown in Figure~\ref{fig:W42_W32_subbins_box_diagram}. The sample was divided into four bins in obscuration corresponding to:
\begin{itemize}
    \item unobscured \newline [$\rm{log}(\nhm{})<22$, Panel A]
    \item Compton-thin ``lightly obscured'' \newline [$22\leq\rm{log}(\nhm{})<23$, Panel B]
    \item Compton-thin ``moderately obscured'' \newline [$23\leq\rm{log}(\nhm{})<24$, Panel C]
    \item ``heavily obscured'' to Compton-thick \newline [$\rm{log}(\nhm{})\geq~24$, Panel D]
\end{itemize}
We also split the two Compton-thin bins into two sub-bins each in increments of log($\nhm{})=0.5$, as summarized in Table~\ref{table:LXW3_populationstats}.

Contours were computed for each bin and/or sub-bin and in each case are designed to encompass $\sim68\%$ of the population of each respective bin. The binned subplots shown in Figure~\ref{fig:W42_W32_subbins_box_diagram} demonstrate much more clearly that, \textit{in general}, the heavily obscured sources (Panel D) tend to occupy a separate region of space than the unobscured (Panel A) or Compton-thin `lightly obscured' (Panel B) sources. While there is some overlap between the CT and Compton-thin `moderately obscured' populations, this is predominantly due to Compton-thin `moderately obscured' sources with $23.5\leq\rm{log}(\nhm{})<24$. In light of this, appropriate selection criteria can be used to construct diagnostic regions in this parameter space which separate heavily obscured and less obscured AGN populations. Here we define the completeness of selection criteria as `the fraction of true heavily obscured AGNs selected,' while we define purity as `the fraction of selected AGNs which are heavily obscured,' i.e. the fractional contribution of heavily obscured AGNs to a diagnostic region. These definitions can be extended in an analogous fashion to the other column density bins.

Defining a horizontal cut in this parameter space:
\begin{equation}
\label{eq:lxl12_horizontalcut}
\rm{log}(\textit{L}_{X,\,\rm{Obs.}}/\textit{L}_{12\,\mu m}) < -1.3
\end{equation}
which is shown as a black dashed line in Figure~\ref{fig:W42_W32_subbins_box_diagram}, provides a simple yet robust method of differentiating between the most heavily obscured AGNs and less obscured AGNs in the \swift{} sample. We report the sample statistics for this cut in Table~\ref{table:LXW3_populationstats}. To derive the population statistics for Table~\ref{table:LXW3_populationstats} (and all percentages quoted hereafter), we calculated the median ($50^{\rm{th}}$ percentile) value for each population in question, while the uncertainties on the fractions are the $16^{\rm{th}}$ and $84^{\rm{th}}$ quantiles of a binomial distribution, all computed following \citet{cameron2011}. The criterion in Equation~\ref{eq:lxl12_horizontalcut}  yields $88.1^{+4.5}_{-5.7}\%$ completeness for the heavily obscured AGNs, a $60.5^{+6.3}_{-6.5}$\% pure sample, and a mean column density of log($\nhm{})=24.0\pm0.1$ for the diagnostic region. It is important to note that the majority of impurities selected with Equation~\ref{eq:lxl12_horizontalcut} arise from AGNs with column densities $23.5\leq\rm{log}(\nhm{})<24.0$; the diagnostic region is in fact $\sim88$\% pure for AGNs with $\rm{log}(\nhm{})\geq23.5$ and suffers minimal impurities from AGNs with lower column densities. 

\begin{table}[t!]
\caption{log(\lxobs{}/\ltwel{})~$\leq1.3$ Diagnostic Cut}
\begin{center}
\begin{tabular}{ccc}
\hline
\hline
\noalign{\smallskip}
\noalign{\smallskip}
$\rm{log}(\nhm{})$ & Completeness  & Purity \\
\noalign{\smallskip}
\noalign{\smallskip}
\hline
\noalign{\smallskip}
$\geq24.0$      & $88.1^{+4.5}_{-5.7}$ & $60.5^{+6.3}_{-6.5}$ \\
$<24.0$         & $5.4 ^{+1.2}_{-1.0}$ & $39.5^{+6.5}_{-6.3}$  \\
$[23.0,\;24.0)$ & $16.3^{+3.7}_{-3.3}$ & $30.8^{+6.2}_{-5.8}$ \\
$[23.5,\;24.0)$ & $30.5^{+6.6}_{-6.1}$ & $27.3^{+6.1}_{-5.6}$ \\
$[23.0,\;23.5)$ & $4.6 ^{+3.2}_{-2.2}$ & $ 4.7^{+3.3}_{-2.2}$ \\
$[22.0,\;23.0)$ & $5.0 ^{+2.6}_{-1.9}$ & $ 8.1^{+4.0}_{-3.1}$ \\
$[22.5,\;23.0)$ & $5.3 ^{+3.7}_{-2.5}$ & $ 4.7^{+3.3}_{-2.2}$ \\
$[22.0,\;22.5)$ & $6.0 ^{+4.2}_{-2.9}$ & $ 4.7^{+3.3}_{-2.2}$ \\
$<22.0$         & $0.8 ^{+0.7}_{-0.4}$ & $ 2.9^{+2.7}_{-1.7}$ \\
\noalign{\smallskip}
\hline
\end{tabular}
\end{center}
\tablecomments{Statistics derived from the $\rm{log}(L_{X}^{\rm{Obs.}}/L_{12\mu m})<-1.3$ threshold, defined in Section~\ref{sec:analysis} (Equation~\ref{eq:lxl12_horizontalcut}), for various \nh{} bins and sub-bins. Column 1: \nh{} bin. Column 2: Completeness, or the fraction of AGNs selected (per column density bin). Column 3: Purity of the sample, or the percentage contribution to the diagnostic box.}
%STATS UPDATED TO FINAL VERSION 2 JULY 2020 
\label{table:LXW3_populationstats}
\end{table}

\begin{table}[t!]
\caption{log(\lfour{}/\ltwo{})~$\geq0.1$ Diagnostic Cut}
\begin{center}
\begin{tabular}{ccc}
\hline
\hline
\noalign{\smallskip}
\noalign{\smallskip}
$\rm{log}(\nhm{})$ & Completeness  & Purity \\
\noalign{\smallskip}
\noalign{\smallskip}
\hline
\noalign{\smallskip}
$\geq24.0$      & $88.1^{+4.5}_{-5.7}$ & $13.1^{+2.1}_{-2.0}$ \\
$<24.0$         & $55.0^{+2.4}_{-2.4}$ & $86.9^{+2.0}_{-2.1}$ \\
$[23.0,\;24.0)$ & $75.4^{+3.9}_{-4.3}$ & $30.8^{+2.9}_{-2.8}$ \\
$[23.5,\;24.0)$ & $81.2^{+5.0}_{-5.8}$ & $15.7^{+2.3}_{-2.1}$ \\
$[23.0,\;23.5)$ & $69.7^{+5.7}_{-6.2}$ & $15.3^{+2.3}_{-2.1}$ \\
$[22.0,\;23.0)$ & $56.4^{+5.0}_{-5.1}$ & $19.8^{+2.5}_{-2.4}$ \\
$[22.5,\;23.0)$ & $60.9^{+6.7}_{-6.9}$ & $11.6^{+2.0}_{-1.9}$ \\
$[22.0,\;22.5)$ & $51.1^{+7.4}_{-7.4}$ & $8.5^{+1.8}_{-1.6}$ \\
$<22.0$         & $44.3^{+3.3}_{-3.3}$ & $36.8^{+3.0}_{-2.9}$ \\
\noalign{\smallskip}
\hline
\end{tabular}
\end{center}
\tablecomments{Statistics derived from the mid-IR log(\lfour{}/\ltwo{})~$\geq0.1$ criteria (without invoking any cut in $\rm{log}[L_{X}^{\rm{Obs.}}/L_{12\mu m}]$) defined in Section~\ref{sec:analysis} (Equation~\ref{eq:just_mir_cut}), for various \nh{} bins and sub-bins. Columns 1-3: The same as Table~\ref{table:LXW3_populationstats}.}
\label{table:W4W2_verticalcut}
\end{table}

In a similar fashion, we could define a vertical cut based on the \WISE{} colors in this space:
\begin{equation}
    \rm{log}(L_{22\,\rm{\mu m}}/L_{4.6\,\rm{\mu m}})>0.1
    \label{eq:just_mir_cut}
\end{equation}
and while this criterion also yields a highly complete sample ($88^{+4.5}_{-5.7}$\%) of heavily obscured AGN, the selected sample is only $13.1^{+2.1}_{-2.0}$\% pure for heavily obscured AGNs and is greatly contaminated by moderately obscured, lightly obscured, and even unobscured AGNs. This is not at all surprising, given the scatter in the mid-IR ratio as demonstrated in Figures~\ref{fig:W4W2fullsamplediagram} and \ref{fig:LXLMIR_v_NH_fit}. Mid-IR selection alone is therefore not sufficient when attempting to select \emph{both} a highly complete and fairly pure sample of heavily obscured sources.

Next we defined a slightly more stringent box region (gray dash-dotted line and black dotted line in Panels A--D of Figure~\ref{fig:W42_W32_subbins_box_diagram}), which encompasses the majority of the most heavily absorbed sources with minimal overlap with the unobscured and Compton-thin bins, using the following relations:
\begin{equation}
    \begin{aligned}
        0.1 < \rm{log}(\textit{L}_{22\,\mu m}/\textit{L}_{4.6\,\mu m}) < 1.0  \\
        -2.8 < \rm{log}(\textit{L}_{X,\,\mathrm{Obs.}}/\textit{L}_{12\,\mu m}) < -1.3\\
    \end{aligned}
\label{eq:w4w2eq}
\end{equation}\\
We report the population statistics for this diagnostic box in Table~\ref{table:W4W2_populationstats}. This box offers a completeness of $83.0^{+5.4}_{-6.4}\%$ for the heavily obscured AGN population, a $62.4^{+6.5}_{-6.8}\%$ pure sample, and a mean column density of log($\nhm) = 24.1\pm0.1$. As with Equation~\ref{eq:lxl12_horizontalcut}, the largest source of impurities within this region are AGNs with $23.5\leq\rm{log}(\nhm{})<24$; the region is $\sim90$\% pure for AGNs with log($\nhm{})\geq23.5$ and suffers few impurities from AGNs of lower column densities.

\begin{comment}
\begin{figure*}[t!]
    \centering
    \includegraphics[width=0.85\linewidth]{W42_fullsample_box_subbins2x2_paper.pdf}
    \caption{\lfour{}/\ltwo{} diagnostic for sources in the \swift{} sample binned by \nh{}, where we have unobscured (Panel A; $\log(\nhm{})<22$), Compton-thin `lightly obscured' (Panel B; $22\leq \log(\nhm{})<23$), Compton-thin `moderately obscured' (Panel C; $23\leq \log(\nhm{})<24$), and ``heavily obscured'' to Compton-thick (Panel D; $\log(\nhm{})>24$). The Compton-thin bins are broken into two sub-bins each: $22\leq \log(\nhm{})<22.5$ (black points), $22.5\leq \log(\nhm{})<23$ (blue points), $23\leq \log(\nhm{})<23.5$ (black points), and $23.5\leq \log(\nhm{})<24$ (orange points). Contours were computed to encompass $\sim68\%$ of the population for each respective bin or sub-bin. Note \textit{in general} that the most obscured sources tend to populate a distinctly different region of the parameter space than do the unobscured sources and Compton-thin `lightly obscured' sources, while there is some overlap with Compton-thin `moderately obscured' sources.}
    \label{fig:W42_fullsample_subbins_box_diagram}
\end{figure*}
\end{comment}

\begin{table}[t!]
\caption{\lfour{}/\ltwo{} Diagnostic Box Statistics}
\begin{center}
\begin{tabular}{ccc}
\hline
\hline
\noalign{\smallskip}
\noalign{\smallskip}
$\rm{log}(\nhm{})$ & Completeness & Purity \\
\noalign{\smallskip}
\noalign{\smallskip}
\hline
\noalign{\smallskip}
$\geq24.0$      & $83.0^{+5.4}_{-6.4}$ & $62.4^{+6.5}_{-6.8}$  \\
$<24.0$         & $4.7^{+1.1}_{-1.0}$ & $37.6^{+6.8}_{-6.5}$  \\
$[23.0,\;24.0)$ & $15.4^{+3.6}_{-3.2}$ & $31.8^{+6.6}_{-6.1}$  \\
$[23.5,\;24.0)$ & $28.6^{+6.5}_{-6.0}$ & $28.0^{+6.4}_{-5.9}$  \\
$[23.0,\;23.5)$ & $4.6^{+3.2}_{-2.2}$ & $5.1^{+3.6}_{-2.5}$ \\
$[22.0,\;23.0)$ & $3.9^{+2.3}_{-1.7}$ & $7.0^{+4.0}_{-3.0}$ \\
$[22.5,\;23.0)$ & $5.3^{+3.7}_{-2.5}$ & $5.1^{+3.6}_{-2.5}$ \\
$[22.0,\;22.5)$ & $3.8^{+3.5}_{-2.2}$ & $3.2^{+3.0}_{-1.8}$ \\
$<22.0$         & $0.3^{+0.5}_{-0.2}$ & $1.3^{+2.1}_{-1.0}$ \\
\noalign{\smallskip}
\hline
\end{tabular}
\end{center}
\tablecomments{Statistics derived from the diagnostic box developed for the \lfour{}/\ltwo{} luminosity ratio (defined by Equation~\ref{eq:w4w2eq} in Section~\ref{sec:analysis}) for various \nh{} bins and sub-bins. Columns 1-3: The same as Table~\ref{table:LXW3_populationstats}.}
%STATS UPDATED TO FINAL VERSION 2 JULY 2020
\label{table:W4W2_populationstats}
\end{table}

We repeated this analysis for the alternative \ltwel{}/\ltwo{} luminosity ratio, and we show the contoured populations binned by column density along with an alternative  diagnostic box for heavily absorbed sources in Panels E--H of Figure~\ref{fig:W42_W32_subbins_box_diagram}. We construct this box with the following relations:
\begin{equation}
    \begin{aligned}
        0.0 < \rm{log}(\textit{L}_{12\,\mu m}/\textit{L}_{4.6\,\mu m}) < 0.75 \\
        -2.8 < \rm{log}(\textit{L}_{X,\,\rm{Obs.}}/\textit{L}_{12\,\mu m}) < -1.3\\
    \end{aligned}
\label{eq:w3w2eq}    
\end{equation}\\
and we find that this box yields a completeness of $80.5^{+5.7}_{-6.7}\%$ for heavily absorbed AGNs, a purity of $59.4^{+6.5}_{-6.8}$\%, and a median column density of log($\nhm) = 24.0\pm0.1$. Again, AGNs with column densities $23.5\leq\rm{log}(\nhm{})<24$ contribute the most to the impurity of the sample, while AGNs with lower column densities do not contribute as significantly.

\begin{comment}
\begin{figure*}[t!]
    \centering
    \includegraphics[width=0.85\linewidth]{W32_fullsample_box_subbins2x2.pdf}
    \caption{Illustrated in an identical fashion to that shown in Figure~\ref{fig:W42_fullsample_subbins_box_diagram}, this four-panel figure shows our \ltwel{}/\ltwo{} diagnostic for the sources in the \swift{} sample binned by \nh{}. As with Figure~\ref{fig:W42_fullsample_subbins_box_diagram}, the most obscured sources tend to populate a distinctly different region of the \lxobs{}/\ltwel{}) vs. \ltwel{}/\ltwo{} parameter space than do the unobscured sources and Compton-thin `lightly obscured' sources, although there is some overlap with Compton-thin `moderately obscured' sources.}
    \label{fig:W32_fullsample_subbins_box_diagram}
\end{figure*}
\end{comment}

The diagnostic metrics defined above --- developed using the well-constrained X-ray and mid-IR properties previously found for \swift{} AGNs (e.g. \citealp{ricci2017apjs,ichikawa2017}) --- carve out parameter spaces that yield fairly complete and fairly pure samples of heavily obscured AGNs, offering an efficient and effective method for identifying heavily obscured or CT AGN candidates, particularly in large samples of AGNs.

\section{Discussion}
\label{sec:discussion}

\subsection{The physical origin of the trend in \WISE{} ratios as a function of \nh{}}
\label{sec:physorigin}

The observed trend of the $22\,\rm{\mu m}/4.6\,\rm{\mu m}$ and $12\,\rm{\mu m}/4.6\,\rm{\mu m}$ \WISE{} ratios increasing with column density can be readily understood from dust absorption and emission properties and basics of the radiation transfer. For media optically thin to the mid-IR radiation, the shape of the resulting SED will be determined predominantly by the dust temperature and its gradient throughout the dusty structure. If the dusty medium is optically thick to its own radiation, the outgoing emission will be reshaped due to a number of reasons: (i) Warm dust emission at shorter wavelengths will be absorbed and re-emitted at longer wavelengths. (ii) Dust emission at shorter wavelengths will suffer more extinction than the long-wavelength emission, owing to the wavelength-dependent extinction for typical AGN dust \citep{laor1993}. (iii) Warm dust emission originates closer to the inner rim of the torus, while colder emission originates farther out. As a consequence, longer wavelength mid-IR radiation has to travel a shorter path through the dust before reaching us and thus, suffers even less extinction than the emission of shorter wavelengths. These three effects result in an increased ratio of longer-to-shorter wavelength mid-IR emission and scale with the column density of the medium through which the X-ray radiation is traversing. (iv) Additionally, a disk-like molecular structure in hydrostatic equilibrium is expected to have a vertical gradient \citep{hoenig2019}. In this case, the observed trend of increasing \WISE{} ratios with increasing \nh{} can be explained simply as an inclination effect: the closer our viewing angle is to the equator, the higher is the column density along the line-of-sight and, at the same time, the dust emission becomes ``redder''. All these effects contribute to the observed trend of increasing \WISE{} ratios with \nh{} and also explain why the effect is more pronounced in the $22\,\rm{\mu m}/4.6\,\rm{\mu m}$ than at $12\,\rm{\mu m}/4.6\,\rm{\mu m}$ luminosity ratio \citep[for illustration, see torus model SEDs in, e.g.,][]{honig2010, stalevski2012}.

There are a few caveats: The dust emission is often degenerate, as SEDs of similar shape can be produced by different combinations of the geometrical and physical parameters of the torus, some of which can conspire to work against or hide the trend in luminosity ratios. However, the above reasoning should hold in general since it relies on universal radiative transfer effects. Another deviation can be introduced by the presence of silicate dust grains which exhibit a strong increase of absorption efficiency around $10$\ and $18$\ $\,\rm{\mu m}$. The apparent strength of these features appear in an SED depends on several factors, including the amount of silicates, grain size distribution, and radiative transfer effects.

We illustrate these effects in Figure 6 with a black dashed line, which represents a theoretical expectation for a very simple model: monochromatic radiation passing through a homogeneous screen of dust. 
For this example, we assumed a typical Galactic interstellar dust mixture of silicates and graphite (e.g., \citealt{stalevski2016}). 
The grain size distribution are from \cite{mathis1977} and optical properties from \cite{laor1993} and \cite{li2001}.
The conversion between the optical depth and \nh{} assumes Galactic relation between extinction and column density found by \cite{predehl1995}.
We see that the theoretical curve is following the trend of the data at lower column densities, but is reaching the breaking point sooner. 
This is because the simple dust screen model does not account for a number of radiative transfer effects (self-consistent absorption and re-emission of the thermal IR radiation), which together with geometry of the dusty medium and orientation shape the resulting SED, and thus, the observed trend of luminosity ratios with column density.

\subsection{Comparison to \citet{asmus2015}}
\label{sec:asmus_comp}
Using sub-arcsecond resolution mid-IR observations -- which enabled the isolation of the nuclear mid-IR emission ($F^{\rm{nuc}}_{12\,\rm{\mu m}}$) -- \citet{asmus2015} found a significant correlation between $\rm{log}(F^{\rm{nuc}}_{12\,\rm{\mu m}}\,/\,F^{\rm{obs}}_{2-10\,\rm{keV}})$ and $\rm{log}(\nhm{})$ for 53 AGNs with reliable X-ray observations and column densities log($\nhm{})>22.8$, which was expressed as (see Equation 6 in  \citealp{asmus2015}):
\begin{equation}
\label{eq:asmusrelation}
    \begin{aligned}
        \rm{log}\left(\frac{\textit{N}_{\rm{H}}}{22.8\ \rm{cm}^{-2}}\right) = (0.14\pm0.11)\ \\+\  (0.67\pm0.12)\ \rm{log}\left(\frac{\textit{F}^{\,\rm{nuc}}_{12\,\rm{\mu m}}}{\textit{F}^{\,obs}_{2-10\,keV}}\right)
    \end{aligned}
\end{equation}
This expression is plotted in Figure~\ref{fig:LXLMIR_v_NH_fit} (dashed red line) along with our Equation~\ref{eq:lxl12_nh_eq_fNH} (solid orange line) and the \swift{} sample. Along with values of log($\nhm{}$) derived using Equation~\ref{eq:lxl12_nh_eq_fNH} in Section~\ref{sec:nh_v_lxl12}, we use the relation from \citet{asmus2015} to derive values of log($\nhm{}$) and the uncertainties for specific values of log(\lxobs{}/\ltwel{}) in order to compare to our own results. Despite the fact that \citet{asmus2015} removed AGNs with $\rm{log}(\nhm{})<22.8$ and utilized subarcsecond-resolution mid-IR emission (whereas in this work we utilized lower angular resolution mid-IR photometry for the \swift{} sample), it does appear that the two relations generally agree (within the uncertainties) for ratios of $-0.75\lesssim \rm{log}$(\lxobs{}/\ltwel{}$)\;\lesssim-3.0$. The two relations differ more severely for higher ratios of log(\lxobs{}/\ltwel{}), though this is expected due to (1) the wide range of ratios that unobscured AGNs exhibit and (2) the fact that our relation turns over to account for less obscured sources while the \citet{asmus2015} relation does not take into account less obscured sources.

\subsection{Diagnosing Column Densities with Uncertain Dust Heating Sources}
\label{subsec:sfcontam}
While the diagnostic boxes defined in Section~\ref{sec:analysis} provide a reliable way to identify the most heavily obscured AGNs, star formation activity can contribute non-negligibly to the mid-IR colors of an AGN host. The mid-IR colors assumed to originate from the AGN itself could therefore be overestimated without performing detailed spectral energy decomposition (SED) fitting to differentiate between the AGN and star formation contributions to the mid-IR continuum. Furthermore, \citet{satyapal2018} demonstrated, using \textit{Cloudy} \citep{ferland2013,ferland2017} radiative transfer models, that heavily obscured star formation activity can actually mimic the mid-IR colors of AGNs. These two points suggest that our diagnostic boxes defined in Section~\ref{sec:analysis} may (1) misdiagnose the column density of an AGN if significant star formation is present, as the contaminating stellar emission could lead to much redder colors than the AGN intrinsically exhibits, or (2) mislead us to think an AGN is present in cases where the dominant dust heating sources are actually stellar-related rather than AGN-related \citep{satyapal2018}. To investigate this potential contamination of the diagnostic boxes, we constructed a catalog of optically-selected galaxies whose optical spectroscopic line ratios suggest star formation dominates the observed emission, and we examined methods -- for example mid-IR or X-ray selection criteria -- through which this contamination could be mitigated.

Beginning with the MPA-JHU catalog of galaxy properties \citep[from the SDSS data release 8,][]{aihara2011}, we first selected systems with redshifts $z<0.1$ and included only systems which are classified as star forming systems (``BPTClass'' = 1) based upon their Baldwin, Phillips, Telervich (BPT; \citealp{baldwin1981}) optical spectroscopic emission line ratios. We also removed any systems with QSO and AGN flags within the ``TARGETTYPE,'' ``SPECTROTYPE,'' and ``SUBCLASS'' columns, and then narrowed the sample to only systems with \WISE{} counterparts  and X-ray counterparts from the 4XMM point source catalog (Webb et al., submitted). These criteria yielded a full parent sample of 448 galaxies which we assume are `purely' star forming systems based upon optical spectroscopic measurements. We make no distinction between morphological classes of galaxies.

\begin{figure}
    \centering
    \subfloat{\includegraphics[width=1\linewidth]{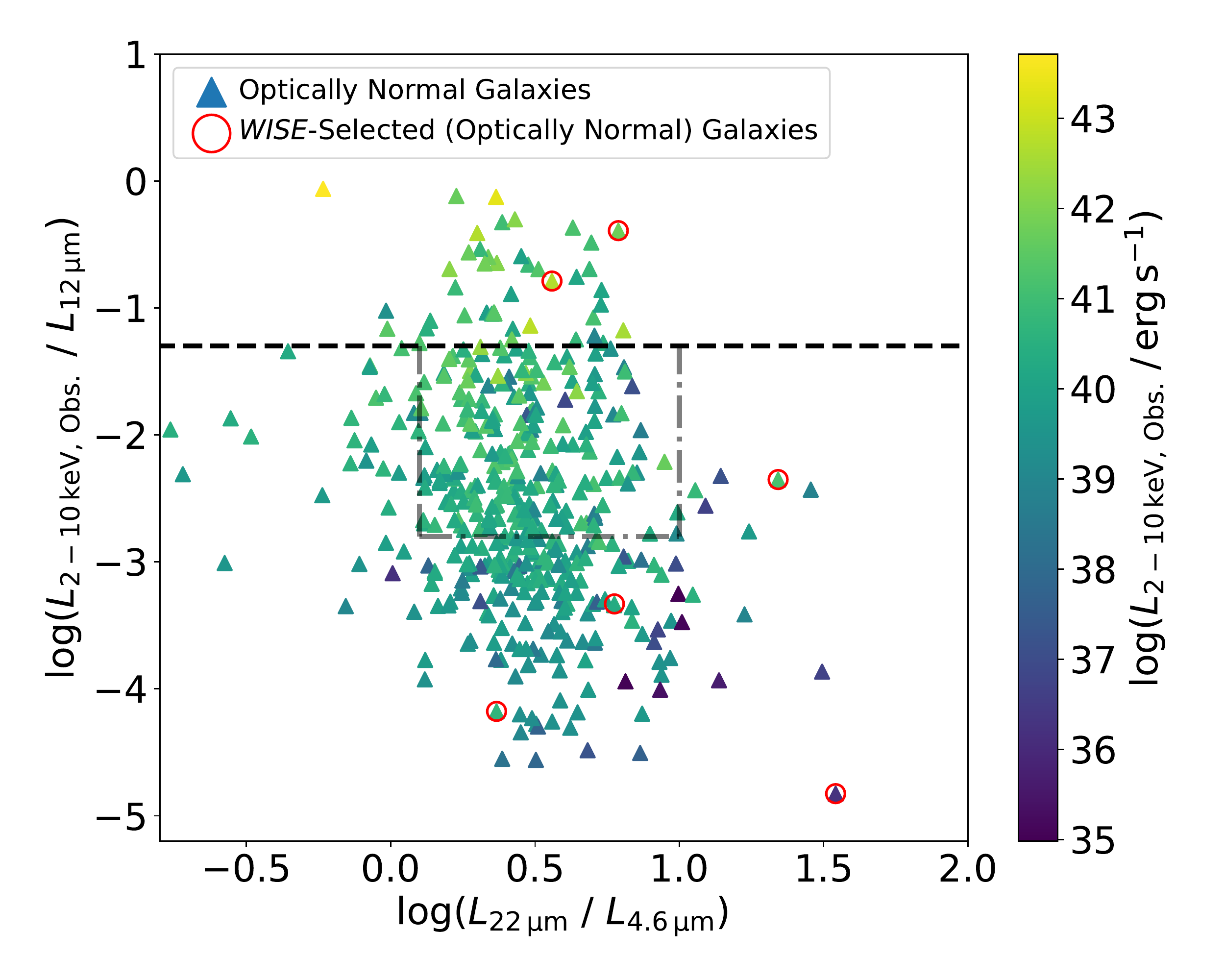}}\\
    \vspace{-0.5cm}
    \subfloat{\includegraphics[width=1\linewidth]{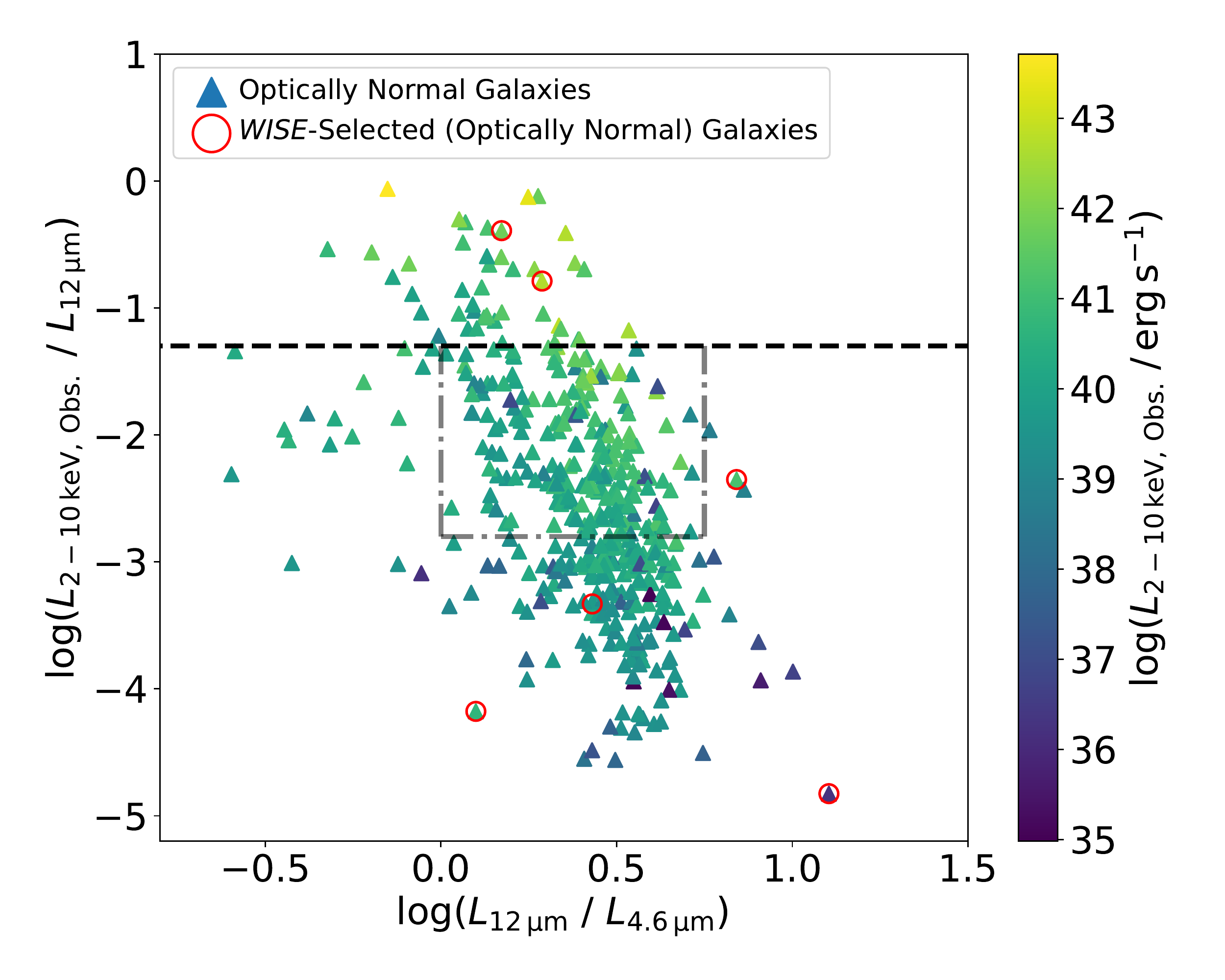}}
    \caption{Top (bottom): Diagnostic box defined in Figure~\ref{fig:W42_W32_subbins_box_diagram} and Equation~\ref{eq:w4w2eq} (Equation~\ref{eq:w3w2eq}) for the \lfour{}/\ltwo{} (\ltwel{}/\ltwo{}) ratio with a population of star forming galaxies (see Section~\ref{subsec:sfcontam}) overlaid as triangles. The horizontal dashed black line is given by Equation~\ref{eq:lxl12_horizontalcut}. The observed X-ray luminosity is denoted on the auxiliary axis. The diagnostics presented here \textit{cannot} unambiguously differentiate between AGNs and star forming systems since a significant fraction of the star forming galaxy population falls within the absorption diagnostic box. This contamination can be mitigated with a mid-IR \WISE{} cut of $W1-W2>0.8$ \citep{stern2012}; only six optically normal galaxies satisfy this mid-IR criteria (red open circles), and these systems fall outside of the diagnostic box defined by Equation~\ref{eq:w4w2eq} (Equation~\ref{eq:w3w2eq}).}
    \label{fig:SFcontam}
\end{figure}

%purity = # true CT classified as CT / (# CT classified as CT + # not obsc. %classified as CT)
%
%(total # classified as CT)

We plot our population of optically-selected star forming galaxies (color coded according to the observed X-ray luminosity) along with the \lfour{}/\ltwo{} and \ltwel{}/\ltwo{} diagnostic boxes in Figure~\ref{fig:SFcontam}. While star formation dominated galaxies tend to exhibit lower ratios of log(\lxobs{}/\ltwel{}) than the majority of the \swift{} sample, they do tend to exhibit similar X-ray deficits as well as \lfour{}/\ltwo{} and \ltwel{}/\ltwo{} mid-IR colors as those exhibited by the heavily obscured \swift{} AGNs; in fact, $44.4\pm2.3\,\%$ of this star forming population overlaps the \lfour{}/\ltwo{} diagnostic region, while $47.1^{+2.3}_{-2.3}\,\%$ of the population overlaps the \ltwel{}/\ltwo{} region. We therefore caution that this diagnostic is emphatically \textit{not} designed to differentiate between star forming and AGN-dominated systems and should \textit{not} be used as a diagnosis of the dominant dust heating source. Nevertheless, the contamination from optically-selected star forming systems can be mitigated through the use of reliable mid-IR and X-ray selection criteria traditionally used for identifying AGNs. 

We applied the two band \WISE{} AGN selection cut ($W1[3.4\,\rm{\mu m}]$--$W2[4.6\,\rm{\mu m}]>0.8$) from \citet{stern2012} to the sample of star forming galaxies, which removed all but six systems (red empty circles, shown in Figure~\ref{fig:SFcontam}). As expected, requiring traditional mid-IR AGN selection criteria eliminates virtually all contamination by optically-selected star forming systems within the diagnostic boxes. While six star forming galaxies ($1.5^{+0.6}_{-0.5}$\% of the total population) succeed in meeting the \citet{stern2012} cut, these do still fall outside of our diagnostic boxes.\footnote{4/6 sources do fall below our cut in \lxobs{}/\ltwel{}: (1) a compact star forming region in a galaxy $\sim318$ Mpc away, and (2) a galaxy at $z=0.07$, (3) a pair of merging galaxies at $z=0.058$, (4) a pair of merging galaxies which actually  host a candidate dual AGN at $z=0.055$ \citep{pfeifle2019}.} Thus, use of mid-IR AGN selection tools could be used to avoid misdiagnosing the dominant photoionization process of the sources within the diagnostic regions. However, it is important to bear in mind that the relation between \lxobs{}/\ltwel{}, \lfour{}/\ltwo{} (and \ltwel{}/\ltwo{}), and \nh{} holds true for \textit{both} \WISE{} \textit{and} non-\WISE{} selected AGNs (see Appendix), and therefore requiring a \WISE{} cut could in general remove true AGNs as well as star forming systems. For example, imposing the $W1$--$W2>0.8$ cut on the \swift{} sample examined in this work would remove 240 systems (or $52.6^{+2.3}_{-2.3}$\% of the parent sample of 456 AGNs); in the parent sample of 456, there are 71 AGNs which possess column densities in excess of $5\times10^{23}$~cm$^{-2}$, and 40 of these would be removed with this mid-IR cut.

We also found that requiring an observed X-ray luminosity of \lxobs{}$>10^{42}$~erg~s$^{-1}$ removes nearly all of the optically star forming population from the diagnostic region (5 galaxies, or $1.3^{+0.6}_{-0.5}$\%, remain within the \lfour{}/\ltwo{} diagnostic box), though this method must also be used judiciously to avoid removing heavily obscured AGNs, which could exhibit lower X-ray luminosities.

Ideally, the usage of this diagnostic should be limited to systems whose dominant photoionization processes are unambiguous or for which detailed SED fitting can be performed to differentiate between AGN and host emission. Otherwise, we recommend proceeding cautiously, taking into account the various caveats outlined above to avoid inaccurate estimations of the obscuration  along the line-of-sight. 

\subsection{Mid-IR Emission Contributions from Galaxies Hosting Obscured AGNs}
\label{sec:hostgal_contributions}
\begin{figure}[t]
    \centering
    \includegraphics[width=1.\linewidth]{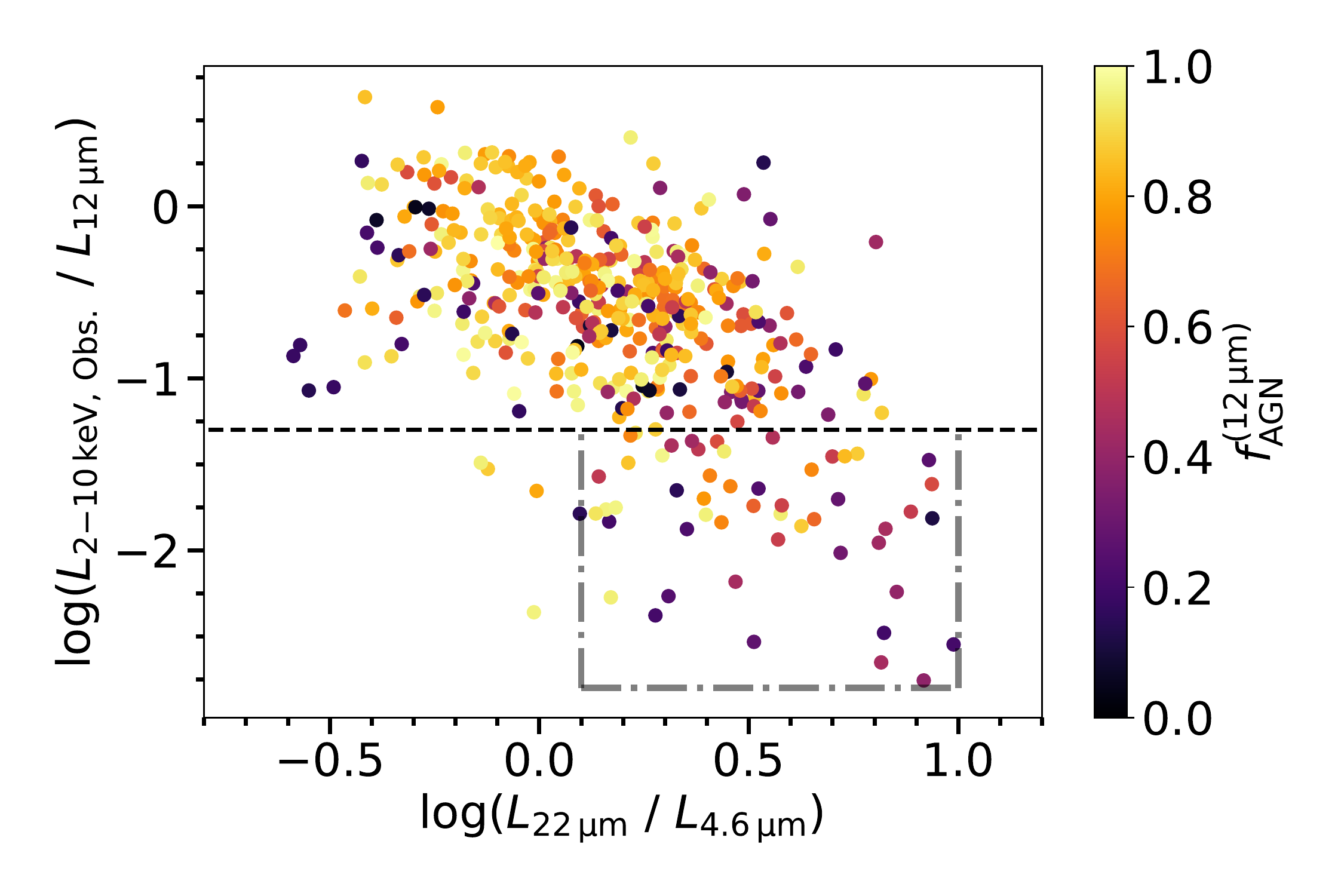}
    \caption{The logarithmic \lxobs{}/\ltwel{} vs. \lfour{}/\ltwo{} ratios of the \swift{} AGN sample, where the data are color-coded according to the fractional contribution of the AGN $12\,\mu m$ emission to the total observed $12\,\rm{\mu m}$ emission ($f^{12\,\rm{\mu m}}_{\rm{AGN}}$). While there is no clear offset in this parameter space between AGN-dominated and host-dominated systems, several of the most heavily obscured AGNs (i.e. sources with significant X-ray deficits and log(\lfour{}/\ltwo{}) $>0.1$) do reside in systems where the host dominates the $12\,\rm{\mu m}$ emission.}
    \label{fig:12um_contributions}
\end{figure}

The realization in Section~\ref{subsec:sfcontam} that optically star forming galaxies, which presumably do not host AGNs, can exhibit luminosity ratios similar to those exhibited by more heavily obscured or CT \swift{} AGNs raises the intriguing point of how much host galaxies may contribute to the observed luminosity ratios derived for the \swift{} AGNs. Figure~\ref{fig:12um_contributions} shows the log(\lxobs{}/\ltwel{}) and log(\lfour{}/\ltwo{}) ratios of the \swift{} AGNs and is color-coded according to the fractional contribution by the AGN to the observed 12\,$\mu \rm{m}$ emission ($f^{12\,\mu \rm{m}}_{\rm{AGN}}$) derived through detailed SED fitting in \citet{ichikawa2019}. While host-dominated systems at 12\,$\mu \rm{m}$ ($f^{12\,\mu \rm{m}}_{\rm{AGN}}<0.5$) can be found across this parameter space, a significant fraction ($43.3^{+6.9}_{-6.7}$\%, 22 out of 51) of AGNs within the diagnostic region (Equation~\ref{eq:w4w2eq}) reside in host-dominated systems. This suggests that the host galaxies could contribute significantly to the observed mid-IR colors of the heavily obscured AGN population in particular, presumably via dust emission heated through star formation.

\begin{figure}[t!]
    \centering
    \subfloat{\includegraphics[width=1\linewidth]{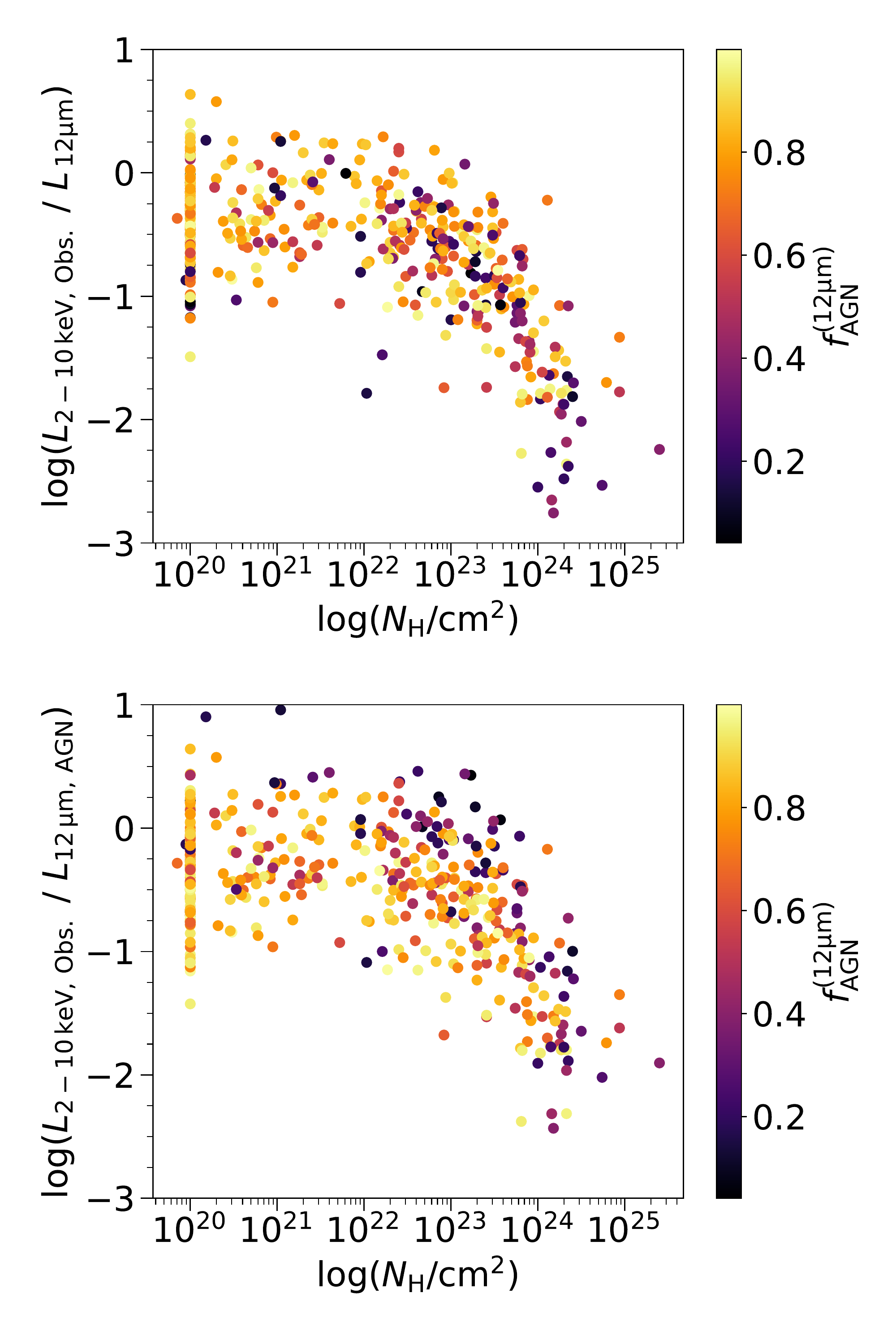}}\\
    %\vspace{-0.5cm}
    %\subfloat{\includegraphics[width=1\linewidth]{fig.pdf}}
    \caption{log$\nhm{}$) vs.  log(\lxobs{}/\ltwel{}) for (top) the total 12\,$\mu \rm{m}$ emission and (bottom) the decomposed AGN 12\,$\mu \rm{m}$ emission from \citet{ichikawa2019}. The auxiliary axes represent the fractional contribution by the AGN to the total observed 12\,$\mu \rm{m}$ emission. (Top) As is already known, log(\lxobs{}/\ltwel{}) decreases with increasing column density, however a significant number of the obscured and CT AGNs contribute less than $50\%$ of the total observed 12\,$\mu \rm{m}$ ($f^{12\,\mu \rm{m}}_{\rm{AGN}}$). (Bottom) The relationship between log(\lxobs{}/\ltwel{}) and log($\nhm{}$) is still present when recalculating the ratio using the decomposed AGN 12\,$\mu \rm{m}$ luminosity ($L_{12\,\rm{\mu m,\,AGN}}$), although obscured and CT AGNs exhibit smaller deficits than when using the total 12\,$\mu \rm{m}$ luminosity. Host galaxies can therefore contribute significantly to the observed X-ray to mid-IR ratios of AGNs, especially CT AGNs.}
    \label{fig:l12um_AGN_host_contrib}
\end{figure}

\citet{ichikawa2019} provided decomposed logarithmic AGN 12\,$\mu \rm{m}$ luminosities for the \swift{} AGNs, which we can use here to examine how the log(\lxobs{}/\ltwel{}) ratios may change if we use the AGN 12\,$\mu \rm{m}$ luminosity ($L_{12\,\rm{\mu m,\,AGN}}$) instead of the total observed 12\,$\mu \rm{m}$ luminosity (\ltwel{}).  Figure~\ref{fig:l12um_AGN_host_contrib} (top panel) shows that host-dominated systems are predominantly occupied by AGNs with log($\nhm{})\gtrsim22.5$; half of the CT \swift{} AGNs reside in host dominated systems. After recalculating the luminosity ratio using $L_{12\,\rm{\mu m,\,AGN}}$ instead (bottom panel), host-dominated systems exhibit a shift toward higher luminosity ratios, with an average difference in ratio of $\Delta$log(\lxobs{}/\ltwel{}$)\approx0.5$ for AGNs with $f^{12\,\mu \rm{m}}_{\rm{AGN}}<0.5$, although we note that these shifts are \emph{not} limited only to heavily obscured AGNs. Here we have assumed that the X-ray emission is AGN-dominated, rather than host-dominated; in reality, if some portion of the X-ray emission is due to the host as well, the observed ratio shifts will not be as large. 

Figure~\ref{fig:12um_contributions} demonstrates that host galaxies do indeed contribute significantly to the diagnostic ratios probed in this work, at least for systems in which the host dominates the mid-IR emission at 12\,$\mu \rm{m}$. In these cases, the host contribution to the mid-IR leads to a perceived larger X-ray deficit for the AGN at a given column density. It does appear, though, that generally this effect actually works in our favor when attempting to identify CT AGNs, as these more severe X-ray deficits and presumably `redder' mid-IR colors aid in separating this population from less obscured populations in color space. Decomposed AGN 22\,$\mu \rm{m}$ and 4.6\,$\mu \rm{m}$ luminosities were not included in Table~1 of \citet{ichikawa2019} and therefore could not be examined in a similar fashion here. While it is beyond the scope of this paper, an analysis of the interplay between the mid-IR colors, host galaxy emission, and AGN emission with regard to the selection diagnostics presented in this work should be performed more rigorously in a future study.

\subsection{Comparison to \citet{kilercieser2020}}
\label{sec:eser_comparison}

In a very recent study, \citet{kilercieser2020} selected a subsample of the 105 month \swift{} catalog \citep{oh2018} and proposed a new selection method for CT AGNs using mid-IR and far-IR photometry. They report, as we do here in Section~\ref{sec:analysis} of this study, a shift in infrared colors (specifically mid-IR \textit{and} far-IR) toward `redder' colors with increasing column density, and they define a physically motivated color-color diagram (see Figure~11, henceforth $F11$, in \citealp{kilercieser2020}) and selection method using the [$9\,\rm{\mu m}$]--[$22\,\rm{\mu m}$] and [$22\,\rm{\mu m}$]--[$90\,\rm{\mu m}$] colors. Of the 32 CT \swift{} AGN for which there exists the relevant photometry, this selection criteria identifies four CT AGNs (a success rate of $14.0^{+6.6}_{-5.2}\%$). However, it is evident from $F11$ that these color cuts cannot reliably distinguish between unobscured, obscured, and CT AGNs, as the AGNs from these three different obscuration bins largely occupy the same [$9\,\rm{\mu m}$]--[$22\,\rm{\mu m}$] and [$22\,\rm{\mu m}$]--[$90\,\rm{\mu m}$] parameter space. In the case of \swift{}, the color-color criteria proposed by \citet{kilercieser2020} yields a far lower success rate of identifying heavily obscured and CT AGNs than the criteria set forth in this work (e.g. Equations~\ref{eq:w4w2eq}~and~\ref{eq:w3w2eq}; see Tables~\ref{table:LXW3_populationstats},~\ref{table:W4W2_populationstats}, and \ref{table:W3W2_populationstats}).
%(see Section~2 of \citealp{kilercieser2020} for further details on the construction of their sample)
%\footnote{Based on $F11$, there does not appear to be a clear trend between [$9\,\rm{\mu m}$]--[$22\,\rm{\mu m}$] and \nh{}. While there is a plot of [$22\,\rm{\mu m}$]--[$90\,\rm{\mu m}$] vs. \nh{} in \citet{kilercieser2020}, there is no plot showing [$9\,\rm{\mu m}$]--[$22\,\rm{\mu m}$] vs. \nh{}.}
%and compared the infrared properties and column densities of the AGNs in an effort to identify

To further test their diagnostic, they applied this color selection criteria to the \textit{AKARI} infrared galaxy catalog developed in \citet{kilercieser2018},  which contains over 17,000 galaxies, and recover one known CT AGN \citep[NGC\,4418, e.g.][]{sakamoto2013}.\footnote{The selection criteria also recovers two other sources: NGC\,7714, an unobscured AGN \citep{gonzalez-delgado1995,smith2005}, and NGC\,1614, which has no clear evidence of an AGN \citep[e.g.][]{xu2015,pereira-santaella2011,herrero-illana2014}.} The remainder of the infrared galaxy sample of \citet{kilercieser2018} is represented with blue contours in $F11$ that partially overlap a significant number of \swift{} CT, obscured, and unobscured AGNs, suggesting that some portion of these infrared galaxies may in fact host heavily obscured AGNs. Despite finding a few cases of CT AGNs between the \swift{} and \citet{kilercieser2018} samples, the diagnostic criteria set forth in \citet{kilercieser2020} does not appear to provide a complete or reliable (see Section~5.4 of \citealp{kilercieser2020}) method of selecting CT AGNs. 
%, which are displayed in $F11$ as black dots
%\footnote{There is no description of how this sample was derived in \citet{kilercieser2020}, and therefore the origin of this additional sample is ambiguous. No AGN selection criteria were applied to this sample.}
%\footnote{A note of caution for the reader: despite the fact that these remaining 338 sources are definitively not AGNs since they are Galactic sources, they are still denoted as ``CT AGN candidates'' in $F11$.}
%A sample of 341 \textit{AKARI} and \WISE{} sources are also selected by the color criteria defined by \citet{kilercieser2020}; only one of these sources is a confirmed CT AGN, while at least 338 out of 341 of these sources are Galactic IR sources (see Section~5.4 of \citealp{kilercieser2020}).

\begin{figure}
    \centering
    \includegraphics[width=1.\linewidth]{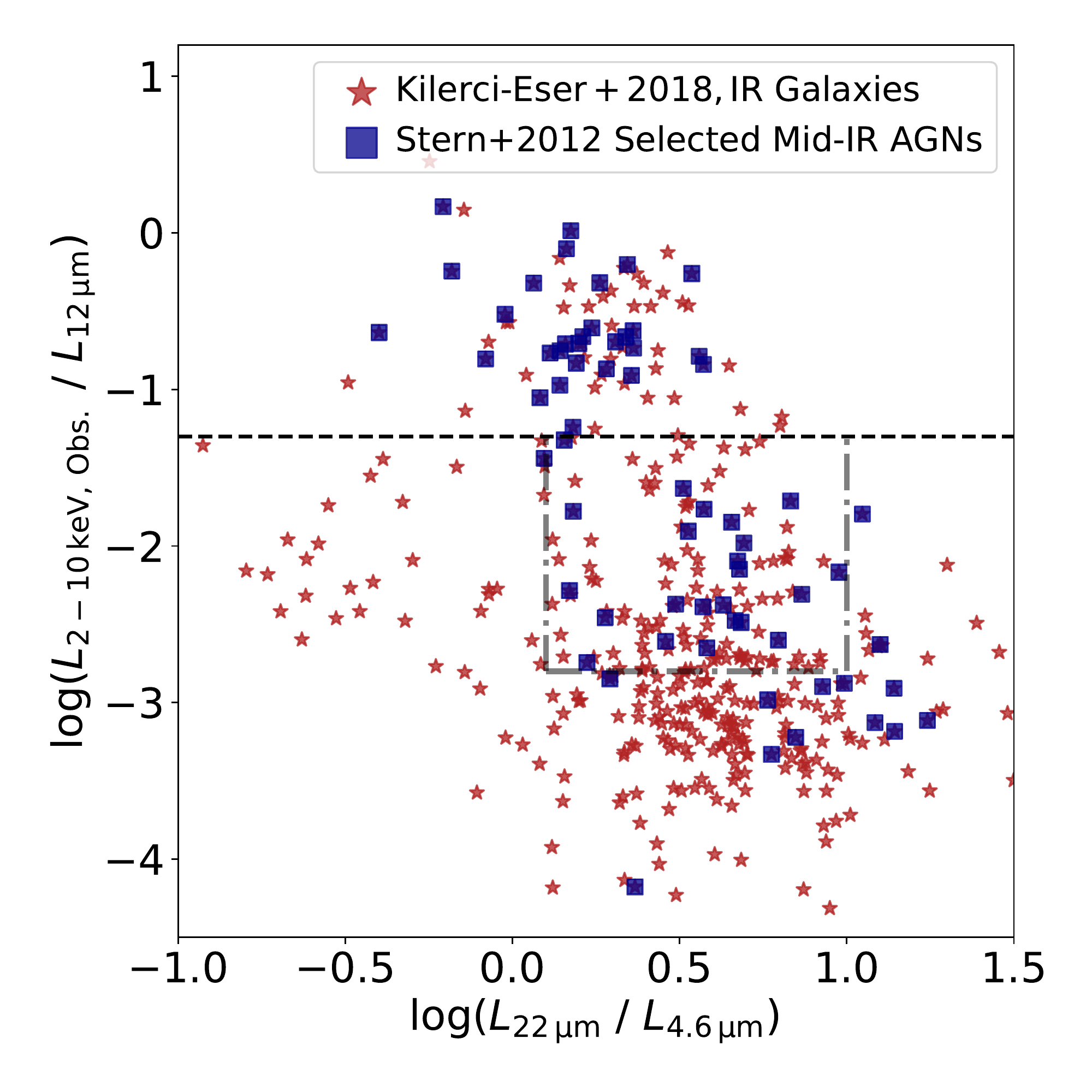}
    \caption{The \lxobs{}/\ltwel{} vs. \lfour{}/\ltwo{} ratios for the infrared galaxies (red stars) cataloged by \citet{kilercieser2018} along with the obscuration diagnostics established in Equations~\ref{eq:lxl12_horizontalcut} and \ref{eq:w4w2eq}. After applying the \citet{stern2012} mid-IR cut to search for AGNs within the sample, we find a significant population of candidate heavily obscured or CT AGNs contained within the \citet{kilercieser2018} catalog, a result not found using the mid-IR to far-IR color-color criteria proposed by \citet{kilercieser2020}. See Table~\ref{table:appendix_candCT_KE} for more details on these AGNs.}
    \label{fig:eser_fig}
\end{figure}

As an additional comparison between our selection method and that proposed by \citet{kilercieser2020}, we turned our attention to the infrared galaxy catalog from \citet{kilercieser2018}. We matched this sample to the AllWISE catalog and the 4XMM DR9 \xmm{} Serendipitous Source Catalog (Webb et al. submitted), using a match radius of 10\arcsec{} for each, which yielded a sample of 401 local ($z<0.1$) infrared galaxies with \xmm{} and \WISE{} detections. In Figure~\ref{fig:eser_fig}, we show the resulting sample of infrared galaxies (red stars), along with our diagnostic criteria from Equations~\ref{eq:lxl12_horizontalcut}~and~\ref{eq:w4w2eq}. As in Section~\ref{subsec:sfcontam}, it is impossible to discern whether or not any of these galaxies host AGNs without a reliable method for removing star formation dominated systems. We tried four different mid-IR AGN selection criteria, defined in \citet{jarrett2011}, \citet{stern2012}, \citet{assef2018}, and \citet{satyapal2018} (with the understanding that some heavily obscured AGNs will be missed with this simple  approach), and in all four cases we recover a significant number of candidate heavily obscured or CT AGNs. We show mid-IR AGNs selected as a result of the \citet{stern2012} cut in Figure~\ref{fig:eser_fig} (blue squares); these candidate CT AGNs likely inhabited the blue contoured prominence that overlapped the \swift{} AGNs in $F11$, but were missed due to fact that they did not satisfy the criteria proposed by \citet{kilercieser2020}.

Table~\ref{table:appendix_candCT_KE} in the Appendix lists the 36 candidate CT AGNs selected from \citet{kilercieser2018} using the \lfour{}/\ltwo{} diagnostic region and at least one of the mid-IR selection cuts listed above. We include in the table the source coordinates, redshifts, luminosity ratios, and alternative identifiers, and we also categorize the column densities of the sources (using the column density bins defined in Section~\ref{sec:diagnosticregions}) based upon any available measurements in the literature. Of the candidates that have inferred or directly measured column densities in the literature (24/36), we find eight CT AGNs, six moderately obscured AGNs, two lightly obscured AGNs, and three unobscured AGN, while a remaining five AGNs have conflicting measurements of \nh{} in the literature (all five of which have been reported as CT at least once in the past).\footnote{There are other moderately and heavily obscured \WISE{} AGNs in this sample which did exhibit log(\lx{}/\ltwel{})$<-1.3$ but fall outside of our more stringent diagnostic region.} Therefore, we conclude that our diagnostic criteria proposed in Equations~\ref{eq:lxl12_horizontalcut},~\ref{eq:w4w2eq}, and~\ref{eq:w3w2eq} offer a more reliable method for identifying candidate CT AGNs than the mid-IR to far-IR color criteria proposed by \citet{kilercieser2020}.

In light of Section~\ref{sec:hostgal_contributions}, we caution that some portion of these AGNs may not dominate the observed $12~\rm{\mu m}$ emission and, therefore, may exhibit larger X-ray deficits and redder mid-IR colors than might be expected for the AGN alone due to additional mid-IR contributions from the host galaxy.

\subsection{Diagnosis of \nh{} in the XXM-XXL Field}
\label{sec:xxl}

To test the power of our absorption diagnostic, we turn our attention to its application in the \textit{XMM} XXL North field \citep{pierre2016,pierre2017}. \citet{menzel2016} presented a rigorous multiwavelength analysis of 8445 X-ray sources detected by \xmm{} in an 18 deg$^2$ area of the \textit{XMM} XXL North field (hereafter XXL-N), with a limiting flux of $F_{0.5-10\ \rm{keV}}>10^{-15}$~erg~cm$^{-2}$~s$^{-1}$, providing optical Sloan Digital Sky Survey (SDSS) and mid-IR \WISE{} counterparts to the \xmm{} sources. In a complementary investigation, \citet{zliu2016} presented a thorough X-ray spectral analysis of the 2512 XXL-N AGNs, deriving the spectral properties (e.g. photon index $\Gamma$, \nh{}, \lxobs{}) for those AGNs using a Bayesian statistical approach contained within the Bayesian X-ray Astronomy (BXA) software package \citep{buchner2014}. 

\begin{figure*}[t]
    \centering
    \subfloat{\includegraphics[width=0.5\linewidth]{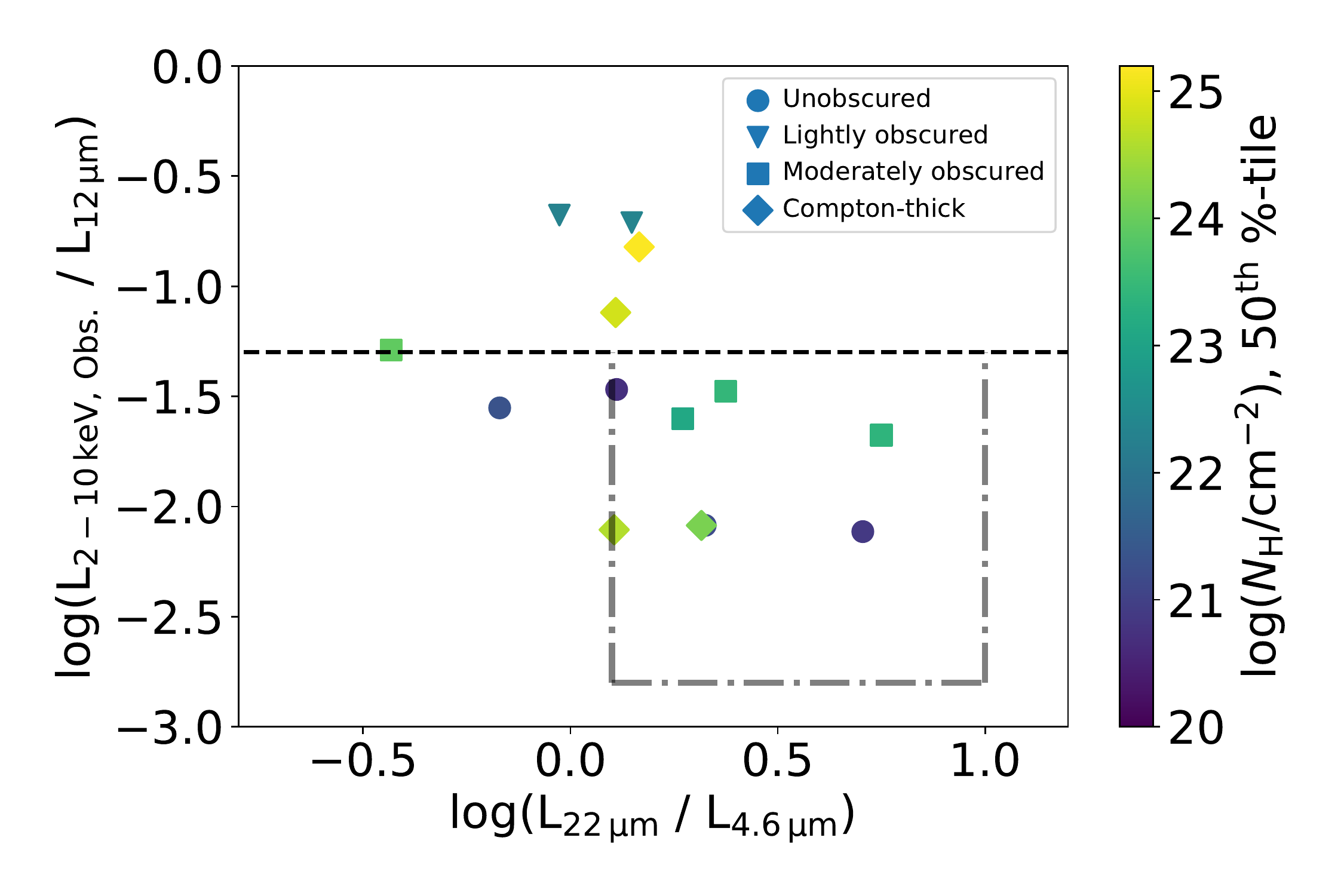}}
    \subfloat{\includegraphics[width=0.5\linewidth]{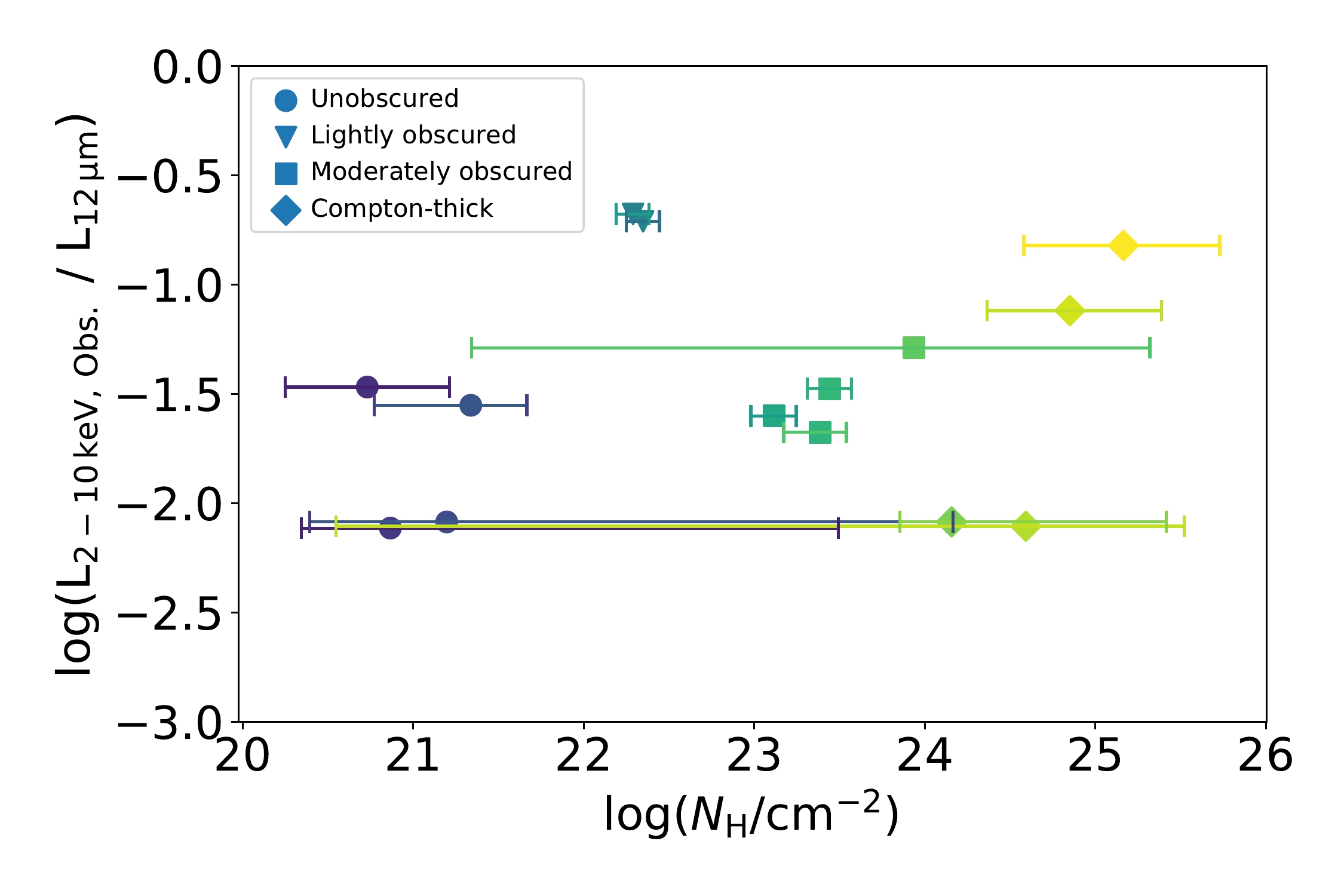}}\\
    \caption{(Left) \lxobs{}/\ltwel{} vs. \lfour{}/\ltwo{} logarithmic ratios with each point color-coded to indicate the $50^{\rm{th}}$ percentile log($\nhm{}$) value. (Right) Logarithmic \lxobs{}/\ltwel{} ratio vs. log($\nhm{}$), where the $50^{\rm{th}}$ percentile log($\nhm{}$) values and error values ($16^{\rm{th}}$ and $84^{\rm{th}}$ percentiles) are drawn from \citet{zliu2016}, and the data points are color-coded according to the auxiliary axis to the left. To aid the reader, we use different markers to denote the obscuration bin for each source. Unlike the \swift{} AGNs, there is not as clear of relationship between the luminosity ratios and column density for the XXL-N AGNs, though this comparison should be viewed with caution because of the large uncertainties on log($\nhm{}$) and because these AGNs were selected with a softer energy band (0.3--10 keV).}
    \label{fig:xxl_localz}
\end{figure*}

We combined the catalogs from \citet{menzel2016} and \citet{zliu2016} to obtain the observed 2--10~keV luminosities and mid-IR \WISE{} magnitudes, from which we derived the relevant luminosity ratios examined in Section~\ref{sec:analysis}. Initially we limited the XXL-N sample to only local ($z<0.1$) AGNs, and as with the \swift{} AGNs, we did not employ any mid-IR or X-ray selection criteria. We plot the resulting luminosity ratios of the low redshift AGNs from the XXL-N field in the left panel of Figure~\ref{fig:xxl_localz} along with our \lfour{}/\ltwo{} diagnostic box (Equation~\ref{eq:w4w2eq}) and horizontal cut in \lxobs{}/\ltwel{} (Equation~\ref{eq:lxl12_horizontalcut}). The data and auxiliary axis in Figure~\ref{fig:xxl_localz} are color coded to represent the derived $50^{\rm{th}}$-percentile \nh{} values from the \citet{zliu2016} catalog and the markers denote the obscuration bin for each AGN. In the right panel of Figure~\ref{fig:xxl_localz} we plot the \lxobs{}/\ltwel{} ratio against the derived log($\nhm{}$) values from \citet{zliu2016}, where the error bars represent the $16^{\rm{th}}$ and  $84^{\rm{th}}$ percentiles and the data points use the same marker and color scheme as the left panel. A dearth of low redshift AGNs in the XXL-N field is immediately apparent, and while it appears that most of the more obscured AGNs do exhibit ``redder'' colors, there is some overlap in \lxobs{}/\ltwel{} ratios exhibited by AGNs of starkly different obscuration bins, e.g. heavily obscured and unobscured AGNs. Unfortunately, we cannot draw definitive conclusions about the reliability of our diagnostic for the XXL-N field with such poor AGN statistics. 

We repeated this analysis for the entire sample of XXL-N AGNs, breaking the sample into redshift bins of $\Delta z=0.5$ each. While a small number of obscured AGNs overlap with the diagnostic region defined by Equation~\ref{eq:w4w2eq}, the majority of the heavily absorbed AGNs still occupy the same parameter space as the Compton-thin and unobscured AGNs, even in the local ($z<0.5$) redshift bin, in stark contrast to the results found with the \swift{} AGNs. We find a very similar result when examining the \ltwel{}/\ltwo{} diagnostic ratio. Due to the fact that at higher redshift the \lfour{}/\ltwo{} and \ltwel{}/\ltwo{} diagnostics do not correspond to the same wavelength ranges as they do at local redshifts, we then turned to the luminosity ratio of $L_{12\,\rm{\mu m}}/L_{3.4\,\rm{\mu m}}$. Yet again the heavily absorbed AGNs generally occupy the same parameter space as unobscured AGNs. The spectral curvature method \citep{kossapj2016} may provide a more effective means of selecting heavily obscured AGNs at higher redshift in the XXL-N field, and indeed \citet{baronchelli2017} demonstrated its effectiveness in selecting high redshift ($z>2$) CT AGNs in both the \chandra{} Deep Field South and the \chandra{} COSMOS legacy survey.

There are a few explanations for why the luminosity ratios of the XXL-N AGNs do not as clearly differentiate between obscuration levels like that seen for the \swift{} AGNs. For one, there may simply not be enough local redshift XXL-N AGNs for a proper statistical comparison to the results found for the \swift{} AGNs. Secondly, it is possible that the chosen redshift bins are not fine enough and we are including AGNs across too large of redshift ranges. After splitting the $z<0.5$ bin into five sub-bins, however, we find the same result as before: the unobscured and heavily obscured AGNs coexist within the parameter space.

Another explanation lies in the reliability of the results of the X-ray spectral fitting performed in \citet{zliu2016}: while the Bayesian statistical framework employed in that work is a powerful method for constraining the spectral properties for AGNs with low counts, our results suggest that the column densities derived for the AGNs may still be inaccurate due to the low counts acquired. For example, for all AGNs detected in the $z<0.5$ bin, the median number of counts detected in the \textsc{epic} \textsc{pn} and in the two \textsc{epic} \textsc{mos} detectors is only $16.8$ and $15.1$ counts, respectively. Furthermore, identification of Compton-thick AGNs via \xmm{} spectroscopy is quite difficult due to the softer X-ray passband (0.3-10 keV) probed with the \xmm{} imaging. It can be very difficult to distinguish between the scenario in which the source is heavily obscured, and the X-ray emission is dominated by reprocessed radiation from the circumnuclear material, and that in which the source is unobscured but the emission is dominated by relativistic reflection from the accretion disk when using low signal-to-noise X-ray spectra alone due to the strong model-dependent degeneracies involved (see e.g., \citealp{gandhi2009}, \citealp{treister2009}). Our mid-IR ratio predictions for the highest X-ray column densities can hence be very complementary to help classify sources in which unobscured and obscured reflection models can fit an observed X-ray spectrum equally well. 

Future, deeper \xmm{} or \nustar{} follow-up observations could improve upon the current photon statistics and could provide robust constraints on the column densities for at least the local XXL-N AGNs which lie within the diagnostic boxes. For now, we identify these systems as candidate heavily obscured or CT AGNs, and we list their spectral properties in  Table~\ref{apptab:CTcandidates}.

\section{Conclusions}
\label{sec:conclusions}
Using the well-studied \swift{} sample of ultra hard X-ray selected AGNs, we have presented an analysis of the AGN \lxobs{}/\ltwel{} luminosity ratio and two different mid-IR luminosity ratios: \lfour{}/\ltwo{} and \ltwel{}/\ltwo{}. Using the well constrained X-ray \citep{ricci2017apjs} and mid-IR \citep{ichikawa2017} properties of the \swift{} AGNs, we probed the utility of these luminosity ratios as tools for inferring line-of-sight column densities and identifying the most heavily obscured AGNs in the local Universe ($z<0.1)$. We summarize the results of our analysis as follows:
\begin{itemize}
    \item We have derived expressions relating the column density \nh{} to both the \lxobs{}/\ltwel{} and \lfour{}/\ltwo{} luminosity ratios, which are defined by Equations~\ref{eq:lxl12_nh_eq_fNH} and \ref{eq:l22l46_nh_eq_fNH}. These expressions can be inverted to give \nh{} as a function of the luminosity ratios. We provide these expressions again here:
    \begin{equation*}
        \begin{aligned}
            \rm{log}(N_{\rm{H}}\,/\,\rm{cm}^{-2}) = 20\; +\; (1.61^{+0.33}_{-0.31})\; \times \\\rm{log}\left(\left|\frac{\rm{log}\left(\frac{L_{X,\,\rm{Obs.}}}{L_{12\,\mu m}}\right) +(0.34^{+0.06}_{-0.06})}{(-0.003^{+0.002}_{-0.005})}\right|\right)
        \end{aligned}
    \end{equation*}
    and
    \begin{equation*}
        \begin{aligned}
            \rm{log}(N_{\rm{H}}\,/\,\rm{cm}^{-2}) = 20\;+\;(3.86^{+1.94}_{-1.00})\; \times \\     \rm{log}\left(\left|\frac{\rm{log}\left(\frac{L_{22\,\rm{\mu m}}}{L_{4.6\,\rm{\mu m}}}\right)    -(0.04^{+0.02}_{-0.02})}{(0.03^{+0.02}_{-0.02})}\right|\right)
        \end{aligned}
    \end{equation*}
    \item We have demonstrated that unobscured and heavily obscured AGNs tend to exhibit different \lxobs{}/\ltwel{}, \lfour{}/\ltwo{}, and \newline \ltwel{}/\ltwo{} luminosity ratios. All three of our diagnostic regions (Equations~\ref{eq:lxl12_horizontalcut}, \ref{eq:w4w2eq}, and \ref{eq:w3w2eq}) identify (in general) the most heavily absorbed AGNs, with average column densities of $\log(\nhm{})\geq24.0$ for each defined parameter space. These regions are all $\gtrsim80$\% complete and $\gtrsim60$\% pure for AGNs with log($\nhm{})\geq24$. The greatest impurities arise due to AGNs with $23.5\lesssim\rm{log(}\nhm{})<24$; these regions are $\gtrsim85$\% pure for AGNs with $\rm{log}(\nhm{})\gtrsim23.5$.
    \item While optically star forming systems can fall within our diagnostic regions, this contamination can be virtually eliminated via mid-IR or X-ray selection criteria. Such selection criteria should be used judiciously to avoid removing non-mid-IR AGNs. These diagnostic regions should \emph{not} be used to differentiate between AGNs and galaxies dominated by star formation.
    \item \swift{} AGNs which do not dominate the total observed $12~\rm{\mu m}$ emission tend to exhibit redder colors and larger X-ray deficits with increasing column density, suggesting that host galaxy contributions to at least the mid-IR emission can be a significant factor in the luminosity ratios examined here, particularly in the case of mildly obscured and CT AGNs selected with our diagnostics. However, it appears that this effect actually aids in the identification of CT AGNs, as the host contributions result in a larger separation in color space between less obscured and more obscured AGNs.
    \item We find that the selection criteria proposed here are more reliable at identifying obscured and CT AGNs than the mid-IR and far-IR selection criteria proposed by \citet{kilercieser2020}. We identify several known obscured and CT AGNs, as well as several candidate CT AGNs, within the IR galaxy catalog of \citet{kilercieser2018} (see Table~\ref{table:appendix_candCT_KE}).
    \item We applied our diagnostics to the \xmm{} XXL-N field and found, in contrast to \swift{} AGNs, that obscured and unobscured XXL-N AGNs do not appear to exhibit distinctly different luminosity ratios. This disparity could be due to poor photon statistics or the softer X-ray energies probed for the XXL-N AGNs, which could lead to inaccurate column density values. Although, given the small number of $z<0.1$ XXL-N AGNs and the large errors associated with several \nh{} values, this comparison should be viewed with caution.
\end{itemize}

Identifying heavily obscured AGNs remains an important yet difficult task, though the study of such AGNs is an important step in the development of our understanding of the evolution of AGNs. In a future study, we could expand our analysis to include diagnostic regions appropriately modified to differentiate between unobscured and heavily obscured sources at higher redshift, although it is difficult to speculate at the moment how the emission ratios may change with redshift, as both star formation activity and AGN activity are expected to increase with redshift.

The selection criteria presented here offers a complementary approach to the spectral curvature method developed in \citet{kossapj2016}, which is very effective at selecting heavily obscured AGNs at local-$z$, with the caveats that one must already have hard X-ray ($>10$ keV) measurements with \nustar{} or \swift{} and that it is most effective for brighter AGNs. Softer X-ray missions can only be utilized for higher redshift sources ($z\sim3$) for which the hard X-ray emission has shifted into the rest frame 10--30 keV passband. The diagnostics presented here, on the other hand, do not require higher energy passbands in order to select local-$z$ sources and can take advantage of softer X-ray missions such as \chandra{} and \xmm{}. The synergy between these two approaches is best summed up by the fact that they select many of the same sources using \emph{different} passbands, and that they select heavily obscured AGNs \emph{missed by one another} (see Figure~\ref{fig:app_sccomp} in the Appendix), yielding a more complete census of heavily obscured \swift{} AGNs overall.

The diagnostic regions proposed in this study, as well as the expressions derived relating \nh{} to the \lxobs{}/\ltwel{} and \lfour{}/\ltwo{} luminosity ratios, could be used to differentiate between unobscured and heavily obscured AGNs in future, large samples of AGNs, such as those now being detected by the eROSITA all-sky survey \citep{predehl2010,merloni2012}. In particular, the eROSITA survey will provide the first all-sky X-ray imaging survey at energies up to 10~keV, yielding a highly complementary catalog to those of other all-sky missions, such as \WISE{}. Future works could cross-match the \WISE{} and eROSITA catalogs and use the diagnostics presented here to identify many more cases of CT AGN candidates, select targets for deeper follow-up multiwavelength observations, and to compute the CT fraction for the future sample, all of which will be crucial in the quest to construct a more complete census of CT AGNs and gain a better understanding of obscured AGNs.

\acknowledgements
We thank L.\init Shao and R.\init Boissay-Malaquin for their helpful comments on the draft. R.\init W.\init P.\init thanks B.\init L.\init Cale for helpful discussions regarding data fitting with Python. C.\init R.\init acknowledges support from Fondecyt Iniciacion grant 11190831. P.\init G.\init B.\init acknowledges financial support from the STFC and the Czech Science Foundation project No. 19-05599Y. D.\init A.\init acknowledges funding through the European Union’s Horizon 2020 and Innovation programme under the Marie Sklodowska-Curie grant agreement no. 793499 (DUSTDEVILS). M.\init K.\init acknowledges support from NASA through ADAP award NNH16CT03C. F.\init R.\init acknowledges support from FONDECYT postdoctorado 3180506. This work is partially supported by Japan Society for the Promotion of Science (JSPS) KAKENHI (18K13584 and 20H01939; K.~Ichikawa). K.\init O.\init acknowledges support from the National Research Foundation of Korea (NRF-2020R1C1C1005462). M.\init S.\init acknowledges support by the Ministry of Education, Science and Technological Development of the Republic of Serbia through the contract no. 451-03-68/2020/14/20002 and the Science Fund of the Republic of Serbia, PROMIS 6060916, BOWIE.

\textit{Facilities:} Chandra, GALEX, NuSTAR, SDSS, Suzaku, Swift, WISE, XMM-Newton.
\software{odr \citep{boggs1990,virtanen2020}, pandas \citep{mckinney2010}, scipy \citep{virtanen2020}, numpy \citep{oliphant2006,walt2011,oliphant2015}, matplotlib \cite{hunter2007}, BXA \citep{buchner2014}}.

This publication makes use of data products from the \textit{Wide-field
Infrared Survey Explorer}, which is a joint project of the University
of California, Los Angeles, and the Jet Propulsion
Laboratory/California Institute of Technology, funded by the National
Aeronautics and Space Administration.
Funding for SDSS-III has been provided by the Alfred P. Sloan Foundation, the Participating Institutions, the National Science Foundation, and the U.S. Department of Energy Office of Science. The SDSS-III web site is \url{http://www.sdss3.org/}.

This research has made use of the NASA/IPAC Extragalactic Database (NED) which is operated by the Jet Propulsion Laboratory, California Institute of Technology, under contract with the National Aeronautics and Space Administration. 

\clearpage

\section{Appendix}
\label{sec:appendix}
\subsection{Exploring the Origin of the Scatter in the \lfour{}/\ltwo{} vs. \nh{} Correlation}

\begin{figure*}

\begin{minipage}[t]{0.46\linewidth}
    \centering
    \subfloat{\includegraphics[width=0.9\linewidth]{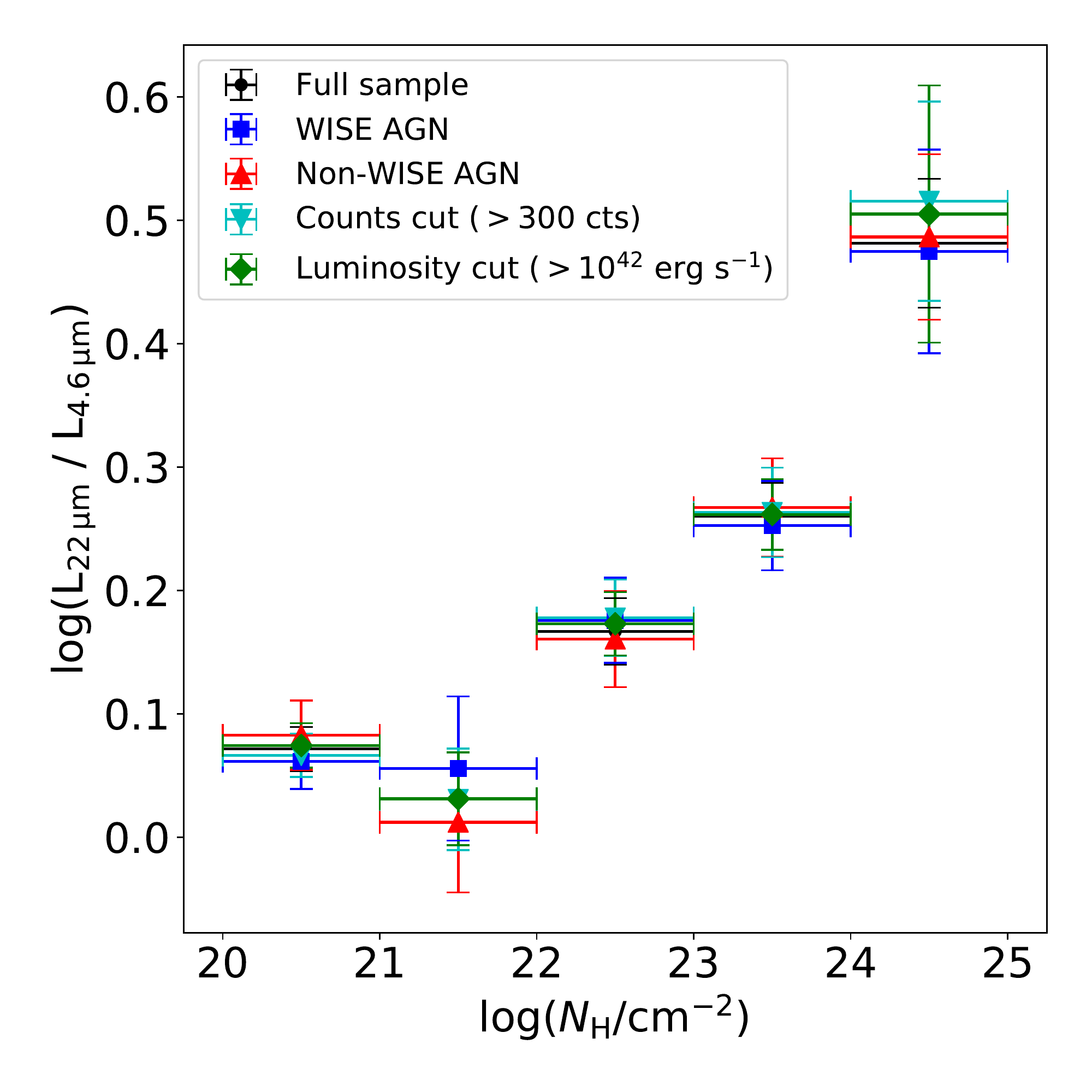}} \vspace{-6mm}\\
    \subfloat{\includegraphics[width=0.9\linewidth]{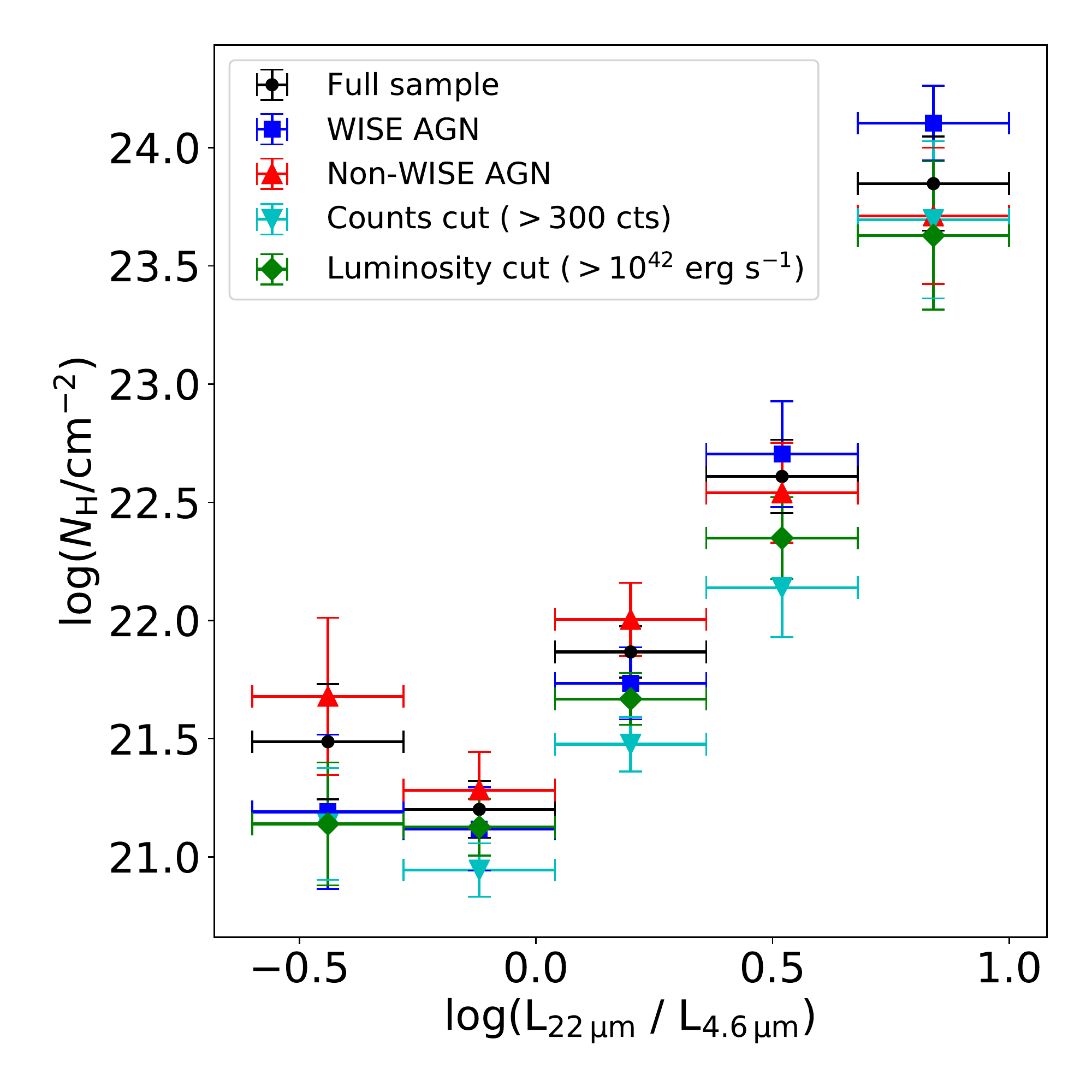}}
    \caption{We applied four different cuts to our full \swift{} sample to  explore the origin of the scatter observed in Figure~\ref{fig:W4W2fullsamplediagram}. Here we present two different comparisons of the derived values for \nh{} from \citet{ricci2017apjs} and the \lfour{}/\ltwo{} mid-IR ratios, where we have (top panel) binned by log(\nh{}) and (bottom panel) binned by the \lfour{}/\ltwo{} ratio. We observe more scatter when binning by the mid-IR ratio than we do when binning instead by \nh{}, but nonetheless the general trend is the same: we observe increasing mid-IR ratios of \lfour{}/\ltwo{} with increasing column density regardless of the sub-sample.}
    \label{fig:W4W2_subsamp_compare}
\end{minipage}
%\hspace{0.1cm}
\hfill
\begin{minipage}[t]{0.46\linewidth}
    \centering
    \subfloat{\includegraphics[width=0.9\linewidth]{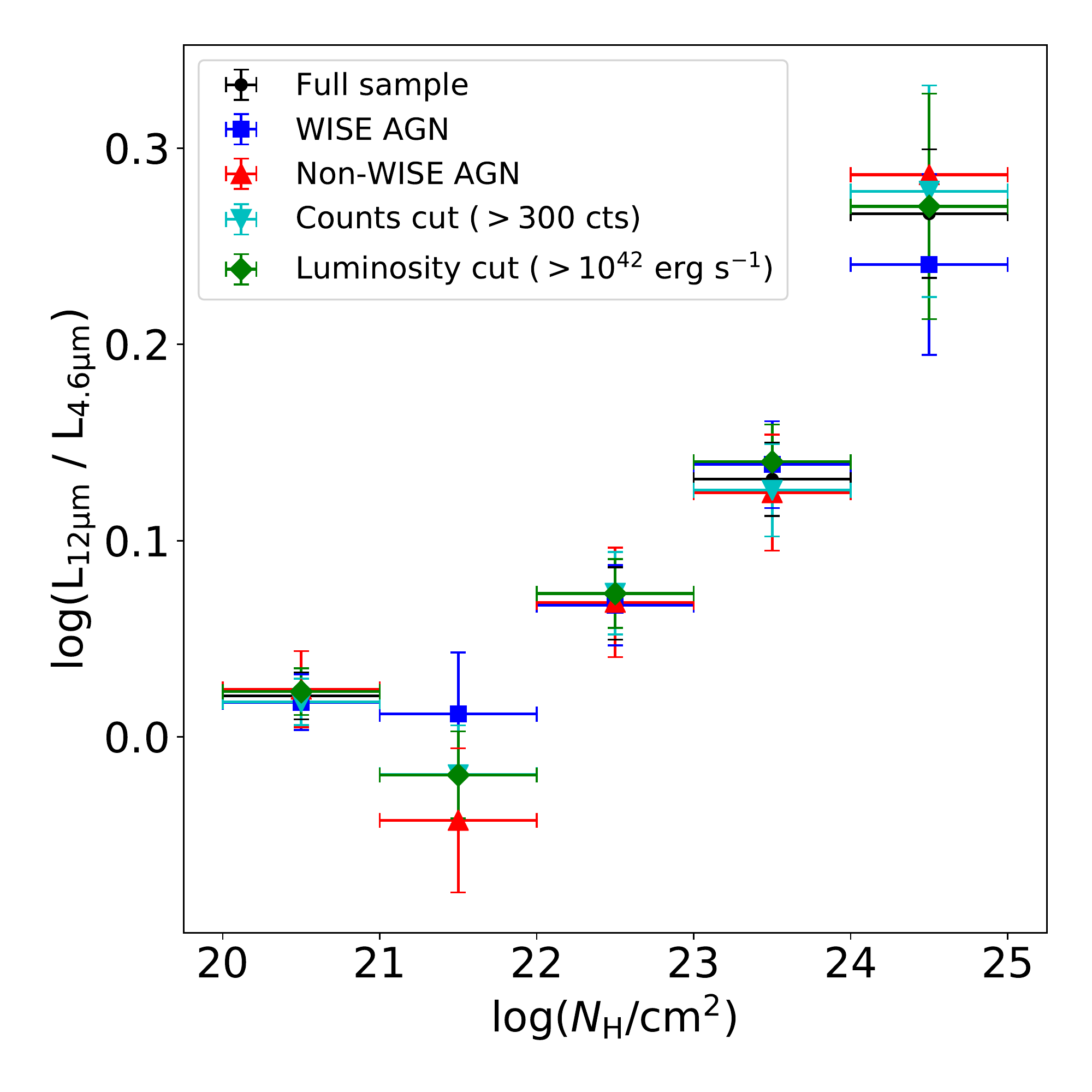}} \vspace{-6mm}\\
    \subfloat{\includegraphics[width=0.9\linewidth]{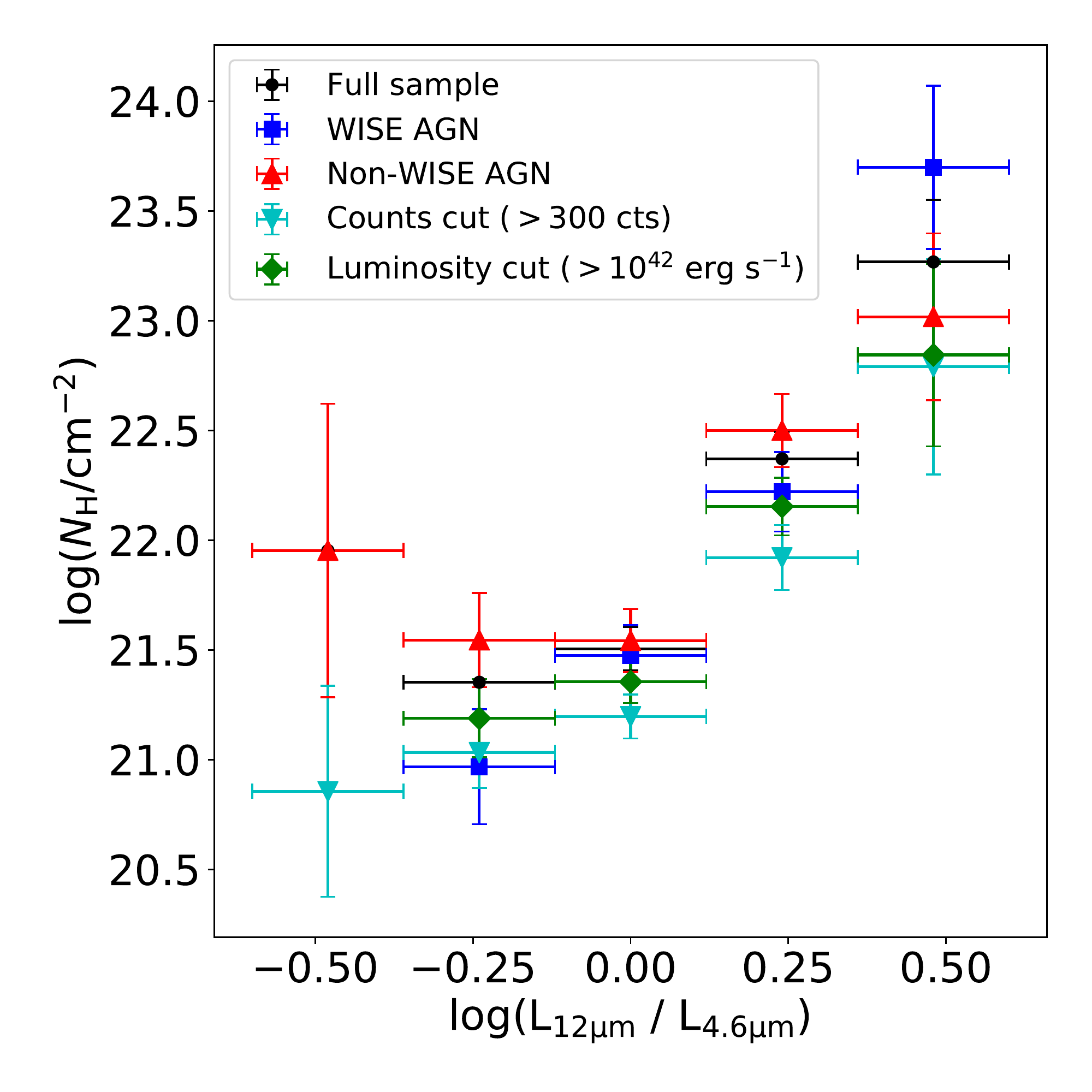}}
    \caption{Analogous to Figure~\ref{fig:W4W2_subsamp_compare} except here we examine the alternative mid-IR diagnostic ratio which depends upon \ltwel{} and \ltwo{}. We observe an increase in the mid-IR ratio of \ltwel{}/\ltwo{} with column density as was observed with \lfour{}/\ltwo{}, although with a large amount of scatter in the \nh{} values for the lowest mid-IR ratio bin. However, given that our focus is on the most heavily obscured sources ($>$ a few times $10^{23}$~cm$^{-2}$) this scatter is not a concern.}
    \label{fig:W3W2_subsamp_compare}
\end{minipage}

\end{figure*}

In the process of our analysis, we explored whether any quality cuts or selection cuts could be applied to the data to reduce the scatter observed in Panel A of both Figures~\ref{fig:W4W2fullsamplediagram} and~\ref{fig:W3W2fullsamplediagram}. Our parent sample was divided into four sub-samples as shown in Figure~\ref{fig:W4W2_subsamp_compare}: 
\begin{itemize}
    \item AGN-dominated systems with $W1$--$W2>0.8$ \citep[blue squares]{stern2012}.
    \item AGNs not selected using the aforementioned \WISE{} cut, i.e. $W1$--$W2<0.8$ (red triangles).
    \item AGNs with $>300$ spectral counts in the X-ray spectra, which provides a statistically significant number of counts to constrain \nh{} in the X-ray spectral fitting analysis \citep[inverted cyan triangles]{ricci2017apjs}.
    \item AGNs with observed 2--10~keV luminosities in excess of $10^{42}$~erg~s$^{-1}$ (green diamonds).
\end{itemize}
In Figures~\ref{fig:W4W2_subsamp_compare} and~\ref{fig:W3W2_subsamp_compare} we compare these sub-samples for the \lfour{}/\ltwo{} and \ltwel{}/\ltwo{} luminosity ratios, respectively, and how they correlate with \nh{}. The resulting mean values per bin for each different sub-sample are consistent with the values (the error bars represent the standard deviation of each subsample) originally found for the parent sample; interestingly, the observed correlation between \WISE{} color and column density holds for AGNs that satisfy the \citet{stern2012} criterion \emph{as well as} AGNs which do not satisfy that criterion. We observe a larger difference in the \lfour{}/\ltwo{} ratios when moving to higher column densities than for the \ltwel{}/\ltwo{} ratios. Additionally, as shown previously in Figure~\ref{fig:W3W2fullsamplediagram}, while there is a large amount of scatter in the lowest-\nh{} bin of the bottom panel of Figure~\ref{fig:W3W2_subsamp_compare}, this scatter is likely due to the low number of sources within that bin, and this still does not overlap the bin probing the highest obscuring columns. Due to the consistency between the results for the sub-samples and that found for the parent sample, we do not implement any of these cuts during our analysis. 

From Figure~\ref{fig:W4W2_subsamp_compare} and \ref{fig:W3W2_subsamp_compare}, it becomes  clear  that there a number of \WISE{} AGNs (W1-W2 > 0.8) within the \swift{} sample which exhibit extremely red mid-IR colors, with log(\lfour{}/\ltwo{}) $>0.5$, suggesting significant obscuration. We tabulate these \WISE{} AGNs in Table~\ref{apptab:WISEAGNs}. Indeed, the majority of these AGNs (18/23) are moderately to heavily obscured with log($\nhm{}$) $>23$, though there are a few exceptions, notably:

\begin{itemize}
    \item HS0328+0528, an unobscured Seyfert 1.
    \item IRAS05189-2524, a lightly obscured Seyfert 2.
    \item 2MASXJ09172716-6456271, an unobscured Seyfert 2.
    \item MCG-1-24-12, a lightly obscured Seyfert 2.
    \item NGC4253, an unobscured Seyfert 1.
\end{itemize}

\WISE{} selection based on the cut defined in \citet{stern2012} is not, however, a necessarily good method for selecting obscured over unobscured AGNs. As we discussed in Section~\ref{subsec:sfcontam}, 40/71 of AGNs with log($\nhm{}$) $>5.0\times10^{23}$ cm$^{-2}$ in the \swift{} sample studied here \emph{do not} meet a color cut of W1-W2 > 0.8, reinforcing our choice to not invoke such a cut on the parent sample.

\begin{table}[t!]
\caption{\ltwel{}/\ltwo{} Diagnostic Box Statistics}
\begin{center}
\hspace{-2.2cm}
\begin{tabular}{ccc}
\hline
\hline
\noalign{\smallskip}
\noalign{\smallskip}
$\rm{log}(\nhm{})$ & Completeness & Purity \\
\noalign{\smallskip}
\noalign{\smallskip}
\hline
\noalign{\smallskip}
$\geq24.0$      & $80.5^{+5.7}_{-6.7}$ & $59.4^{+6.5}_{-6.8}$  \\
$<24.0$         & $5.2^{+1.1}_{-1.0}$ & $40.6^{+6.8}_{-6.5}$  \\
$[23.0,\;24.0)$ & $16.3^{+3.7}_{-3.3}$ & $33.1^{+6.6}_{-6.2}$  \\
$[23.5,\;24.0)$ & $30.5^{+6.6}_{-6.1}$ & $29.4^{+6.4}_{-5.9}$  \\
$[23.0,\;23.5)$ & $4.6^{+3.2}_{-2.2}$ & $5.0^{+3.5}_{-2.4}$ \\
$[22.0,\;23.0)$ & $5.0^{+2.6}_{-1.9}$ & $8.8^{+4.3}_{-3.3}$ \\
$[22.5,\;23.0)$ & $5.3^{+3.7}_{-2.5}$ & $5.0^{+3.5}_{-2.4}$ \\
$[22.0,\;22.5)$ & $6.0^{+4.2}_{-2.9}$ & $5.0^{+3.5}_{-2.4}$ \\
$<22.0$         & $0.3^{+0.5}_{-0.2}$ & $1.3^{+2.1}_{-1.0}$ \\
\noalign{\smallskip}
\hline
\end{tabular}
\end{center}
\tablecomments{A breakdown of the statistics derived from the diagnostic box developed for the \ltwel{}/\ltwo{} luminosity ratio (defined by Equation~\ref{eq:w3w2eq} in Section~\ref{sec:analysis}) for various \nh{} bins and sub-bins. Columns 1-4: The same as Table~\ref{table:LXW3_populationstats}.}
\label{table:W3W2_populationstats}
\end{table}

\begin{table*}[h!]
\caption{Mid-IR AGNs from \citet{kilercieser2018} Selected via Equation~\ref{eq:w4w2eq}}
\begin{center}
%\begin{adjustbox}{max width=1.\textwidth}
\hspace{-2.6cm}
%\resizebox{\textwidth}{!}{
\scalebox{0.75}{
\begin{tabular}{cccccccccc}
\hline
\hline
\noalign{\smallskip}
\vspace{-3mm}\\
\noalign{\smallskip}
\textit{AKARI} I.D. & RA & Dec & $z$ & Selection & log$\left(\frac{L_{22\,\rm{\mu m}}}{L_{4.6\,\rm{\mu m}}}\right)$ &  log$\left(\frac{L_{\rm{X,\,Obs.}}}{L_{12\,\rm{\mu m}}}
\right)$ & Alternate & Obscuration & \nh{} \\
\noalign{\smallskip}
 & & & & Method & & & I.D. & Class & Ref. \\
\noalign{\smallskip}
\hline
\noalign{\smallskip}
0041533+402120 &  10.473 &  40.355 &   0.071 &  3, 4  &  0.62 &  -1.52 & {Mrk 957} & \dots & \dots \\
%\href{http://simbad.u-strasbg.fr/simbad/sim-id?Ident=\%401556378&Name=Mrk\%20\%20957&submit=submit}
0138053-125210 & 24.522  & -12.87  & 0.04  & 1, 2, 3, 4 & 0.47 & -2.47 &  {IRAS\,01356-1307} & Heavily Obscured & 1 \\
%\href{http://simbad.u-strasbg.fr/simbad/sim-coo?Coord=01+38+05.27+-12+52+11.9&CooFrame=FK5&CooEpoch=2000&CooEqui=2000&CooDefinedFrames=none&Radius=2&Radius.unit=arcmin&submit=submit+query&CoordList=}
%\citep{terashima2015apj}
0143576+022059 & 25.991  & 2.35    & 0.017 & 1, 2, 3, 4 & 0.35 & -2.29 &  {Mrk\,573 / UGC 1214} & Heavily Obscured & 2 \\
%\href{http://ned.ipac.caltech.edu/cgi-bin/nph-objsearch?objname=MRK\%20573&extend=no&out_csys=Equatorial&out_equinox=J2000.0&obj_sort=RA+or+Longitude&of=pre_text&zv_breaker=30000.0&list_limit=5&img_stamp=YES}%\citep{guainazzi2005}
0150029-072549 & 27.511  & -7.43   & 0.018 & 1, 2, 3, 4 & 0.64 & -1.44 &  {IRAS 01475-0740} & Unobscured or Heavily Obscured & 3, 4  \\ 
%\href{http://simbad.u-strasbg.fr/simbad/sim-coo?Coord=01+50+02.6+-07+25+48.50&NbIdent=1&Radius=1.0&Radius.unit=arcmin&CooFrame=FK5&CooEpoch=2000&CooEqui=2000&output.max=all&o.catall=on&output.mesdisp=N&Bibyear1=1983&Bibyear2=2005&Frame1=FK5&Frame2=FK4&Frame3=G&Equi1=2000.0&Equi2=1950.0&Equi3=2000.0&Epoch1=2000.0&Epoch2=1950.0&Epoch3=2000.0}
0222435-084305 & 35.682  & -8.719  & 0.045 & 1, 2, 3, 4 & 0.37 & -1.72 &  {NGC\,905} &  \dots & \dots \\
%\href{http://simbad.u-strasbg.fr/simbad/sim-coo?Coord=02+22+43.5+-08+43+08.50&NbIdent=1&Radius=1.0&Radius.unit=arcmin&CooFrame=FK5&CooEpoch=2000&CooEqui=2000&output.max=all&o.catall=on&output.mesdisp=N&Bibyear1=1983&Bibyear2=2005&Frame1=FK5&Frame2=FK4&Frame3=G&Equi1=2000.0&Equi2=1950.0&Equi3=2000.0&Epoch1=2000.0&Epoch2=1950.0&Epoch3=2000.0}
0325256-060832 &   51.356 &  -6.144 &   0.034 &  4  &  0.36 &  -1.45  & {Mrk 609} & Unobscured\footnote{Despite a lack of broad optical lines, Mrk 609 shows no sign of obscuration at X-ray wavelengths \citep{lamassa2014}} & 5 \\ %\citep{lamassa2014}
%\href{http://simbad.u-strasbg.fr/simbad/sim-coo?Coord=51.356++-6.144&CooFrame=FK5&CooEpoch=2000&CooEqui=2000&CooDefinedFrames=none&Radius=0.5&Radius.unit=arcmin&submit=submit+query&CoordList=}
0330407-030814 &   52.670 &  -3.138 &   0.021 &  4  &  0.52 &  -1.73  & {Mrk 612} & Moderately Obscured & 2 \\ % \citet{guainazzi2005}
%\href{http://simbad.u-strasbg.fr/simbad/sim-coo?Coord=52.670+-3.138&CooFrame=FK5&CooEpoch=2000&CooEqui=2000&CooDefinedFrames=none&Radius=0.5&Radius.unit=arcmin&submit=submit+query&CoordList=}
0452447-031256 & 73.186  & -3.216  & 0.016 & 1, 2, 3, 4 & 0.64 & -2.35 &  {IRAS\,04502-0317} &  \dots & \dots \\ % but Terashima should have it...
%\href{http://simbad.u-strasbg.fr/simbad/sim-id?Ident=\%40748400&Name=IRAS\%2004502-0317&submit=submit}
0453257+040341 &   73.357 &   4.062 &   0.029 &  1, 2, 3, 4  &  0.18 &  -1.78 & {2MASX J04532576+0403416} & Moderately or Heavily Obscured & 6, 7 \\ %\citet{marchesi2018,ricci2017apjs}
%\href{http://simbad.u-strasbg.fr/simbad/sim-id?Ident=\%40747153&Name=2MASX\%20J04532576\%2b0403416&submit=submit}
0518178-344536 &   79.575 & -34.761 &   0.066 &  4  &  0.49 &  -1.43 & {IRAS 05164-3448} & \dots & \dots \\
%\href{http://simbad.u-strasbg.fr/simbad/sim-coo?Coord=79.575+-34.761&CooFrame=FK5&CooEpoch=2000&CooEqui=2000&CooDefinedFrames=none&Radius=0.5&Radius.unit=arcmin&submit=submit+query&CoordList=}
0521013-252146 & 80.256  & -25.363 & 0.043 & 1, 2, 3, 4 & 0.36 & -1.44 &  {IRAS 05189-2524} &  Lightly obscured\footnote{Based upon current measurements, it is believed that IRAS 05189-2524 is currently lightly obscured, although it is possible that it may have been heavily obscured in the past \citep{teng2015apj}.} & 8, 7  \\ %\citep{teng2015apj,ricci2017apjs}
%\href{http://simbad.u-strasbg.fr/simbad/sim-id?Ident=\%40789602&Name=2MASX\%20J05210136-2521450&submit=submit}
0525179-460023 & 81.325  & -46.006 & 0.042 & 1, 2, 3, 4 & 0.42 & -1.56 &  {ESO\,253-3} & Moderately obscured & 9 \\ %\citep{asmus2015}
%\href{http://simbad.u-strasbg.fr/simbad/sim-id?Ident=\%402965741&Name=ESO\%20253-3&submit=submit}
0742406+651031 & 115.674 & 65.177  & 0.037 & 1, 2, 3, 4 & 0.49 & -1.58 &  {Mrk\,78} & Heavily or moderately obscured & 10, 7, 11  \\ %\citep{severgnini2012,ricci2017apjs,gilli2010}
%\href{http://simbad.u-strasbg.fr/simbad/sim-coo?Coord=07+42+41.7+\%2B65+10+37.40&NbIdent=1&Radius=1.0&Radius.unit=arcmin&CooFrame=FK5&CooEpoch=2000&CooEqui=2000&output.max=all&o.catall=on&output.mesdisp=N&Bibyear1=1983&Bibyear2=2005&Frame1=FK5&Frame2=FK4&Frame3=G&Equi1=2000.0&Equi2=1950.0&Equi3=2000.0&Epoch1=2000.0&Epoch2=1950.0&Epoch3=2000.0}
0759401+152314 & 119.917 & 15.387  & 0.016 & 4 & 0.41 & -2.74 &  {UGC\,4145} & \dots & \dots \\
%\href{http://simbad.u-strasbg.fr/simbad/sim-coo?Coord=07+59+40.1+\%2B15+23+12.50&NbIdent=1&Radius=1.0&Radius.unit=arcmin&CooFrame=FK5&CooEpoch=2000&CooEqui=2000&output.max=all&o.catall=on&output.mesdisp=N&Bibyear1=1983&Bibyear2=2005&Frame1=FK5&Frame2=FK4&Frame3=G&Equi1=2000.0&Equi2=1950.0&Equi3=2000.0&Epoch1=2000.0&Epoch2=1950.0&Epoch3=2000.0}
0807411+390015 &  121.921 &  39.004 &   0.023 &  4  &  0.83 &  -2.04 & {Mrk 622} & Heavily Obscured & 7 \\ %\citep{ricci2017apjs}
%\href{http://simbad.u-strasbg.fr/simbad/sim-coo?Coord=121.921+39.004&CooFrame=FK5&CooEpoch=2000&CooEqui=2000&CooDefinedFrames=none&Radius=0.5&Radius.unit=arcmin&submit=submit+query&CoordList=}
0810401+481233 &  122.668 &  48.209 &   0.077 &  3, 4  &  0.74 &  -2.11 & {2MASX J08104028+4812335} & Moderately Obscured & 1 \\ %\citep{terashima2015apj}
%\href{http://simbad.u-strasbg.fr/simbad/sim-coo?Coord=122.668+48.209&CooFrame=FK5&CooEpoch=2000&CooEqui=2000&CooDefinedFrames=none&Radius=0.5&Radius.unit=arcmin&submit=submit+query&CoordList=}
0904011+012733 & 136.004 & 1.458   & 0.054 & 4 & 0.6  & -2.54 &  {IRAS\,09014+0139} & \dots & \dots \\
%\href{http://simbad.u-strasbg.fr/simbad/sim-coo?Coord=09+04+01+\%2B01+27+29.10&NbIdent=1&Radius=1.0&Radius.unit=arcmin&CooFrame=FK5&CooEpoch=2000&CooEqui=2000&output.max=all&o.catall=on&output.mesdisp=N&Bibyear1=1983&Bibyear2=2005&Frame1=FK5&Frame2=FK4&Frame3=G&Equi1=2000.0&Equi2=1950.0&Equi3=2000.0&Epoch1=2000.0&Epoch2=1950.0&Epoch3=2000.0}
0935514+612112 & 143.965 & 61.353  & 0.039 & 1, 2, 3, 4 & 0.12 & -2.28 &  {UGC\,5101} & Heavily Obscured & 7, 12  \\
%\citep{ricci2017apjs,oda2017apj}
%\href{http://simbad.u-strasbg.fr/simbad/sim-id?Ident=\%40465776&Name=UGC\%20\%205101&submit=submit}
1010432+061157 & 152.681 & 6.2     & 0.098 & 1, 2, 3, 4 & 0.57 & -2.39 &  {2MASS\,J10104334+0612013} & \dots & \dots \\
%\href{http://simbad.u-strasbg.fr/simbad/sim-coo?Coord=10+10+43.3+\%2B06+12+01.40&NbIdent=1&Radius=1.0&Radius.unit=arcmin&CooFrame=FK5&CooEpoch=2000&CooEqui=2000&output.max=all&o.catall=on&output.mesdisp=N&Bibyear1=1983&Bibyear2=2005&Frame1=FK5&Frame2=FK4&Frame3=G&Equi1=2000.0&Equi2=1950.0&Equi3=2000.0&Epoch1=2000.0&Epoch2=1950.0&Epoch3=2000.0}
1021428+130655 &  155.428 &  13.115 &   0.076 &  4  &  0.92 &  -2.71 & {3XMM J102142.6+130654} & Unobscured & 13, 1 \\ %\citep{teng2010,terashima2015apj}
%\href{http://simbad.u-strasbg.fr/simbad/sim-id?Ident=\%401711649&Name=3XMM\%20J102142.6\%2b130654&submit=submit}
1034080+600152 & 158.536 & 60.031  & 0.051 & 1, 2, 3, 4 & 0.47 & -1.97 &  {Mrk\,34} & Heavily Obscured & 14 \\ %\citep{gandhi2014apj}
%\href{http://simbad.u-strasbg.fr/simbad/sim-coo?Coord=10+34+08.5+\%2B60+01+52.10&NbIdent=1&Radius=1.0&Radius.unit=arcmin&CooFrame=FK5&CooEpoch=2000&CooEqui=2000&output.max=all&o.catall=on&output.mesdisp=N&Bibyear1=1983&Bibyear2=2005&Frame1=FK5&Frame2=FK4&Frame3=G&Equi1=2000.0&Equi2=1950.0&Equi3=2000.0&Epoch1=2000.0&Epoch2=1950.0&Epoch3=2000.0}
1034381+393820 &  158.661 &  39.641 &   0.043 &  1, 2, 3, 4  &  0.15 &  -1.32 & {7C 103144.10+395402.00} & Unobscured & 15 \\
%\href{http://simbad.u-strasbg.fr/simbad/sim-coo?Coord=158.661+39.641&CooFrame=FK5&CooEpoch=2000&CooEqui=2000&CooDefinedFrames=none&Radius=0.5&Radius.unit=arcmin&submit=submit+query&CoordList=}
%\citep{gonzalezmartin2018}
1100183+100255 &  165.075 &  10.049 &   0.036 &  3, 4  &  0.68 &  -2.28 & {LEDA 200263} & Lightly Obscured &  16 \\ %\citep{dutta2018}
%\href{http://simbad.u-strasbg.fr/simbad/sim-id?Ident=\%401768892&Name=LEDA\%20\%20200263&submit=submit}
1219585-355743 &  184.996 & -35.960 &   0.058 &  1, 2, 3, 4  &  0.22 &  -2.74 & {6dFGS gJ121959.0-355735} & \dots & \dots \\
%\href{http://simbad.u-strasbg.fr/simbad/sim-coo?Coord=184.996+-35.960&CooFrame=FK5&CooEpoch=2000&CooEqui=2000&CooDefinedFrames=none&Radius=0.5&Radius.unit=arcmin&submit=submit+query&CoordList=}
1307059-234033 & 196.775 & -23.677 & 0.01  & 4 & 0.55 & -2.46 &  {NGC\,4968} & Heavily Obscured & 17 \\ 
%\href{http://simbad.u-strasbg.fr/simbad/sim-id?Ident=\%402059829&Name=NGC\%20\%204968&submit=submit}
% \citep{lamassa2019} %(Lamassa+2019, Lamassa+2017, Brightman+2011
1344421+555316 & 206.175 & 55.887  & 0.037 & 1, 2, 3, 4 & 0.89 & -1.97 &  {Mrk\,273} & Moderately obscured & 18, 8 \\
%\href{http://simbad.u-strasbg.fr/simbad/sim-id?Ident=\%40529836&Name=Mrk\%20\%20273&submit=submit}
%\citep{brightman2011,teng2015apj}
1347044+110626 & 206.768 & 11.106  & 0.023 & 1, 2, 3, 4 & 0.5  & -1.81 &  {Mrk\,1361} &  \dots & \dots \\
%\href{http://simbad.u-strasbg.fr/simbad/sim-id?Ident=\%402197137&Name=Mrk\%201361&submit=submit}
1356027+182222 &  209.012 &  18.372 &   0.051 &  1, 2, 3, 4  &  0.17 &  -2.29 & {Mrk 463} & Moderately Obscured & 19 \\ %\citep{bianchi2008}
%\href{http://simbad.u-strasbg.fr/simbad/sim-id?Ident=\%402215975&Name=Mrk\%20\%20463&submit=submit}
1550415-035314 & 237.673 & -3.888  & 0.03  & 1, 2, 3, 4 & 0.53 & -1.97 &  {IRAS\,15480-0344} & Heavily Obscured & 18, 10 \\
%\href{http://simbad.u-strasbg.fr/simbad/sim-id?Ident=\%402621890&Name=IRAS\%2015480-0344&submit=submit}
%\citep{brightman2011,severgnini2012}
1651053-012747 & 252.774 & -1.463  & 0.041 & 4 & 0.38 & -1.58 &  {LEDA\,1118057} &  \dots & \dots \\
%\href{http://simbad.u-strasbg.fr/simbad/sim-id?Ident=\%404057069&Name=2MFGC\%2013496&submit=submit}
1847441-630920 & 281.934 & -63.157 & 0.015 & 3, 4 & 0.52 & -2.43 &  {IC\,4769 } &  Heavily Obscured & 10 \\ %\citep{severgnini2012}
%\href{http://simbad.u-strasbg.fr/simbad/sim-id?Ident=\%403433055&Name=IC\%204769&submit=submit}
1931212-723919 & 292.839 & -72.656 & 0.062 & 1, 2, 3, 4 & 0.52 & -2.23 &  {`Superantennae'} & Unobscured or Heavily Obscured & 20, 8 \\
%\href{http://simbad.u-strasbg.fr/simbad/sim-id?Ident=\%403368053&Name=NAME\%20Superantennae&submit=submit}
%\citep{braito2009,teng2015apj}
2019593-523716 & 304.996 & -52.622 & 0.017 & 4 & 0.34 & -1.8  &  {IC\,4995} & Moderately or Heavily Obscured & 2, 21, 10 \\ %\citep{guainazzi2005,noguchi2009,severgnini2012}
%\href{http://simbad.u-strasbg.fr/simbad/sim-id?Ident=\%403467679&Name=IC\%204995&submit=submit}
2059127-520024 & 314.804 & -52.006 & 0.05  & 1, 2, 3, 4 & 0.33 & -2.48 &  {ESO\,235-26} &  \dots & \dots \\
%\href{http://simbad.u-strasbg.fr/simbad/sim-coo?Coord=20+59+12.95+-52+00+21.6&CooFrame=FK5&CooEpoch=2000&CooEqui=2000&CooDefinedFrames=none&Radius=2&Radius.unit=arcmin&submit=submit+query&CoordList=}
2316006+253326 & 349.003 & 25.557  & 0.027 & 1, 2, 3, 4 & 0.73 & -2.23 &  {IC\,5298} & Moderately Obscured & 22 \\ %\citep{torres-alba2018}
%\href{http://simbad.u-strasbg.fr/simbad/sim-coo?Coord=23+16+00.6+\%2B25+33+23.90&NbIdent=1&Radius=1.0&Radius.unit=arcmin&CooFrame=FK5&CooEpoch=2000&CooEqui=2000&output.max=all&o.catall=on&output.mesdisp=N&Bibyear1=1983&Bibyear2=2005&Frame1=FK5&Frame2=FK4&Frame3=G&Equi1=2000.0&Equi2=1950.0&Equi3=2000.0&Epoch1=2000.0&Epoch2=1950.0&Epoch3=2000.0}
2351135+201349 &  357.808 &  20.230 &   0.044 &  4  &  0.63 &  -1.37 & {MCG+03-60-031} & \dots & \dots \\
%\href{http://simbad.u-strasbg.fr/simbad/sim-coo?Coord=357.808+20.230&CooFrame=FK5&CooEpoch=2000&CooEqui=2000&CooDefinedFrames=none&Radius=0.5&Radius.unit=arcmin&submit=submit+query&CoordList=}
\noalign{\smallskip}
\hline
\end{tabular}
}
%\end{adjustbox}
\end{center}
\tablecomments{Column 1: \textit{AKARI} ID from \citet{kilercieser2018}. Columns 2-4: right ascension, declination, and redshift of the AGN. Column 5: selection method satisfied by this AGN, where 1, 2, 3, and 4 correspond to the selection criteria in \citet{stern2012}, \citet{jarrett2011}, \citet{assef2018}, and \citet{satyapal2018}. 6-7: log(\lfour{}/\ltwo{}) and log(\lxobs{}/\ltwel{}) luminosity ratios.  Column 8: alternative identifier for candidate CT AGN from other literature studies. Column 9: Obscuration class, assigned based on column density found in the literature. Class labels are adopted from Section~\ref{sec:analysis}. Column 9: Reference(s) for column densities (used to classify the objects here): (1) \citet{terashima2015apj}, (2) \citet{guainazzi2005}, (3) \citet{huang2011}, (4) \citet{brightman2008}, (5) \citet{lamassa2014}, (6) \citet{marchesi2018}, (7) \citet{ricci2017apjs}, (8) \citet{teng2015apj}, (9) \citet{asmus2015}, (10) \citet{severgnini2012}, (11) \citet{gilli2010}, (12) \citet{oda2017apj}, (13) \citet{teng2010}, (14) \citet{gandhi2014apj}, (15) \citet{gonzalezmartin2018}, (16) \citet{dutta2018}, (17) \citet{lamassa2019}, (18) \citet{brightman2011}, (19) \citet{bianchi2008}, (20) \citet{braito2009}, (21) \citet{noguchi2009}, (22) \citet{torres-alba2018}. }
\label{table:appendix_candCT_KE}
\end{table*}

%(1) \citet{torres-alba2018}, (2) \citet{terashima2015apj}, (3) \citet{guainazzi2005}, (4) \citet{huang2011}, (5) \citet{brightman2008}, (6) \citet{teng2015apj}, (7) \citet{ricci2017apjs}, (8) \citet{asmus2015}, (9) \citet{severgnini2012}, (10) \citet{gilli2010}, (11) \citet{oda2017apj}, (12) \citet{nucita2017}, (13) \citet{gandhi2014apj}, (14) \citet{lamassa2009}, (15) \citet{kolourdis2016}, (16) \cite{lamassa2019}, (17) \citet{brightman2011}, (18) \citet{evans2008}, (19) \citet{braito2009}, (20) \citet{noguchi2009}.

\begin{table*}[h!]
\caption{Low-Redshift Candidate Heavily Obscured AGNs from \textit{XMM}-XXL North Selected via Equation~\ref{eq:w4w2eq}}
\begin{center}
\begin{tabular}{cccccccccccc}
\hline
\hline
\noalign{\smallskip}
\vspace{-3mm}\\
\noalign{\smallskip}
UXID & $\alpha$ & $\delta$ & z & $F_{2-10~\rm{keV}}$ & log(\nh{}/cm$^{-2}$) & log(\lxobs{}/\ltwel{}) & log(\lfour{}/\ltwo{}) \\
\noalign{\smallskip}
 & & & & (erg cm$^{-2}$ s$^{-1}$) &  & & \\
\noalign{\smallskip}
\hline
\noalign{\smallskip}
%N\_42\_10 & 34.7902 & -5.42083 & 0.0987 & $1.14\times10^{-14}$   & $20.87^{+2.63}_{-0.52}$ & -1.74 & 0.62\\
%N\_97\_13 & 35.7049 & -5.56736 & 0.0687 & $1.16\times10^{-14}$   & $21.20^{+2.97}_{-0.80}$ & -1.71 & 0.24\\
%N\_45\_48 & 36.4573 & -4.00696 & 0.0433 & $8.83\times10^{-14}$   & $23.39^{+0.16}_{-0.21}$ & -1.30 & 0.66\\
%N\_113\_19 & 37.3037 & -5.18955 & 0.0736 &  $4.61\times10^{-14}$ & $24.16^{+1.26}_{-0.30}$ & -1.71 & 0.23\\
N\_96\_28 & 33.3335  & -3.48755 & 0.0754 & $9.8\times10^{-14}$  & $23.12^{+0.13}_{-0.14}$ & -1.6 & 0.27 \\
N\_42\_10 & 34.7902  & -5.42083 & 0.0987 & $1.14\times10^{-14}$ & $20.87^{+2.63}_{-0.52}$ & -2.11 & 0.71 \\
N\_97\_13 & 35.7049  & -5.56736 & 0.0687 & $1.16\times10^{-14}$ & $21.2 ^{+2.97}_{-0.8 }$ & -2.09 & 0.33 \\
N\_35\_14 & 36.0105  & -5.22831 & 0.0843 & $1.48\times10^{-14}$ & $23.44^{+0.13}_{-0.13}$ & -1.48 & 0.37 \\
N\_45\_48 & 36.4573  & -4.00696 & 0.0433 & $8.83\times10^{-14}$ & $23.39^{+0.16}_{-0.21}$ & -1.68 & 0.75 \\
N\_30\_7 & 36.5185   & -4.99197 & 0.0539 & $2.95\times10^{-14}$ & $24.59^{+0.93}_{-4.05}$ & -2.11 & 0.11 \\
N\_113\_19 & 37.3037 & -5.18955 & 0.0736 & $4.61\times10^{-14}$ & $24.16^{+1.26}_{-0.3 }$ & -2.09 & 0.32 \\
N\_105\_14 & 37.5322 & -4.53268 & 0.0444 & $3.83\times10^{-14}$ & $20.73^{+0.48}_{-0.48}$ & -1.47 & 0.11 \\
\noalign{\smallskip}
\hline
\end{tabular}
\end{center}
\tablecomments{All eight low-redshift ($z<0.1$) candidate heavily obscured or CT AGNs within the \textit{XMM}-XXL-N field pulled from the \citet{menzel2016} and \citet{zliu2016} catalogs selected using the diagnostic box defined by Equation~\ref{eq:w4w2eq}. Column 1: unique X-ray identification string used by both \citet{menzel2016} and \citet{zliu2016}. Column 2-3: X-ray source right ascension (RA) and declination (Dec) values (uncorrected for systematic offsets) given in degrees. Column 3: redshift. Column 4: Hard X-ray 2--10~keV fluxes. Column 5: $50^{\rm{th}}$-percentile logarithmic line-of-sight column density derived by \cite{zliu2016}. Error bounds are calculated using the $84^{\rm{th}}$ and $16^{\rm{th}}$-percentile column density values. Column 6-7: values for the log(\lxobs{}/\ltwel{}) and log(\lfour{}/\ltwo{}) luminosity ratios.}
\label{apptab:CTcandidates}
\end{table*}

\begin{table*}[h!]
\caption{\WISE{}-Selected \swift{} AGNs with  log(\lfour{}/\ltwo{}) > 0.5}
\begin{center}
\begin{tabular}{ccccccccc}
\hline
\hline
\noalign{\smallskip}
\vspace{-3mm}\\
\noalign{\smallskip}
\textit{SWIFT} I.D. & RA & Dec & z & log$\left(\frac{L_{22\,\rm{\mu m}}}{L_{4.6\,\rm{\mu m}}}\right)$ & log$\left(\frac{L_{\rm{X,\,Obs.}}}{L_{12\,\rm{\mu m}}}\right)$ & Alternate I.D. & log(\nh{}/cm$^{-2}$) \\
\noalign{\smallskip}
\noalign{\smallskip}
\hline
\noalign{\smallskip}
SWIFTJ0107.7-1137B &  16.9152 & -11.65320 & 0.0475 & 0.51 & -1.10 & 2MASXJ01073963-1139117  & $23.58^{+0.28}_{-0.18}$ \\
 SWIFTJ0122.8+5003 &  20.6435 &  50.05500 & 0.0204 & 0.76 & -1.44 & MCG+8-3-18              & $24.24^{+0.34}_{-0.15}$ \\
 SWIFTJ0308.2-2258 &  47.0449 & -22.96080 & 0.0360 & 0.89 & -1.78 & NGC1229                 & $24.94^{+1.06}_{-0.45}$ \\
 SWIFTJ0331.3+0538 &  52.7174 &   5.64040 & 0.0460 & 0.62 & -0.35 & HS0328+0528             & $20.00^{+0.00}_{-0.00}$ \\
 SWIFTJ0521.0-2522 &  80.2561 & -25.36260 & 0.0426 & 0.51 & -1.74 & IRAS05189-2524          & $22.92^{+0.04}_{-0.03}$ \\
 SWIFTJ0615.8+7101 &  93.9015 &  71.03750 & 0.0135 & 0.82 & -1.20 & Mrk3                    & $24.07^{+0.06}_{-0.04}$ \\
 SWIFTJ0656.4-4921 & 104.0498 & -49.33060 & 0.0410 & 0.58 & -1.79 & LEDA478026              & $24.03^{+0.30}_{-0.10}$ \\
 SWIFTJ0743.0+6513 & 115.6739 &  65.17710 & 0.0371 & 0.66 & -1.82 & Mrk78                   & $24.11^{+0.08}_{-0.12}$ \\
 SWIFTJ0804.2+0507 & 121.0244 &   5.11380 & 0.0135 & 0.77 & -1.09 & Mrk1210                 & $23.40^{+0.06}_{-0.08}$ \\
 SWIFTJ0843.5+3551 & 130.9375 &  35.82830 & 0.0540 & 0.58 & -1.09 & CASG218                 & $23.61^{+0.24}_{-0.21}$ \\
 SWIFTJ0917.2-6457 & 139.3634 & -64.94090 & 0.0860 & 0.55 & -0.07 & 2MASXJ09172716-6456271  & $21.41^{+0.33}_{-0.71}$ \\
 SWIFTJ0920.8-0805 & 140.1927 &  -8.05610 & 0.0196 & 0.54 & -0.28 & MCG-1-24-12             & $22.81^{+0.05}_{-0.03}$ \\
 SWIFTJ1214.3+2933 & 183.5741 &  29.52860 & 0.0632 & 0.53 & -0.94 & Was49b                  & $23.41^{+0.17}_{-0.11}$ \\
 SWIFTJ1218.5+2952 & 184.6105 &  29.81290 & 0.0129 & 0.56 & -0.81 & NGC4253                 & $20.32^{+0.08}_{-0.17}$ \\
 SWIFTJ1225.8+1240 & 186.4448 &  12.66210 & 0.0084 & 0.61 & -0.77 & NGC4388                 & $23.52^{+0.02}_{-0.01}$ \\
 SWIFTJ1238.6+0928 & 189.6810 &   9.46017 & 0.0829 & 0.64 & -0.93 & SDSSJ123843.43+092736.6 & $23.60^{+0.09}_{-0.07}$ \\
 SWIFTJ1322.2-1641 & 200.6019 & -16.72860 & 0.0165 & 0.63 & -1.86 & MCG-3-34-64             & $23.80^{+0.02}_{-0.02}$ \\
 SWIFTJ1717.1-6249 & 259.2478 & -62.82060 & 0.0037 & 0.51 & -1.16 & NGC6300                 & $23.31^{+0.02}_{-0.03}$ \\
 SWIFTJ1800.3+6637 & 270.0304 &  66.61510 & 0.0265 & 0.94 & -1.62 & NGC6552                 & $24.05^{+0.35}_{-0.22}$ \\
 SWIFTJ2052.0-5704 & 313.0098 & -57.06880 & 0.0114 & 0.73 & -1.45 & IC5063                  & $23.56^{+0.07}_{-0.01}$ \\
 SWIFTJ2207.3+1013 & 331.7582 &  10.23340 & 0.0267 & 0.71 & -1.70 & UGC11910                & $24.41^{+0.07}_{-0.07}$ \\
 SWIFTJ2304.9+1220 & 346.2361 &  12.32290 & 0.0079 & 0.82 & -2.65 & NGC7479                 & $24.16^{+0.12}_{-0.13}$ \\
 SWIFTJ2343.9+0537 & 355.9982 &   5.64000 & 0.0560 & 0.52 & -0.62 & LEDA3092070             & $23.26^{+0.19}_{-0.08}$ \\
\noalign{\smallskip}
\hline
\end{tabular}
\end{center}
\tablecomments{\WISE{}-selected \swift{} AGN \citep[$W1-W2>0.8$,][]{stern2012} which exhibit mid-IR ratios of log(\lfour{}/\ltwo{}) $>0.5$. Column 1: \textit{SWIFT} I.D. number. Column 2-4: Right ascension, declination, and redshift. Column 5-6: log(\lxobs{}/\ltwel{}) and log(\lfour{}/\ltwo{}) luminosity ratios. Column 7: Alternative I.D. drawn from \citet{ricci2017apjs}. Column 8: Column  density derived in \citet{ricci2017apjs}.}
\label{apptab:WISEAGNs}
\end{table*}

\begin{figure}
    \centering
    \includegraphics[width=0.9\linewidth]{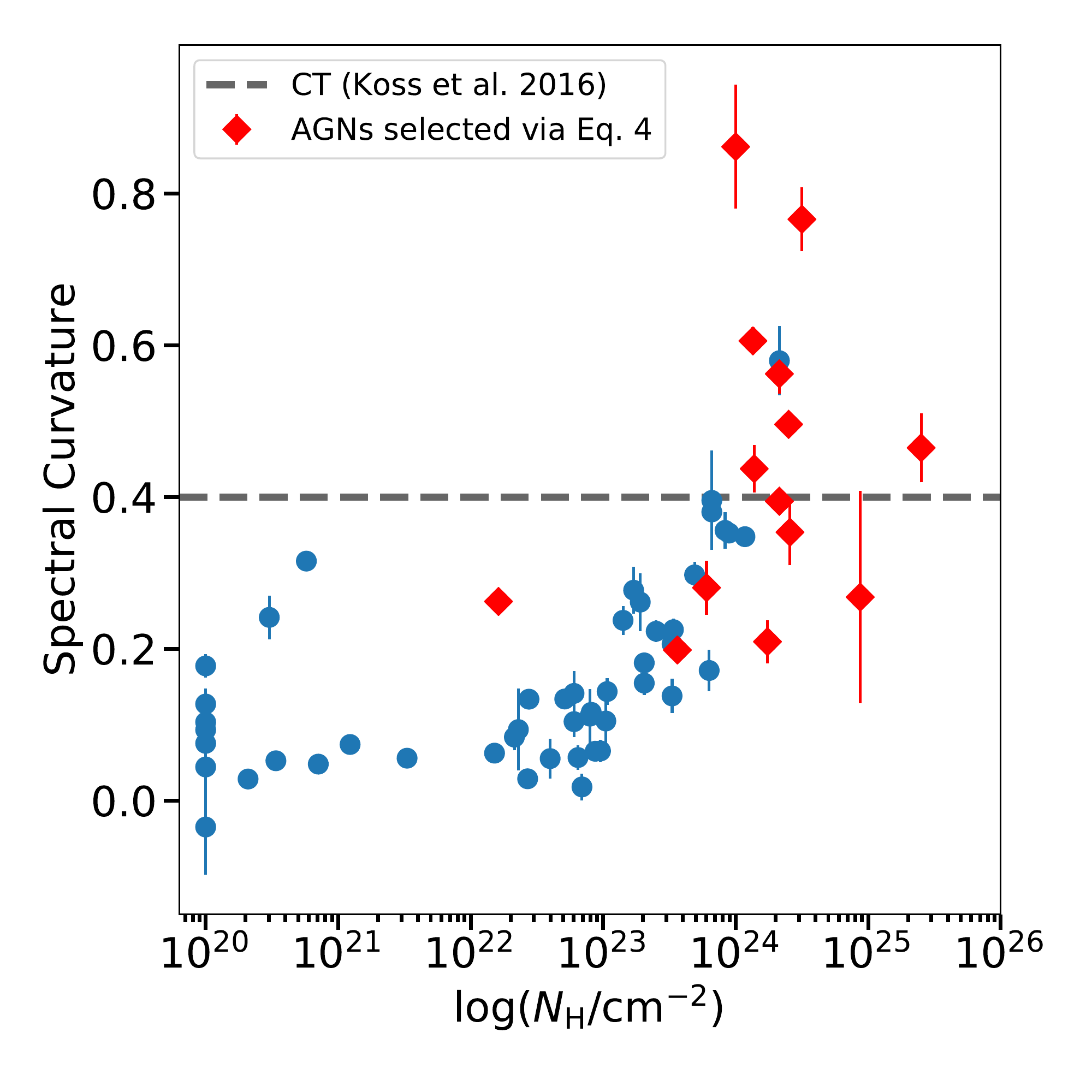}
    \caption{A comparison between the X-ray and mid-IR selection criteria introduced in this work and the spectral curvature method from \citet{kossapj2016}. The spectral curvature of each point was calculated using observations from \emph{NuSTAR}, whereas the column densities come from \citet{ricci2017apjs}. Only AGNs with spectral curvature errors of $<0.2$ are included in this plot. The dashed grey line represents a spectral curvature value of 0.4, above which an AGN is considered to be CT. Red points are AGNs selected using the diagnostic box defined in Equation~\ref{eq:w4w2eq}, whereas blue points are not selected via Equation~\ref{eq:w4w2eq}. The spectral curvature method and the diagnostic defined in this work find several of the same sources, and in fact these two approaches also find CT AGNs missed by one another: a few CT AGNs fall below the 0.4 spectral curvature cutoff, yet the diagnostic presented here selects them, meanwhile the spectral curvature method recovered a CT AGN missed by our new diagnostic. There is one outlier which is selected by Equation~\ref{eq:w4w2eq} and exhibits a relatively high spectral curvature, yet it possesses a column density of only $\sim10^{22}$ cm$^{-2}$. This source, NGC 1365, is a well known variable absorber and has gone through massive absorption transitions. Recent high signal-to-noise observations have shown that the column density remains substantial, above $10^{23}$ cm$^{-2}$ (e.g. \citealp{risaliti2009apj,risaliti2009mnras,risaliti2009apjl, maiolino2010,walton2010,brenneman2013}) and occasionally increases to the extent of becoming CT \citep{risaliti2005}. It is possible that the mid-IR emission is tracing a higher absorption period seen in past observations.}
    \label{fig:app_sccomp}
\end{figure}

%72 SWIFT/BAT AGNs from Koss+2016 that would be selected based on the spectral curvature method. 54/72 are present within the sample in our work (based on the listed SWIFT IDs); of these 54, 20 have column densities of log(NH)>24. Comparing this to the X-ray and mid-IR selection criteria in our work, our criteria finds 18/20 of the CT AGNs found via the spectral curvature method. We note that the two CT AGNs found via the spectral curvature method that are not identified with our color criteria failed to meet the LX/L12 criteria (SWIFTJ0615.8+7101) or the LW4/LW2 criteria (SWIFTJ0111.4-3808). 

\clearpage 

\bibliographystyle{aasjournal}
\bibliography{references}

\end{document}